\documentclass[twocolumn]{aastex62}            
\usepackage{graphicx}
\usepackage{graphicx,times,epsf}             
\usepackage{longtable}
\usepackage{natbib}
\usepackage{amssymb}
\usepackage{amsmath}
\usepackage{psfig}
\usepackage{txfonts}
\usepackage[T1]{fontenc}

\shortauthors{Yang et al.}
\begin{document}

\title{A Catalog of Short Period Spectroscopic and Eclipsing Binaries Identified from the LAMOST $\&$ PTF Surveys}
\email{jfliu@nao.cas.cn}
\email{shansusu@nao.cas.cn}

\author[0000-0002-6039-8212]{fan yang}
\affil{National Astronomical Observatories, Chinese Academy of Sciences, 20A
Datun Road, Chaoyang District, Beijing 100101, China\\}
\affil{IPAC, Caltech, KS 314-6, Pasadena, CA 91125, USA\\}     
\affil{School of Astronomy and Space Science, University of Chinese Academy of Sciences,
Beijing 100049, China\\}

\author[0000-0002-8559-0067]{richard j. long}
\affil{Department of Astronomy, Tsinghua University, Beijing 100084, China\\}
\affil{Jodrell Bank Centre for Astrophysics, Department of Physics and Astronomy, The University of Manchester, Oxford Road, Manchester M13 9PL, UK\\}

\author[0000-0002-5744-2016]{su-su shan}
\affil{National Astronomical Observatories, Chinese Academy of Sciences, 20A
Datun Road, Chaoyang District, Beijing 100101, China\\}
\affil{School of Astronomy and Space Science, University of Chinese Academy of Sciences,
Beijing 100049, China\\}

\author[0000-0002-6434-7201] {bo zhang}
\affil{Department of Astronomy, Beijing Normal University, Beijing 100875,
People's Republic of China}

\author {rui guo}
\affil{National Astronomical Observatories, Chinese Academy of Sciences, 20A
     Datun Road, Chaoyang District, Beijing 100101, China\\}
\affil{School of Astronomy and Space Science, University of Chinese Academy of Sciences,
  Beijing 100049, China\\}

\author[0000-0002-4740-3857]{yu bai}
\affil{National Astronomical Observatories, Chinese Academy of Sciences, 20A
     Datun Road, Chaoyang District, Beijing 100101, China\\}
     
\author{zhongrui bai}
\affil{National Astronomical Observatories, Chinese Academy of Sciences, 20A
     Datun Road, Chaoyang District, Beijing 100101, China\\}

\author {kai-ming cui}
\affil{National Astronomical Observatories, Chinese Academy of Sciences, 20A
     Datun Road, Chaoyang District, Beijing 100101, China\\}
\affil{School of Astronomy and Space Science, University of Chinese Academy of Sciences,
Beijing 100049, China\\}
\author[0000-0003-3116-5038]{song wang}
\affil{National Astronomical Observatories, Chinese Academy of Sciences, 20A
     Datun Road, Chaoyang District, Beijing 100101, China\\}
\author {ji-feng liu}
\affil{National Astronomical Observatories, Chinese Academy of Sciences, 20A
     Datun Road, Chaoyang District, Beijing 100101, China\\}
\affil{School of Astronomy and Space Science, University of Chinese Academy of Sciences,
Beijing 100049, China\\}
\begin{abstract}
  Binaries play key roles in determining stellar parameters and exploring stellar evolution models. We build a catalog of 88 eclipsing binaries with spectroscopic information, taking advantage of observations from both the Large Sky Area Multi-Object fiber Spectroscopic Telescope (LAMOST) and the Palomar Transient Factory (PTF) surveys. A software pipeline is constructed to identify binary candidates by examining their light curves. The orbital periods of binaries are derived from the Lomb-Scargle method. The key distinguishing features of eclipsing binaries are recognized by a new filter \textit{Flat Test}. We classify the eclipsing binaries by applying Fourier analysis on the light curves. Among all the binary stars, 13 binaries are identified as eclipsing binaries for the first time. The catalog contains information: position, primary eclipsing magnitude and time, eclipsing depth, the number of photometry and radial velocity observations, largest radial velocity difference, binary type, the effective temperature of observable star $T_{\rm eff}$, and surface gravity of observable star log \emph{g}. The false-positive probability is calculated by using both a Monte Carlo simulation and real data from the SDSS Stripe 82 Standard Catalog. The binaries in the catalog are mostly with a period of less than one day. The period distribution shows a 0.22-day cut-off which is consistent with the low probability of an eclipsing binary rotating with such a period.
\end{abstract}

\section{Introduction}           
\label{sect:intro}

Binary systems, common in stars, play important roles in, for example, determining stellar parameters, understanding the evolution of stars, measuring distances and tracing black hole candidates. Binaries usually appear as a single point in images because of the distances involved. Some binary stars can be identified from optical spectra \citep{1905LicOB...3..136C,2006ApJS..162..207A} and are termed spectroscopic binaries (SB). Among all the binary stars, a very small fraction, closely aligned to the line of sight, are revealed as eclipsing \citep{2006Ap&SS.304....5G}, the so-called eclipsing binaries (EB). They allow us to derive the fundamental stellar parameters such as mass, radius, and luminosity \cite[see][]{1967Natur.213...21D,1991A&ARv...3...91A,1971ApJ...166..605W,2012AJ....144...73W,2005ApJ...628..426P,2016ApJS..227...29P}. Thus, binary systems serve as testbeds for stellar evolution theories \cite[see][]{2007A&A...472L..17C,2002ApJ...567.1140T}.

Eclipsing binaries with spectroscopic information are a particularly powerful tool. They are used as a standard candle to determine the distances to the Magellanic Clouds, the Andromeda Galaxy (M31) and the Triangulum Galaxy (M33) \cite[see][]{1997vsar.conf..309P,2000AAS...196.1001P,1998ApJ...509L..21G,2001ApJ...559..260W,2006ApJ...652..313B}. Some EBs with both optical and X-ray observations are identified as black hole candidates \cite[see][]{ 2013Natur.503..500L,2015Natur.528..108L,2014Natur.505..378C}.

After two centuries of observations, the EB sample has now reached several hundred thousand. The beginning of EB studies can be attributed to John Goodricke in 1783. \citet{1802RSPT...92..477H} was the first to use the term "binary star" for double stars. The term "spectroscopic binary" comes from work by Edward (1889) and was used explicitly by \citet{1890AN....123..289V}. As a result of multiple research activities\citep[e.g.,][]{1938BHarO.909...14S,1943BAN.....9..337F,1946PASP...58..249B}, the size of the EB sample increased to several thousand. In the past twenty years, due to the results from photometric microlensing surveys (e.g., EROS, Grison et al. 1995; MACHO, Alcock, et al. 1997; OGLE, Udalski, et al. 1998, Wyrzykowski et al. 2004; NSVS, Otero et al. 2004), the sample size grew rapidly to about 15,000. 
\citet{2006MNRAS.368.1311P} then almost doubled the number by identifying and classifying 11,076 eclipsing binaries from the All Sky Automated Survey (ASAS). The work from \citet{2006MNRAS.368.1311P} is particularly important because of their use of Fourier transform in analyzing light curves. In this work, our analysis uses the same Fourier technique. Kepler space mission identified 2165 EBs with precise light curves \citep{2011AJ....141...83P,2012AJ....143..123M}. \citet{2016AcA....66..405S} published a list of over 450,000 eclipsing binary candidates toward the Galactic bulge in OGLE survey.

EB catalogs with spectroscopic information are challenging to produce since the multiple spectroscopic observations required are highly time-consuming activities. The first large sample of spectroscopic binaries was made by \citet{1905LicOB...3..136C}. The ninth edition of this sample was published in \citep{2004A&A...424..727P}, extending the number of spectroscopic binaries to 2386. Catalogs of solar-type spectroscopic stars \citep[e.g.,][]{1975BAAS....7..268A,1991A&A...248..485D,2010ApJS..190....1R} increased the total number of spectroscopic binaries by a few hundred. Among all the binary star observations, the EBs with spectroscopic information, providing the most comprehensive constraints on the binary parameters, are still rare.

This paper describes a catalog of spectroscopic and eclipsing binaries from the Large Sky Area Multi-Observation fiber Spectroscopic Telescope (LAMOST) survey and the Palomar Transient Factory (PTF) data. The combination of a very large number of spectra and photometry light curves enables us to systematically study binary systems \citep{2014ApJ...788L..37G,2016MNRAS.461.2747K}. Spectra tell us more about the observable stars, while the light curves are used to determine the orbital properties of the transit system. We utilize "\textit{Flat Test}" method on the light curves to remove contamination by variable stars such as Cepheid and RR Lyrae stars.

Our catalog contains 88 spectroscopic and eclipsing binaries, among which 13 EBs are newly identified. The sources in the catalog all have good quality light curves. The catalog contains orbital parameters, as well as the stellar parameters of the observable companion. Some parameters like effective temperature of observable star $T_{\rm eff}$, surface gravity of observable star log \emph{g} help to constrain the evolution of binary stars. The binaries in the catalog show the 0.22 day cut-off in period distribution which supports the rareness of binary stars rotating with such a period.

The paper is organized as follows. In section 2, we briefly describe the survey data. The binaries selection algorithm and the catalog are presented in section 3. In section 4, we assess our catalog and discuss the 0.22 day cut-off in the orbital period distribution. In section 5, we give a summary.

\section{The Data used from LAMOST and PTF}
\label{sect:Obs}

The catalog is established from the survey products of LAMOST and PTF. We use LAMOST to derive stellar parameters and the radial velocity of the observable star. PTF data is used to obtain the orbital period and morphological features of the light curve.

\subsection{LAMOST Data}

LAMOST (Large Sky Area Multi-Object fiber Spectroscopic Telescope) employs the Guo Shou Jing Telescope which is a 4 meter aperture Schmidt telescope with a 20 square degree field of view. Taking 4000 spectra per exposure, the limiting magnitude is r = 17.8 at resolution R $\sim$ 1800 \citep [see overview:][]{2012RAA....12..723Z,2012RAA....12.1197C}. The survey contains two parts: the LAMOST Extra Galactic Survey (LEGAS) and the LAMOST Experiment for Galactic Understanding and Exploration (LEGUE) survey \citep{Deng2012}.

Spectroscopic binaries can be classified if the spectral lines from both companions are visible (a double-lined spectroscopic binary) or if the spectral lines from only one star are detectable (a single-lined spectroscopic binary) \citep[e.g.,][]{2010AJ....140..184M,2011AJ....141..200M}. The identification of double-lined spectroscopic binaries in LAMOST spectra is generally difficult due to limited spectral resolution (R$\sim1800$, equivalent to $\sim$150 km s$^{-1}$) unless the radial velocities or the spectral types of the component stars are significantly different \citep[for example, white dwarf-main sequence binaries,][]{ 2014A&A...570A.107R}. The forthcoming LAMOST-II medium-resolution spectroscopic survey \citep[MRS,][]{zhang2019} with resolution increased to R$\sim7500$ will assist in identification. Work is in progress on searching for single-lined and double-lined spectroscopic binaries by modeling spectra (Zhang et al. in prep.). In this research, we identify binary stars via eclipsing and significant radial velocity differences of the same component in multi epoch observations. All the systems identified are regarded as single-lined spectroscopic binaries.

\citet{2017MNRAS.467.1890X} released the fourth edition of the value-added catalog of the LAMOST Spectroscopic Survey of the Galactic Anticenter (LSS-GAC DR4). LSS-GAC is a major part of the LEGUE project. The catalog contains more than 3 million stars down to a limiting magnitude of r $\sim$ 17.8 mag, centered on the Galactic anticentre ($\left|b\right|$ $\leq$ 30$^{\circ}$, 150 $\leq$ $\mathit{l}$ $\leq$ 210$^{\circ}$). The stellar parameters are derived by a pipeline based on minimum chi-square template matching \citep[LSP3;][]{2015MNRAS.448..822X}. The radial velocity (RV, center shifts of the spectra lines) and stellar atmospheric parameters (effective temperature $T_{\rm eff}$, surface gravity log \emph{g}, metallicity) are the free parameters of the template fitting.

\begin{figure}[!h]
\centering
\includegraphics[width=2.5in]{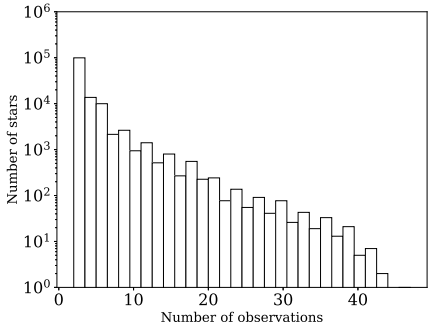}
\caption{The number of observations of LAMOST targets.}
\label{rv}
\end{figure}

We chose sources with multi-epoch observations in LSS-GAC DR4, requiring $\Delta$RV$_{max}$ (the difference between the highest and smallest RV values of the same component of a given target) to be 2 times larger than the largest radial velocity error. The typical spectral signal-to-noise ratio (SNR) is > 10. The radial velocity precision is about a few km s$^{-1}$ \citep{2017MNRAS.467.1890X}.
This procedure helped us select 128,833 objects with 421,436 observations. The number of observations are shown in Figure \ref{rv}. Those large RV variation stars could be binary systems, or variable stars such as RR Lyrae, etc. We then cross matched these candidates with the PTF data.

\subsection{PTF Data}
The Palomar Transient Factory (PTF) employs an 8 square degree camera installed on the 48 inch Samuel Oschin telescope. A 60-inch telescope is used for following-up observations. The survey aims at time-domain events with cadences from 90 seconds to a few days \citep{2009PASP..121.1395L,2009PASP..121.1334R}. With exposure time of 60 seconds, the PTF 5$\sigma$ limiting magnitude is $m_{g^{'}}$ $\sim$ 21.3 and $m_{R} \sim$ 20.6. The PTF collaboration released more than 8 billion light curve tables prior to Sep.1 2016.

\begin{figure}[!h]
\centering
\includegraphics[width=2.5in]{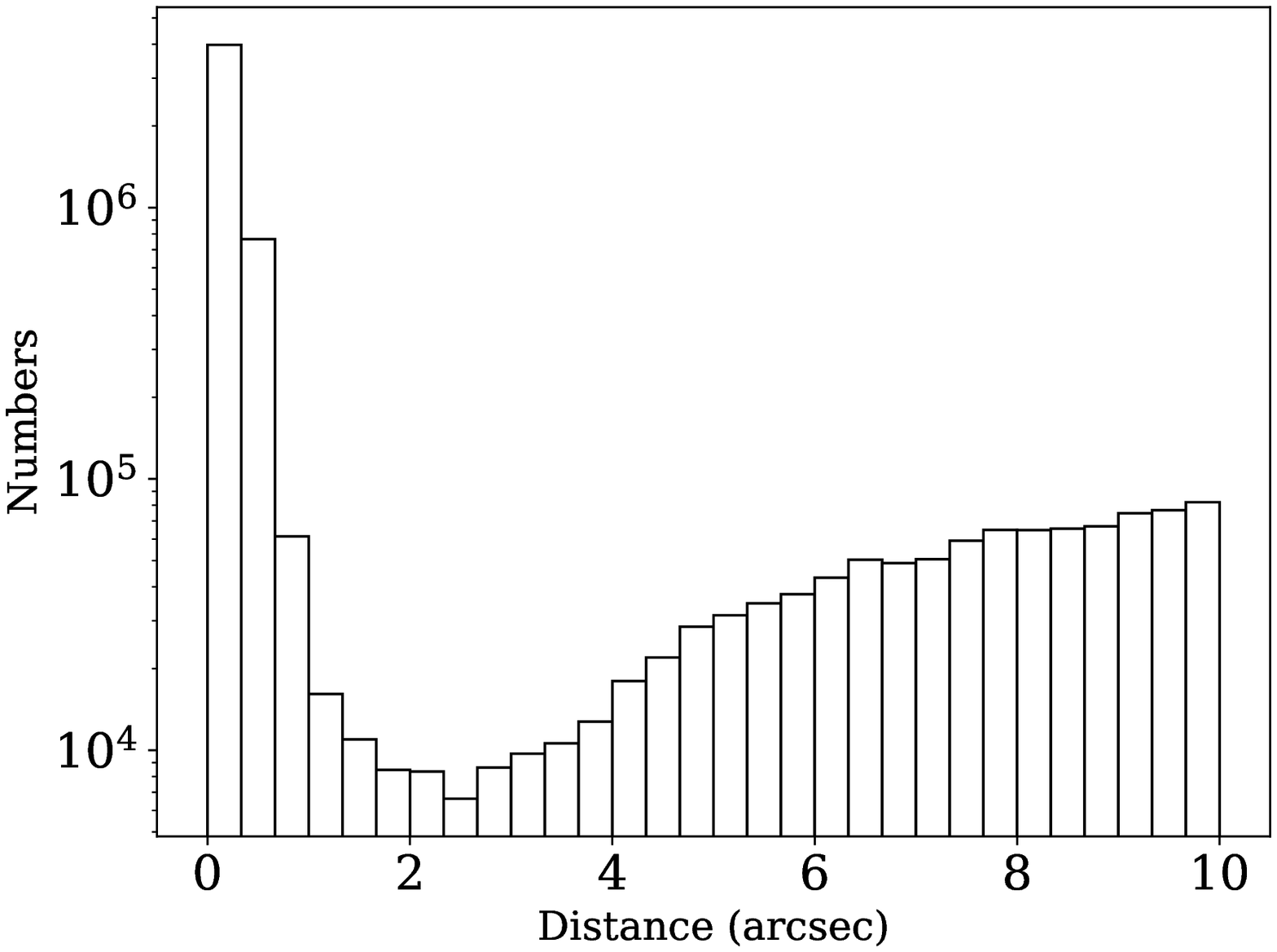}
\includegraphics[width=2.5in]{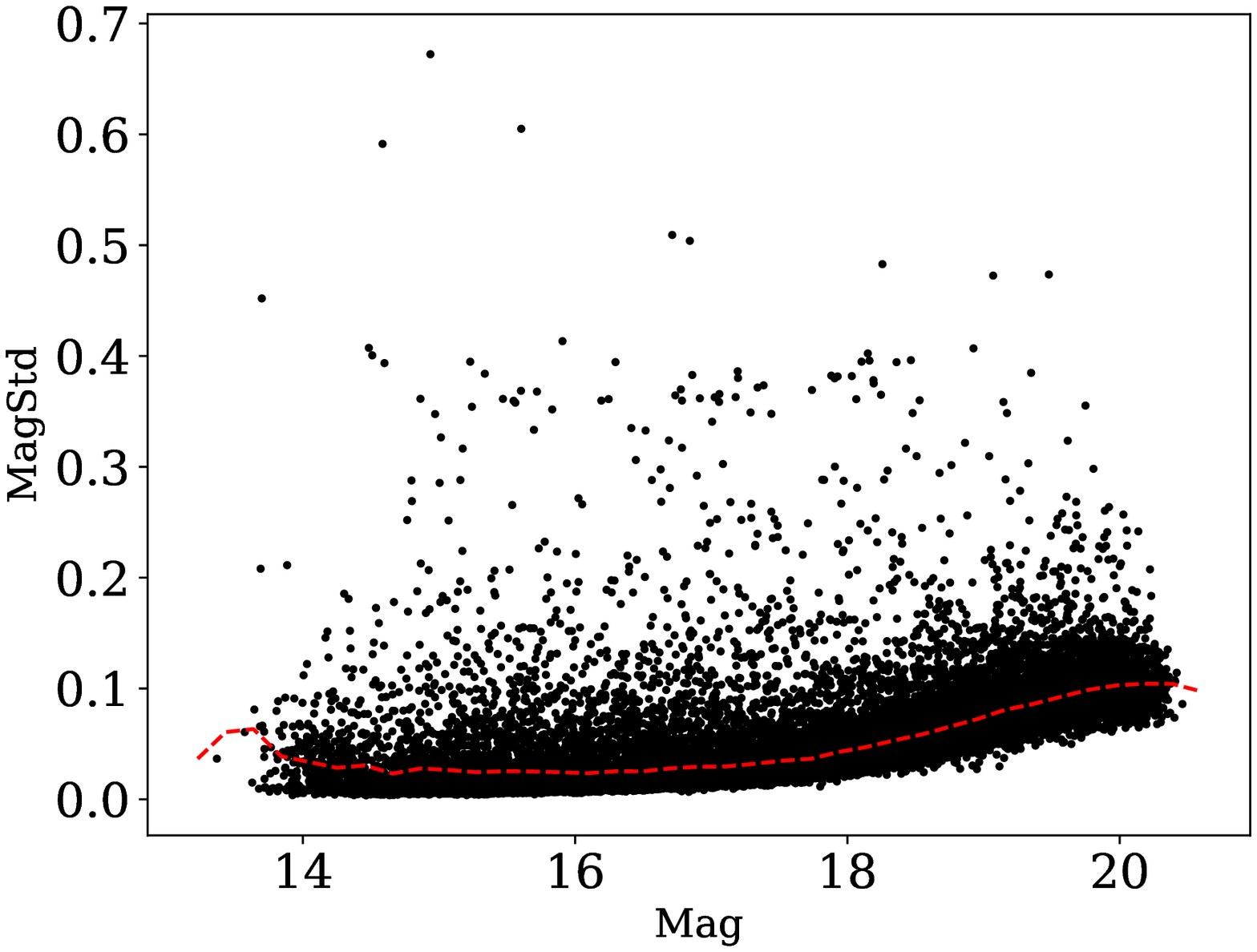}
\caption{PTF Astrometry and Photometry precision. Upper panel: the astrometry precision. The x-axis is the position difference for the same source between SDSS and PTF. The y-axis is the matched number at certain position difference. Lower panel: the photometry precision. The x-axis is magnitude while the y-axis is the mag standard deviation. The red line indicates the median value of the magnitude error. 
    }
\label{accu}
\end{figure}

To select a radius for cross-matching, we randomly chose 100,000 non-variable stars in the SDSS stripe 82 Standard Catalog \citep{2007AJ....134..973I}. The position difference between SDSS and PTF is used to validate the PTF astrometry precision (see Figure \ref{accu}). The sharp decrease in cross-match radius occurs at 2 arcsecs. Considering the uncertainty of LAMOST astrometry, we took 3 arcsecs to be the cross-match radius between the PTF and LAMOST data. 

We also randomly chose 10$^5$ sources in the SDSS Stripe 82 Standard Catalog to evaluate the photometry precision of PTF multi-epoch observations. We used the magnitude standard deviation of the standard stars as the statistical error of PTF photometry (see Figure \ref{accu}). The magnitude standard deviation is less than 0.1 in the magnitude range 14 to 20 which is the magnitude range for our work.

The sources selected from the LAMOST data were matched to the PTF light curve data base\footnote{\url{https://irsa.ipac.caltech.edu/cgi-bin/Gator/nph-dd}}. 39,179 stars with 2,784,673 observation frames, requiring the goodflag=1, were retrieved both in g$^{'}$ and R band. The tag "goodflag" ensures the quality of the photometry in PTF. We analyzed these sources in our binary systems identification pipeline.

\begin{figure}[!h]
\centering
\includegraphics[width=2.5in]{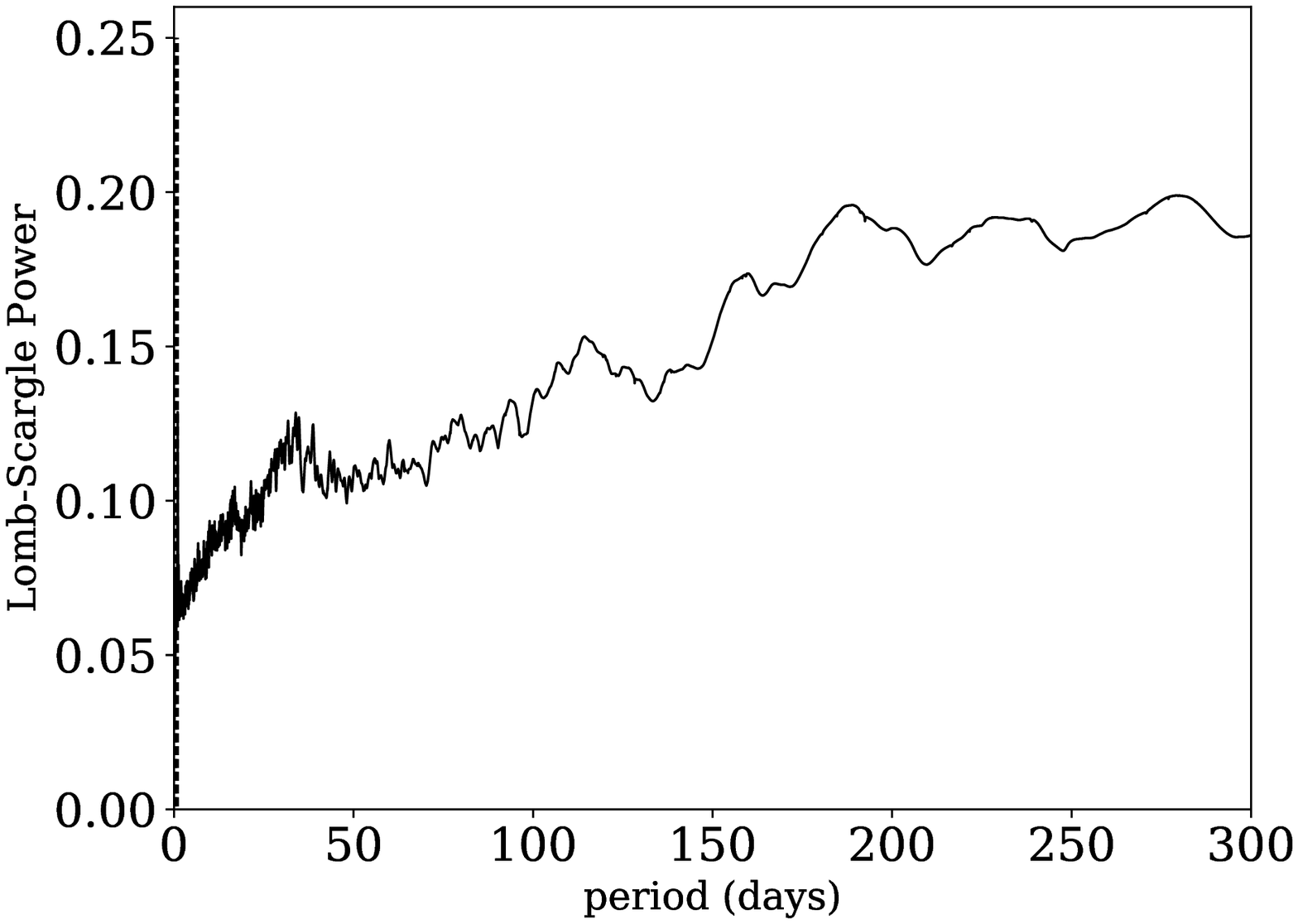}
\includegraphics[width=2.5in]{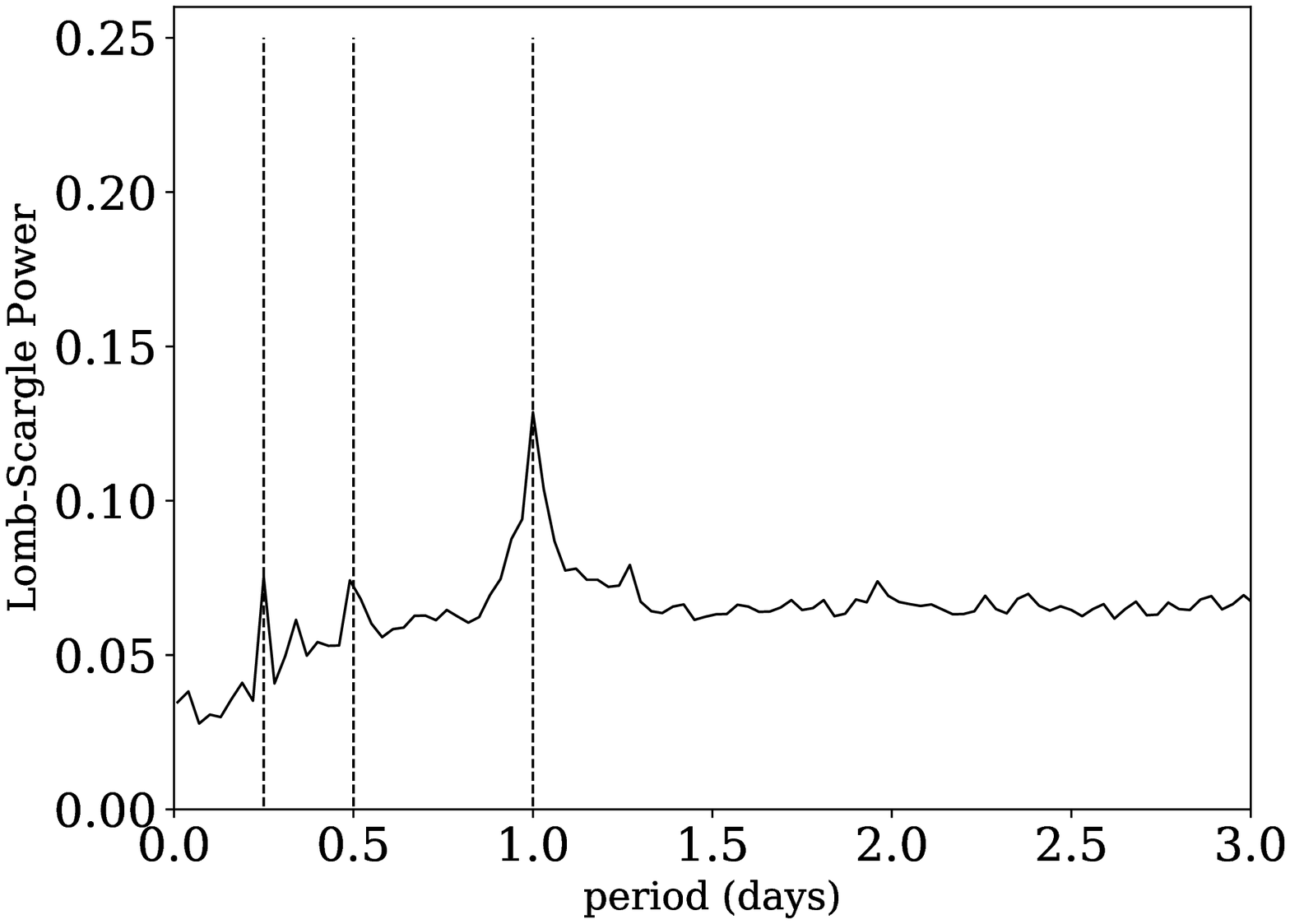}
\caption{The period power distribution. Upper: power distribution of all the periods. Lower: power distribution of the period from 0 to 3 day. The vertical lines show the period gathering at 1 day, and the harmonic periods at 0.25 and 0.5 day.}
\label{all}
\end{figure}

\section{The Catalog of spectroscopic and eclipsing binaries}
\label{sect:Obs}
\subsection{Finding Binaries}
Our method of finding binaries is built into our software pipeline and involves two steps. The first is the Lomb-Scargle periodogram \citep{1976Ap&SS..39..447L,2015ApJ...812...18V,astroML} which searches for the orbital period. The second is the \textit{Flat Test} developed by us which removes artifacts as well as contamination from other variable stars. In addition, every light curve is subject to a visual inspection to assure the purity of the sample.

\begin{figure}[!h]
  \centering
  \includegraphics[width=3.7in]{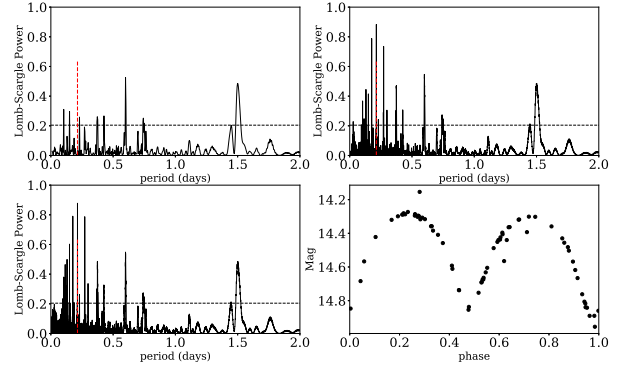}
  \caption{The period finding process, using LPSEB2 as an example. The upper left-hand panel has the largest bin size. The upper right panel has the bin size reduced by a factor of 10 and the lower-left by a further factor 10. The lower right panel shows the light curve folded with the optimized period. The upper left panel has the same bin size as used in Figures \ref{ebex} and \ref{full}. The red dashed line indicates the optimized period which is half of the orbital period. The intensity of the real peak is weak in the upper left panel. Reducing the bin sizes reshapes the peak intensity with the real peak becoming stronger and surrounded by a clump of peaks. The final reduction in bin size in the lower-left panel does not significantly change the intensity distribution.}
  \label{LS}
\end{figure}

We applied the Lomb-Scargle periodogram for PTF data to search for significant periods. The significant period is identified as the period at peak intensity if the peak is higher than a level predicted by false positive probability. The peak here we applied is an optimized peak using the method from \citet{astroML}. The optimization is a two-step grid peak searching. First, it searches a broad grid for some candidates and then zooms in using a narrower grid to find the best estimation of the period peak. This method avoids possible false peak detections due to binning. The real peak is sometimes smoothed to be smaller when the binning size is too large. A solitary peak is treated as improbable if a clump of peaks appears when reducing the binning size. The period finding process is illustrated in Figure \ref{LS}.

The false positive probability we applied contains two parts. The first is the false alarm probability reported by \citet{1976Ap&SS..39..447L}. We required the evidence level of 99.99$\%$. The second is based on the statistical properties of the periodogram. The threshold is defined as the 3$\sigma$ excess to the median value of the power spectral density. The period must simultaneously satisfy both parts of the false probability to be accepted as a significant period.

Then we put all the sources' periodograms together (in Figure \ref{all}) to remove artifact signals. The orbital periods significantly cluster at 1 day and its harmonic periods such as 0.25 or 0.5 day, because of the alternation between night and day for the ground-based telescope. We rejected sources if the most evident power is within 0.05 day of 1 day. We found a long timescale power excess (in Figure \ref{all}), which was due to the difficulty with long-time calibration. We accept only orbital periods between 0.01 to 10 day. After following these procedures, we were left with 275 variable sources with sinusoidal periods.

\begin{figure}[!h]
  \centering
  \includegraphics[width=2.5in]{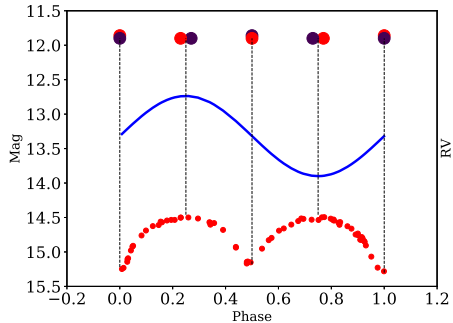}
  \caption{Upper: orbital morphology of EB systems, the circles indicate two components. Middle: The simulated radial velocity of the observable star. Lower: the light curve of the EB system, LPSEB71. The presented RV curve corresponds to the component marked with the darker circle.}
  \label{orb}
\end{figure}

These variable sources were then analyzed using the \textit{Flat Test} method we have developed which is a filter for investigating the characteristics of a light curve. For the close binary eclipsing systems we were interested in, the systems should be highly orbitally circularized \citep{2002MNRAS.329..897H}. In most cases, the binary systems have a shorter phase at eclipse than out of eclipse. Outliers appear among a few overcontact systems like some RR Centauri \citep {2005PASJ...57..983Y}, which might be missed by our detection. Over an orbit, tangential velocity can be taken as a sinusoidal function over time. Radial velocity which is orthogonal to the tangential velocity is more commonly used since it can be calculated from the spectra of the binary. A typical light curve and the associated radial velocity curve are shown in Figure \ref{orb}.

The light curves of eclipsing binaries should be flatter at the top than at the bottom, which enables them to be distinguished from the RRc Lyrae variable stars\citep{book2009}. RRc Lyrae stars sometimes have a symmetrical, sinusoidal light curve, unlike Cepheid and RRa, RRb Lyrae. Moreover, the RR Lyrae stars have a typical radial velocity amplitude of about 30km/s to 50km/s, which locates at a similar place as EBs in the RV distribution. The \textit{Flat Test} helps us remove such periodic light curves which are not binary systems.

\begin{figure}[!h]
  \centering
  
  \includegraphics[width=2.5in]{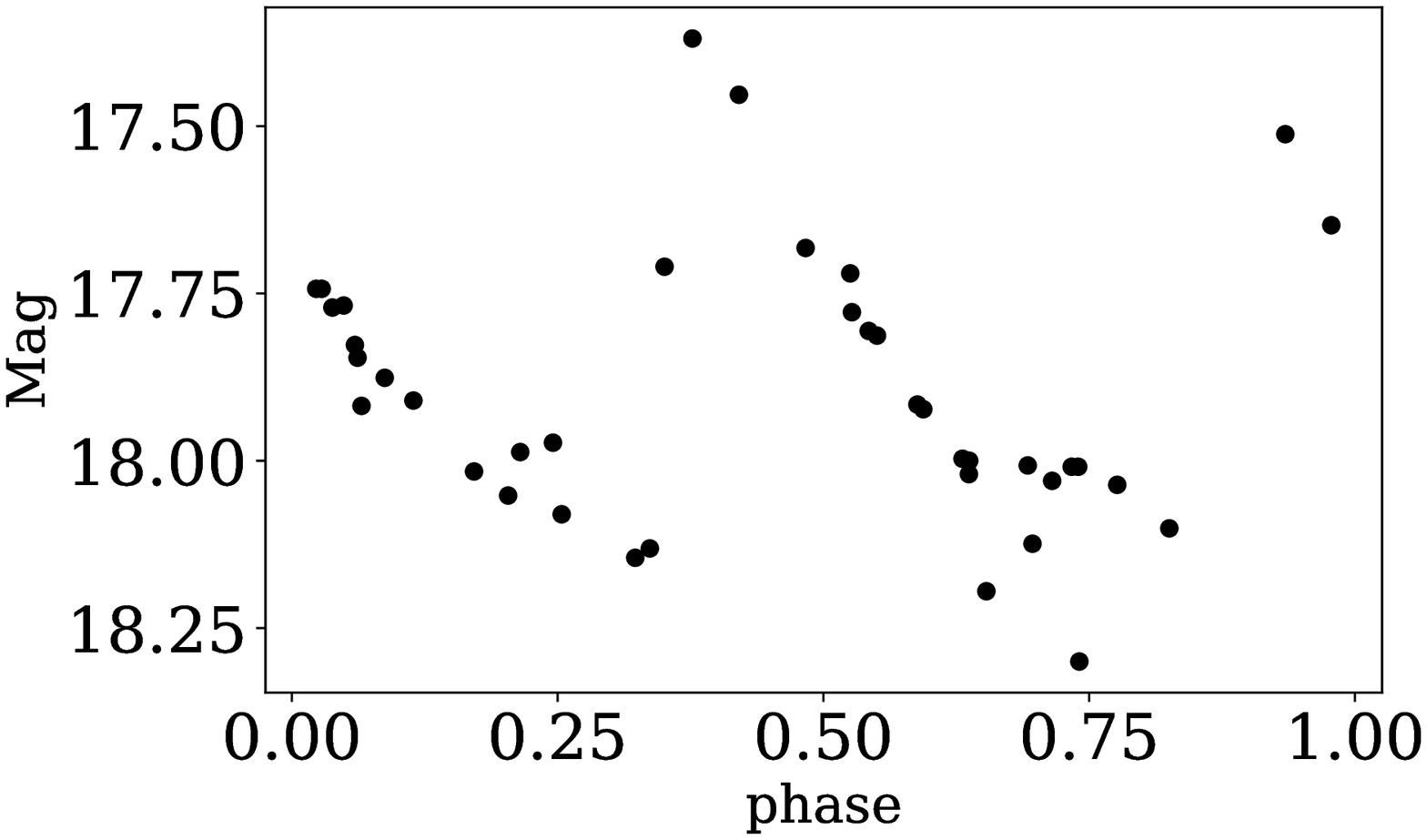}
  \includegraphics[width=2.5in]{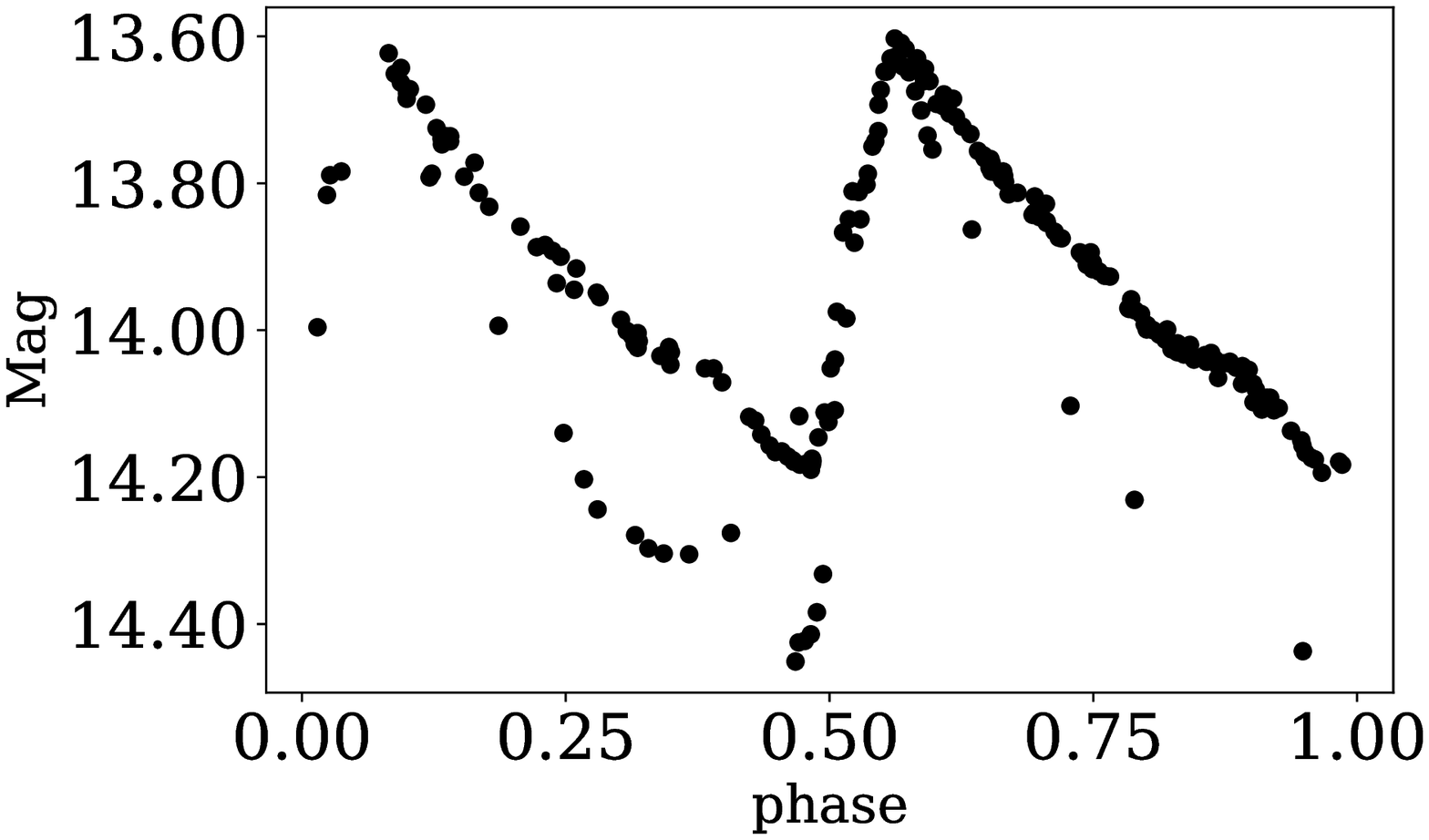}
  \includegraphics[width=2.5in]{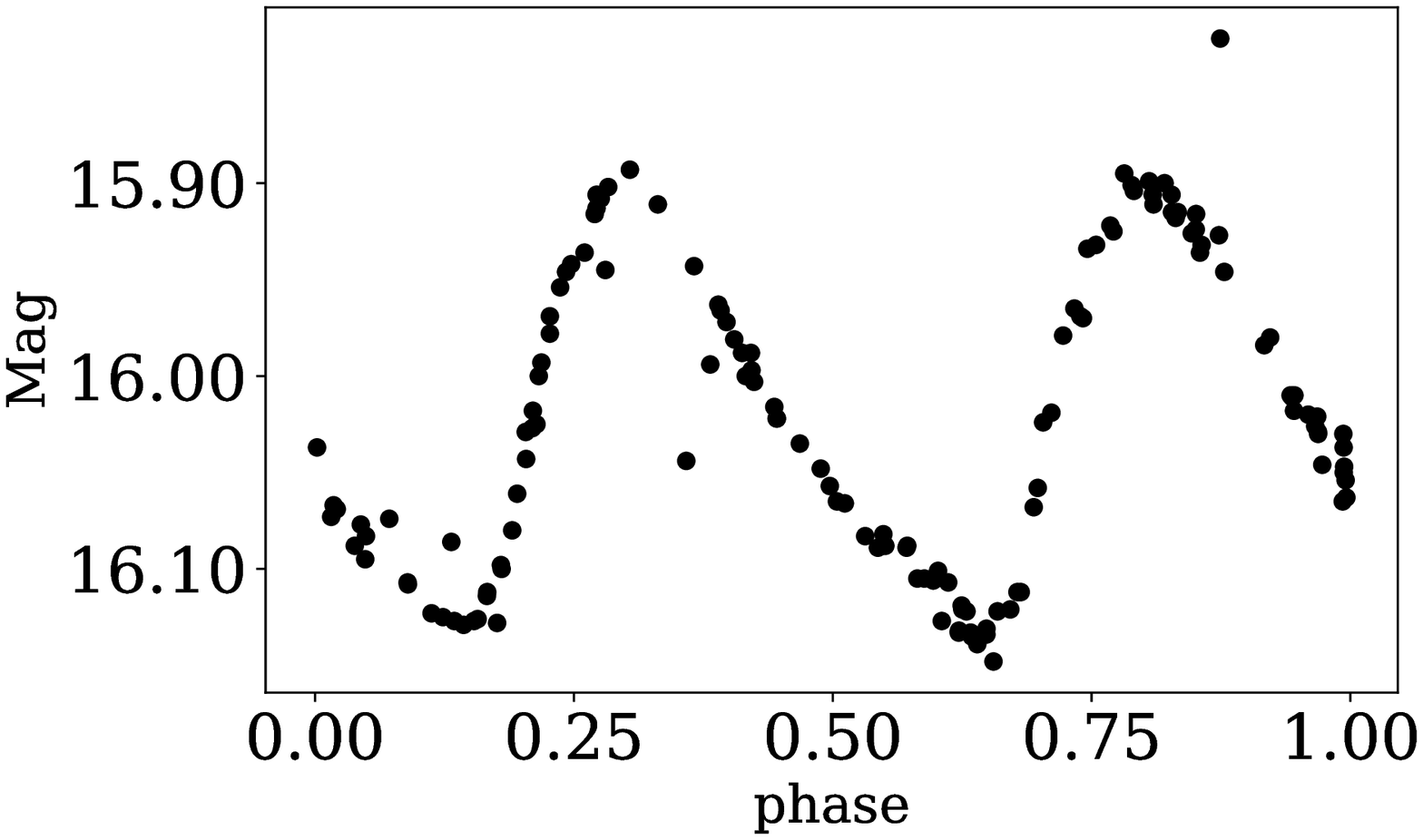}
  \includegraphics[width=2.5in]{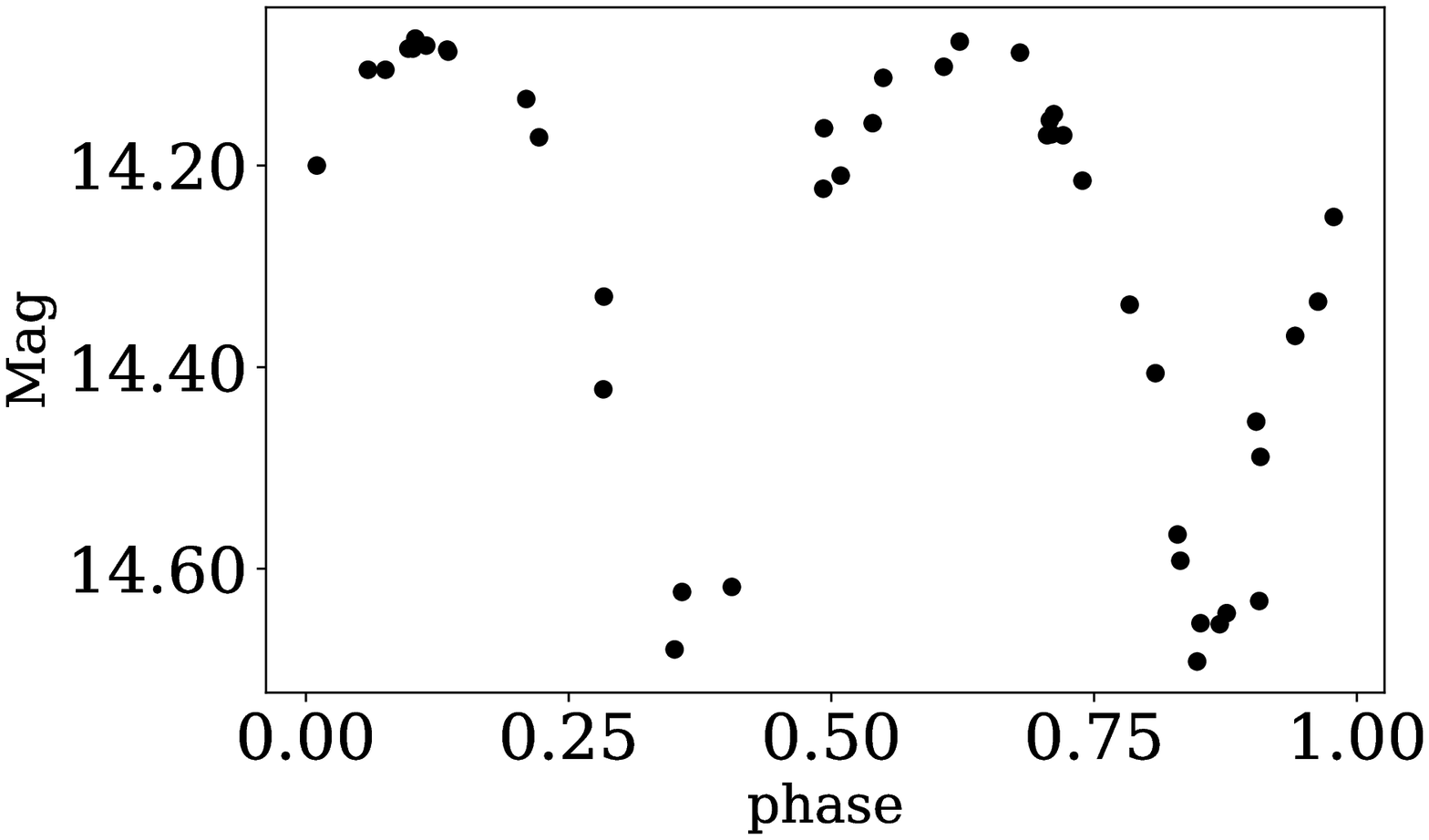}
  
  \caption{Three folded light curves rejected by \textit{Flat Test}. The bottom one is a binary light curve accepted by \textit{Flat Test}, shown for contrast.   }
  \label{rej}
\end{figure}

In applying the \textit{Flat Test}, we calculated the local average magnitude (Mag$_{ave}$) along the direction of the transit phase. We compared the gradients of Mag$_{ave}$ around the maximum and minimum of the light curve. The gradient here is defined as the difference between Mag$_{ave}$ in the transit-phase length of 0.2 and the transit-phase length of 0.1. We defined a threshold to help us assess whether the source is an eclipsing binary or not. The threshold was set in such a way that the gradient around the maximum phase should be 2 times larger than the gradient at the minimum phase. After applying the \textit{Flat Test}, 178 out of 275 sources were eliminated leaving us with 97 candidates. Light curves with a significant period but rejected by \textit{Flat Test} are shown in Figure \ref{rej}.

The light curves of our 97 candidates from our pipeline were subjected to visual inspections of the amplitude and shape of the curves. The amplitude of the light curve is modulated by eclipsing and ellipsoidal variations, limb darkening, reflection and beam effects\citep[see][]{2012ApJ...751..112J,{2010A&A...521L..59M}}. The first two should make the light curve reveal two similar maxima and two minima in one orbit. In contrast, the reflection and beam effects should appear only once in one orbit. In our sample, eclipsing and ellipsoidal modulations dominated the variation of the light, so we expected two maxima and minima in the folded light curve. 88 binary systems are left as our final sample.

The shapes of the light curves can be classified into three types: contact (EC), semi-detached (ESD) and detached (ED) binaries \citep{2006MNRAS.368.1311P,2012AJ....143..123M}. The light curves of EC and ESD types are continuously varying, while ED type is almost flat-topped. The EC binary having a deep common envelope, also known as "W UMa", tends to cause a more ellipsoidal light curve with (but not always) similar eclipsing depths. The contact is not strong enough in ESD: components can still have different temperatures with eclipse depths usually being different, and light curves appearing sharper near eclipse. The light curves of eclipsing binaries should match one of the three types on visual inspection.

In addition, a classification of the binary sample was performed using a Fourier transform of the light curves \citep{1993PASP..105.1433R,1997AJ....113..407R}. The maximum flux of light curves is normalized to be 1. The major eclipse was shifted to phase $\theta$=0.25. Then the light curve ($l(\theta)$) is transformed into 11 components: $l(\theta)$=$\sum_{0}^{10}$a$_{i}$cos(2$\pi$i$\theta$). The coefficients a$_{2}$ and a$_{4}$ are especially sensitive for classification. Using the empirical distribution of coefficients from \citet{2006MNRAS.368.1311P}, we classify our binaries sample as shown in Table 1.

\subsection{Properties of the Derived Catalog}

We now describe our catalog and the properties of the binaries within it. Each system has been given a unique identifier of the form LPSEBX (refer to "LAMOST $\&$ PTF Spectroscopic and Eclipsing Binaries") where X is our identification. In the catalog (see Table 1),
we provide RA (column 2) in degrees, declination (column 3) in degrees, primary eclipsing magnitude (column 4), primary eclipsing mid-day (Ecl$_{mid}$, column 5), primary eclipsing depth (column 6), orbital period (column 7) in days, number of photometric observations (Num$_{phot}$, column 8), largest radial velocity difference (column 9), number of spectroscopic observations (Num$_{spec}$, column 10), binary type (column 11), observational band (column 12; 1=g$^{'}$, 2=R), the effective temperature of the observable star $T_{\rm eff}$ (columns 13) in Kelvin, and surface gravity of observable star log \emph{g} (column 14). The top 13 rows show the newly identified eclipsing binaries. 

\newpage

\setlength\LTcapwidth{\textwidth}
\begin{longtable*}{|c|c|c|c|c|c|c|c|c|c|c|c|c|c|}
\caption{LPSEB catalog.The timestamp of radial velocity is taken from LAMOST data release 7 (\url{http://dr7.lamost.org/v1/search}) when not presented in LSS-GAC. The second observation of LPSEB21, LPSEB66 still misses the information of timestamp after the supplement that we abandon all the information in that epoch.}\label{Tab:publ-works}\\

\hline

ID      &      RA      & Declination        &  Mag$_{max}$   & Ecl$_{mid}$ & Ecl$_{depth}$  &   Period  & Num$_{phot}$  &  $\Delta$RV$_{max}$   &  Num$_{spec}$  & Type & Band & $T_{\rm eff}$ & log \emph{g}    \\
\hline
         &     deg     & deg                &               &   mjd      & mag           &   day    &               &    km s$^{-1}$       &              &       &      &K            & cgs    \\

\endfirsthead
\multicolumn{14}{c}%
{\tablename\ \thetable\ -- \textit{Continued from previous page}} \\
\hline
ID      &      RA      & Declination        &  Mag$_{max}$   & Ecl$_{mid}$ & Ecl$_{depth}$  &   Period  & Num$_{phot}$  &  $\Delta$RV$_{max}$   &  Num$_{spec}$  & Type & Band & $T_{\rm eff}$ & log \emph{g}    \\
\hline
         &     deg     & deg                &               &   mjd      & mag           &   day    &               &    km s$^{-1}$       &              &       &      &K            & cgs    \\

\hline
\endhead
\hline \multicolumn{14}{c}{\textit{Continued on next page}} \\
\endfoot
\hline
\endlastfoot
\hline 

LPSEB1   & 111.405   &  26.749   &   14.07   &56281.259416   &    0.62   &    0.33   &44   &64   &3   &EC   &2   &  5753.1   &     4.2  \\
LPSEB2   &   1.222   &  39.236   &   14.15   &56244.209892   &    0.80   &    0.43   &72   &19   &2   &ESD   &1   &  5692.4   &     4.2  \\
LPSEB3   & 108.699   &  27.831   &   13.89   &56281.456397   &    0.41   &    0.39   &55   &27   &2   &ESD   &2   &  6257.7   &     4.2  \\
LPSEB4   & 115.256   &  20.332   &   14.24   &56281.418748   &    0.52   &    0.37   &117   &27   &2   &ESD   &2   &  6011.2   &     4.0  \\
LPSEB5   & 101.707   &  59.360   &   13.00   &56267.655751   &    0.29   &    0.44   &38   &83   &3   &ESD   &2   &  5784.4   &     4.4  \\
LPSEB6   & 116.317   &  18.073   &   15.96   &56663.360043   &    0.59   &    0.37   &149   &83   &2   &EC   &2   &  6134.5   &     4.3  \\
LPSEB7   & 116.611   &  18.012   &   15.72   &55180.570289   &    0.57   &    0.37   &188   &33   &2   &ESD   &2   &  5557.6   &     3.8  \\
LPSEB8   & 203.086   &  35.480   &   14.24   &54903.754973   &    0.13   &    1.14   &315   &36   &2   &EC   &2   &  7249.5   &     3.9  \\
LPSEB9   &   1.368   &  44.458   &   15.30   &56903.401158   &    0.38   &    0.47   &60   &37   &2   &ESD   &1   &  7699.8   &     3.9  \\
LPSEB10   &  71.979   &  16.744   &   14.08   &55445.599193   &    0.33   &    0.34   &31   &40   &2   &ESD   &2   &  5844.7   &     3.9  \\
LPSEB11   & 114.532   &  19.665   &   15.45   &56281.339489   &    0.38   &    0.47   &58   &100   &3   &ESD   &2   &  7261.9   &     4.0  \\
LPSEB12   & 115.331   &  22.351   &   15.23   &56281.234100   &    0.84   &    0.29   &109   &46   &2   &ESD   &2   &  5568.1   &     4.1  \\
LPSEB13   & 191.932   &  31.921   &   15.09   &54964.537554   &    0.09   &    0.79   &211   &54   &3   &EC   &1   &  7328.3   &     4.1  \\
LPSEB14   &  47.187   &  40.434   &   14.87   &56301.405627   &    0.37   &    0.44   &33   &115   &2   &ESD   &2   &  7128.0   &     4.1  \\
LPSEB15   & 114.907   &  20.351   &   13.45   &56281.336666   &    0.53   &    0.40   &116   &49   &6   &ESD   &2   &  5753.4   &     3.9  \\
LPSEB16   & 118.565   &  19.181   &   14.01   &56663.267514   &    0.46   &    0.41   &154   &51   &3   &ESD   &2   &  6215.9   &     4.3  \\
LPSEB17   &  42.864   &  31.001   &   14.94   &55889.333756   &    0.53   &    0.56   &29   &51   &2   &ESD   &2   &  7781.1   &     4.0  \\
LPSEB18   & 181.380   &  23.028   &   13.79   &55002.176900   &    0.16   &    1.10   &33   &51   &3   &ESD   &2   &  6377.2   &     3.9  \\
LPSEB19   &  43.534   &  26.999   &   14.51   &55419.423323   &    0.86   &    0.31   &67   &124   &3   &ESD   &2   &  5640.5   &     3.6  \\
LPSEB20   & 234.732   &   4.484   &   14.88   &55417.168838   &    0.43   &    0.36   &104   &55   &2   &ESD   &2   &  6532.6   &     4.3  \\
LPSEB21   & 118.024   &  38.319   &   14.41   &55889.677648   &    0.23   &    0.55   &40   &0   &1   &ESD   &2   &  6666.5   &     4.1  \\
LPSEB22   & 116.912   &  22.304   &   14.24   &56281.340582   &    0.20   &    0.47   &117   &57   &5   &ESD   &2   &  7262.5   &     3.8  \\
LPSEB23   & 113.777   &  50.813   &   13.64   &55880.694541   &    0.39   &    0.42   &53   &57   &2   &ESD   &1   &  7247.3   &     4.0  \\
LPSEB24   & 248.511   &  56.371   &   14.28   &54961.330909   &    0.76   &    0.26   &106   &131   &2   &EC   &1   &  5133.5   &     4.3  \\
LPSEB25   & 188.007   &  35.500   &   13.23   &54903.623583   &    0.75   &    0.31   &101   &57   &4   &ESD   &2   &  5399.8   &     3.8  \\
LPSEB26   &  35.066   &  24.674   &   14.04   &55053.642630   &    0.27   &    0.51   &56   &59   &2   &ESD   &2   &  7667.5   &     3.8  \\
LPSEB27   & 350.398   &   5.570   &   15.10   &55007.538635   &    0.16   &    0.68   &36   &60   &3   &ESD   &2   &  6442.3   &     4.1  \\
LPSEB28   & 337.216   &   6.878   &   14.94   &55007.459400   &    0.35   &    0.37   &48   &60   &5   &ESD   &2   &  6552.1   &     4.3  \\
LPSEB29   &   3.613   &  40.348   &   13.37   &56244.216749   &    0.96   &    0.48   &71   &64   &10   &ESD   &1   &  6279.2   &     4.4  \\
LPSEB30   & 117.410   &  20.933   &   14.55   &56281.307494   &    0.49   &    0.37   &99   &13   &3   &EC   &2   &  5568.5   &     4.2  \\
LPSEB31   &  20.297   &  27.493   &   14.73   &55058.637609   &    0.79   &    0.32   &89   &14   &2   &ESD   &2   &  5766.5   &     4.1  \\
LPSEB32   & 116.744   &  22.747   &   13.52   &55931.215038   &    0.98   &    0.22   &365   &66   &4   &EC   &2   &  4616.7   &     4.6  \\
LPSEB33   & 111.745   &  38.808   &   13.68   &56229.613596   &    0.71   &    0.30   &51   &67   &2   &ESD   &2   &  5069.2   &     4.1  \\
LPSEB34   & 215.131   &  47.978   &   14.45   &54962.304145   &    0.46   &    0.24   &95   &16   &2   &ESD   &2   &  4688.5   &     4.6  \\
LPSEB35   & 240.184   &  43.145   &   13.23   &54957.191639   &    0.75   &    0.25   &111   &67   &2   &ESD   &2   &  5145.9   &     4.1  \\
LPSEB36   & 116.864   &  22.647   &   15.01   &55931.204753   &    0.84   &    0.28   &369   &17   &2   &EC   &2   &  5532.0   &     3.9  \\
LPSEB37   & 114.315   &  20.401   &   13.18   &56281.349946   &    0.62   &    0.45   &111   &161   &4   &ESD   &2   &  6267.5   &     4.5  \\
LPSEB38   & 139.061   &  16.257   &   12.15   &55954.260757   &    0.80   &    0.40   &76   &161   &4   &ESD   &2   &  7200.7   &     4.1  \\
LPSEB39   & 135.520   &  52.575   &   13.82   &54905.415970   &    0.56   &    0.41   &67   &21   &2   &ESD   &2   &  6234.3   &     3.7  \\
LPSEB40   & 142.889   &  10.765   &   14.38   &55324.315394   &    0.68   &    0.25   &67   &22   &2   &ESD   &2   &  5384.7   &     3.4  \\
LPSEB41   & 242.588   &  19.036   &   14.40   &54982.317626   &    0.51   &    0.25   &248   &22   &2   &EC   &2   &  5179.2   &     4.1  \\
LPSEB42   & 117.242   &  38.626   &   14.07   &55889.416055   &    0.32   &    0.53   &49   &171   &3   &ESD   &2   &  7210.8   &     4.1  \\
LPSEB43   &  13.320   &  42.534   &   15.24   &55213.131300   &    1.01   &    0.46   &112   &75   &3   &ESD   &1   &  7540.0   &     3.7  \\
LPSEB44   & 116.784   &  28.116   &   14.36   &54907.235921   &    1.06   &    0.35   &149   &41   &3   &ESD   &2   &  5701.5   &     3.9  \\
LPSEB45   & 113.266   &  19.829   &   14.51   &56281.426007   &    0.51   &    0.39   &58   &25   &2   &ESD   &2   &  5882.2   &     4.2  \\
LPSEB46   & 149.922   &  11.733   &   13.96   &56708.151000   &    0.62   &    0.31   &86   &77   &2   &ESD   &2   &  4777.0   &     4.5  \\
LPSEB47   & 102.948   &  59.447   &   12.26   &56267.562676   &    0.54   &    0.47   &51   &77   &6   &ESD   &2   &  5896.5   &     4.2  \\
LPSEB48   & 115.525   &  19.358   &   14.43   &56663.157707   &    0.53   &    0.37   &147   &27   &3   &EC   &2   &  6474.0   &     4.2  \\
LPSEB49   &  31.102   &   6.471   &   14.79   &55402.486390   &    0.37   &    0.42   &36   &27   &2   &ESD   &2   &  6570.0   &     4.3  \\
LPSEB50   & 116.537   &  20.529   &   14.10   &56281.234100   &    0.31   &    0.51   &113   &28   &2   &EC   &2   &  6043.7   &     4.0  \\
LPSEB51   & 115.312   &  18.024   &   16.05   &56663.158787   &    0.75   &    0.31   &155   &79   &4   &EC   &2   &  5649.9   &     4.3  \\
LPSEB52   & 127.808   &  19.831   &   15.57   &55229.348204   &    0.90   &    0.26   &300   &79   &2   &EC   &2   &  5177.9   &     3.8  \\
LPSEB53   &  20.221   &  26.528   &   14.11   &55058.544450   &    0.25   &    0.42   &52   &29   &3   &ESD   &2   &  6552.1   &     4.1  \\
LPSEB54   &  18.955   &  32.020   &   14.95   &55059.510558   &    0.65   &    0.42   &69   &29   &2   &ESD   &1   &  7155.1   &     4.2  \\
LPSEB55   & 174.022   &  43.071   &   16.30   &55143.498277   &    0.80   &    0.30   &116   &29   &2   &ESD   &2   &  5093.2   &     4.0  \\
LPSEB56   & 236.158   &   5.445   &   14.53   &56077.352920   &    0.46   &    0.29   &51   &81   &2   &ESD   &2   &  5631.0   &     3.5  \\
LPSEB57   & 234.123   &  39.936   &   13.88   &54963.242886   &    0.63   &    0.32   &87   &30   &2   &ESD   &2   &  5285.3   &     4.0  \\
LPSEB58   & 204.162   &  34.313   &   13.69   &54903.536745   &    0.94   &    0.28   &299   &201   &3   &EC   &2   &  5681.1   &     4.4  \\
LPSEB59   & 115.849   &  24.314   &   13.42   &55931.170118   &    0.85   &    0.38   &332   &30   &3   &EC   &2   &  6089.6   &     4.2  \\
LPSEB60   & 120.919   &  16.446   &   14.85   &55168.362149   &    0.81   &    0.37   &72   &30   &4   &ESD   &2   &  5723.8   &     3.5  \\
LPSEB61   & 114.987   &  39.081   &   15.96   &55889.495410   &    0.66   &    0.45   &52   &84   &3   &ESD   &2   &  6522.6   &     4.7  \\
LPSEB62   & 183.024   &  55.746   &   14.24   &54961.269211   &    0.62   &    0.34   &113   &32   &2   &EC   &2   &  5325.8   &     4.1  \\
LPSEB63   & 130.034   &  20.338   &   15.00   &55229.235087   &    1.02   &    0.27   &937   &32   &2   &EC   &2   &  4286.3   &     2.1  \\
LPSEB64   &  51.317   &  38.059   &   15.07   &56301.511701   &    0.37   &    0.36   &31   &32   &2   &ESD   &2   &  6173.9   &     4.4  \\
LPSEB65   & 114.604   &  20.146   &   15.36   &56281.320422   &    0.38   &    0.35   &58   &33   &4   &ESD   &2   &  5898.5   &     4.2  \\
LPSEB66   & 125.032   &  21.746   &   13.90   &54907.191682   &    1.07   &    0.31   &525   &0   &1   &EC   &1   &  5462.4   &     3.8  \\
LPSEB67   & 165.364   &  44.264   &   15.28   &55975.488397   &    0.47   &    0.26   &51   &34   &2   &ESD   &2   &  5133.3   &     3.8  \\
LPSEB68   &   0.614   &  41.160   &   16.30   &56244.096200   &    0.48   &    0.35   &67   &36   &2   &ESD   &1   &  6501.0   &     4.2  \\
LPSEB69   & 109.364   &  28.651   &   14.69   &56281.617659   &    0.36   &    0.72   &56   &37   &3   &ESD   &2   &  6906.6   &     3.6  \\
LPSEB70   & 123.441   &  18.670   &   15.98   &55903.513386   &    0.20   &    0.48   &155   &37   &2   &EC   &2   &  7348.0   &     3.6  \\
LPSEB71   & 359.536   &  41.724   &   15.86   &56244.335775   &    0.43   &    0.48   &62   &37   &2   &ESD   &1   &  6482.5   &     4.3  \\
LPSEB72   & 186.290   &  49.392   &   14.49   &54962.274686   &    0.79   &    0.31   &73   &38   &2   &ESD   &2   &  5614.1   &     3.9  \\
LPSEB73   & 114.266   &  24.448   &   13.76   &54904.203500   &    0.57   &    0.64   &47   &39   &2   &ESD   &2   &  7268.4   &     4.1  \\
LPSEB74   & 120.487   &  16.503   &   13.99   &54907.260025   &    1.05   &    0.43   &229   &98   &4   &ESD   &1   &  5923.8   &     4.0  \\
LPSEB75   &   3.501   &  38.696   &   13.93   &56244.328989   &    0.38   &    0.47   &63   &40   &15   &ESD   &1   &  6346.9   &     4.6  \\
LPSEB76   & 191.362   &  27.768   &   15.52   &54903.616144   &    0.57   &    0.38   &207   &40   &2   &EC   &1   &  5867.0   &     4.0  \\
LPSEB77   & 122.059   &  18.826   &   14.62   &55168.565600   &    0.34   &    0.25   &95   &41   &2   &EC   &2   &  4339.6   &     4.8  \\
LPSEB78   & 221.328   &  35.468   &   13.88   &54903.487314   &    0.87   &    0.34   &165   &41   &2   &EC   &1   &  5806.5   &     4.3  \\
LPSEB79   & 118.970   &  14.147   &   14.83   &55168.357706   &    0.61   &    0.42   &40   &49   &2   &ESD   &2   &  7632.2   &     4.0  \\
LPSEB80   & 123.735   &  21.152   &   12.84   &55939.276800   &    0.59   &    0.41   &109   &44   &2   &ESD   &2   &  5772.6   &     4.7  \\
LPSEB81   & 115.691   &  19.995   &   13.68   &56663.364301   &    0.34   &    0.78   &145   &44   &2   &EC   &2   &  5769.5   &     3.6  \\
LPSEB82   &  45.636   &  19.494   &   14.79   &55419.511872   &    0.34   &    0.31   &44   &44   &5   &ESD   &2   &  5196.6   &     4.4  \\
LPSEB83   &  16.162   &  33.557   &   16.50   &55058.589472   &    0.45   &    0.33   &146   &44   &2   &EC   &1   &  6473.9   &     4.2  \\
LPSEB84   & 139.317   &  16.326   &   14.17   &55954.248251   &    0.54   &    0.35   &80   &46   &3   &ESD   &2   &  5873.2   &     4.0  \\
LPSEB85   & 118.040   &  21.110   &   14.45   &56663.339884   &    0.70   &    0.61   &83   &46   &2   &ESD   &2   &  6433.9   &     4.3  \\
LPSEB86   & 209.106   &   4.330   &   15.54   &55022.334781   &    0.52   &    0.25   &64   &47   &2   &ESD   &2   &  5263.8   &     4.1  \\
LPSEB87   &  21.555   &  18.012   &   14.64   &55447.405146   &    0.31   &    0.34   &266   &8   &2   &EC   &2   &  5944.9   &     4.3  \\
LPSEB88   & 178.992   &  33.944   &   14.89   &54903.467600   &    0.38   &    0.32   &66   &48   &2   &ESD   &2   &  6656.6   &     4.1  \\

\hline

 
\end{longtable*}

The EB parameters were derived from different approaches. The maximum magnitude and eclipsing depth are estimated from the folded light curve. The orbital period is twice the peak period in the Lomb-Scargle periodogram. The folded light curves show high consistency when we fold the light curves at 2 times the peak period. The temperature and log \emph{g} are taken from LSS-GAC \citep{2017MNRAS.467.1890X}.



\begin{figure*}[!h]%
  \centering
  \includegraphics[width=4.5in]{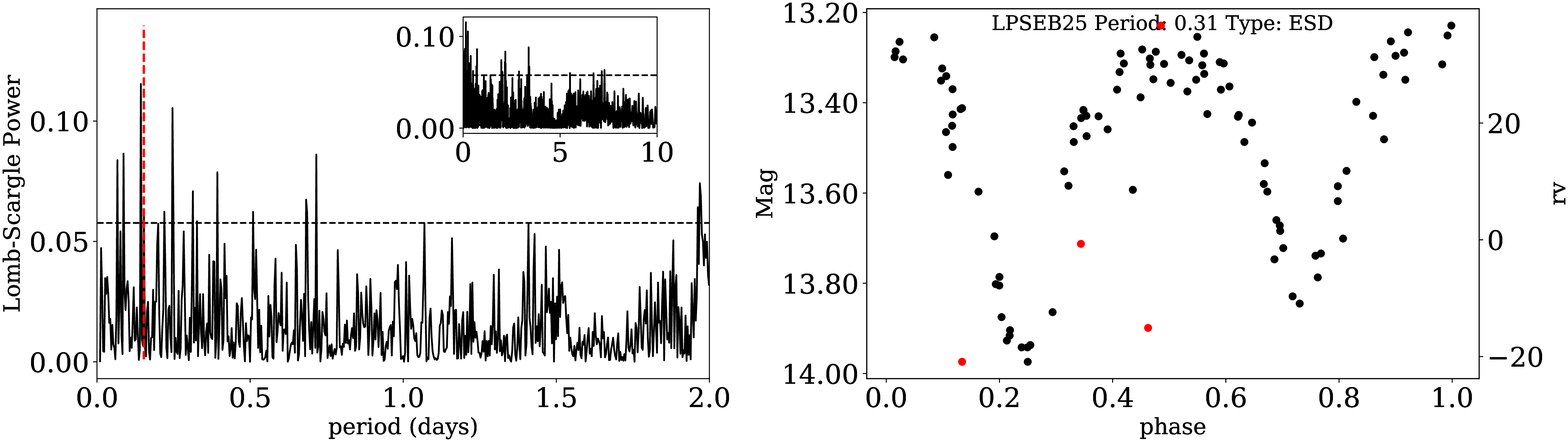}
  \includegraphics[width=4.5in]{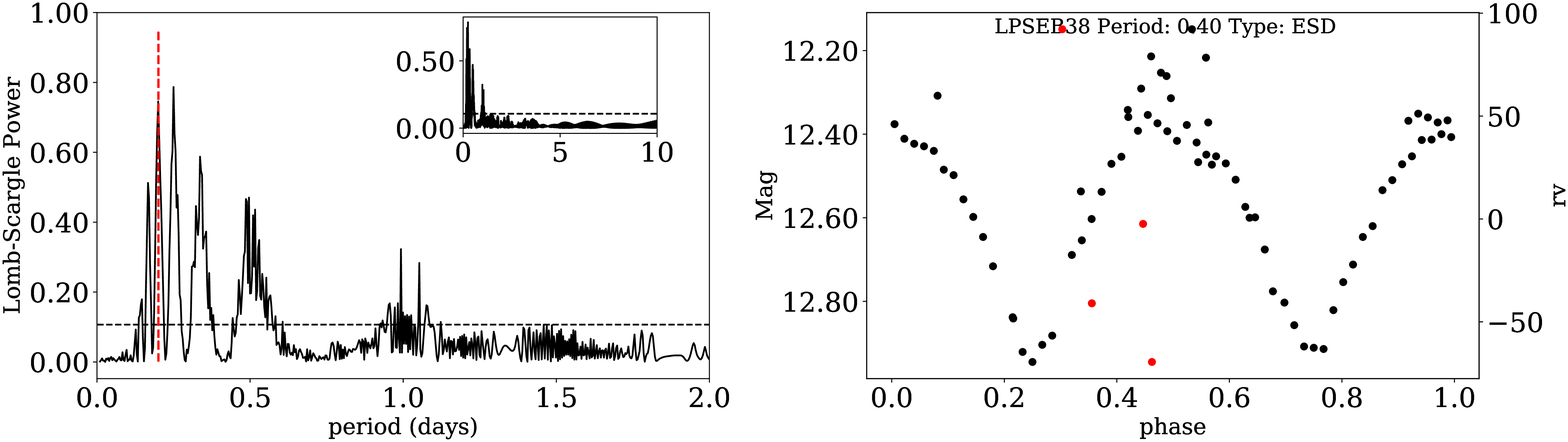}
  \includegraphics[width=4.5in]{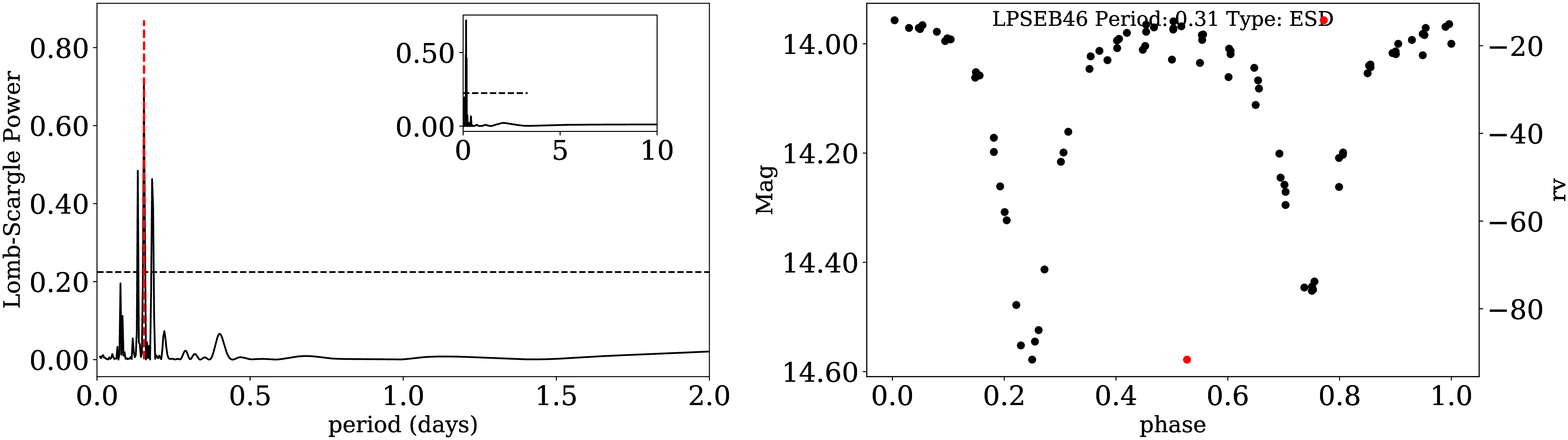}
  \includegraphics[width=4.5in]{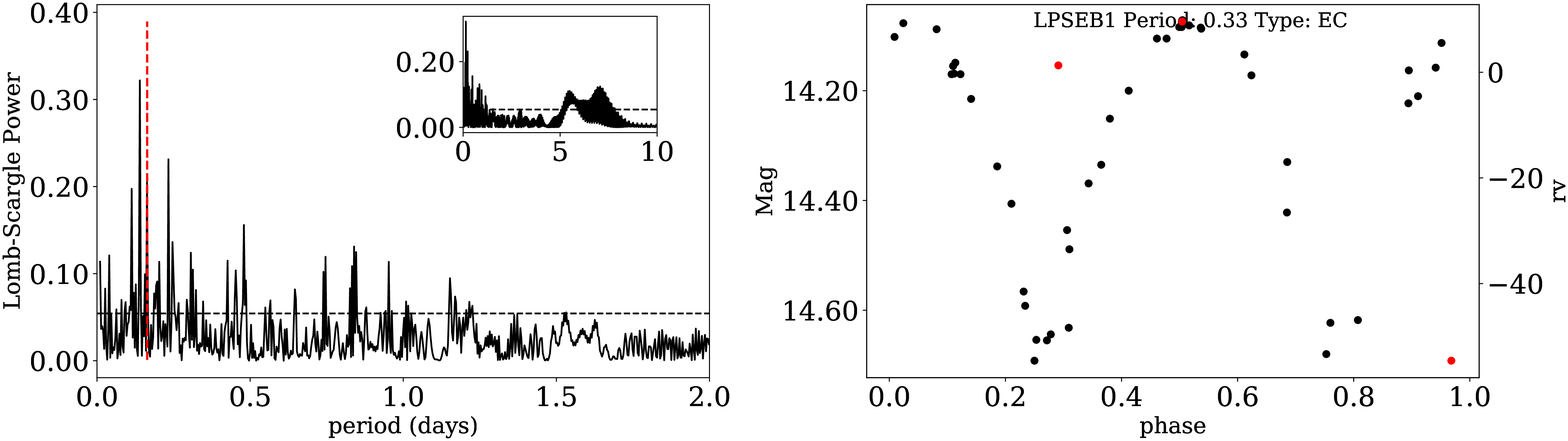}
  \includegraphics[width=4.5in]{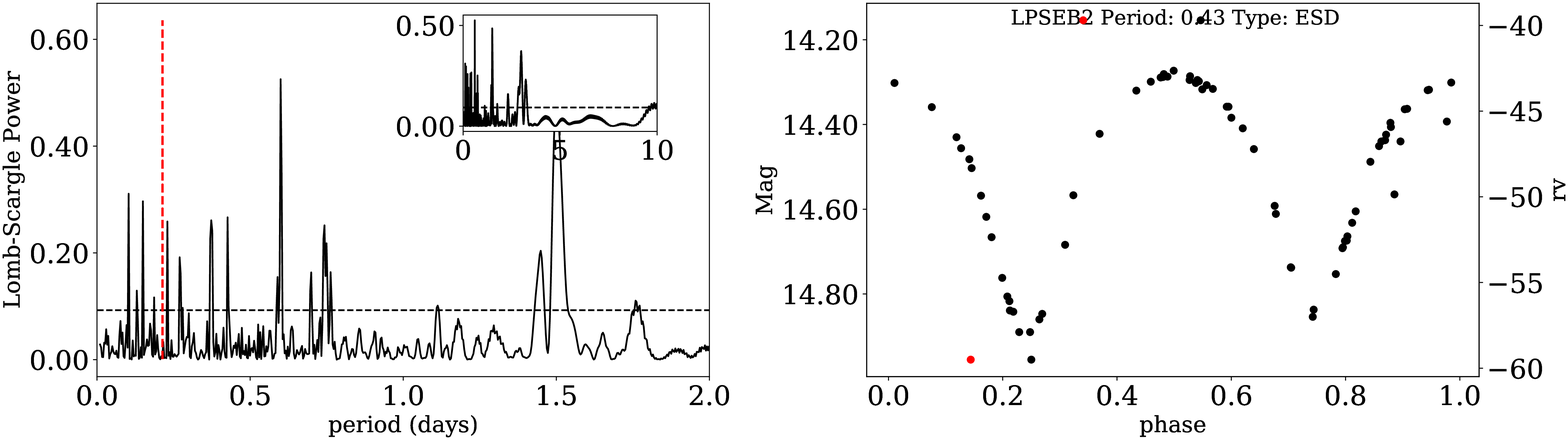}
  \includegraphics[width=4.5in]{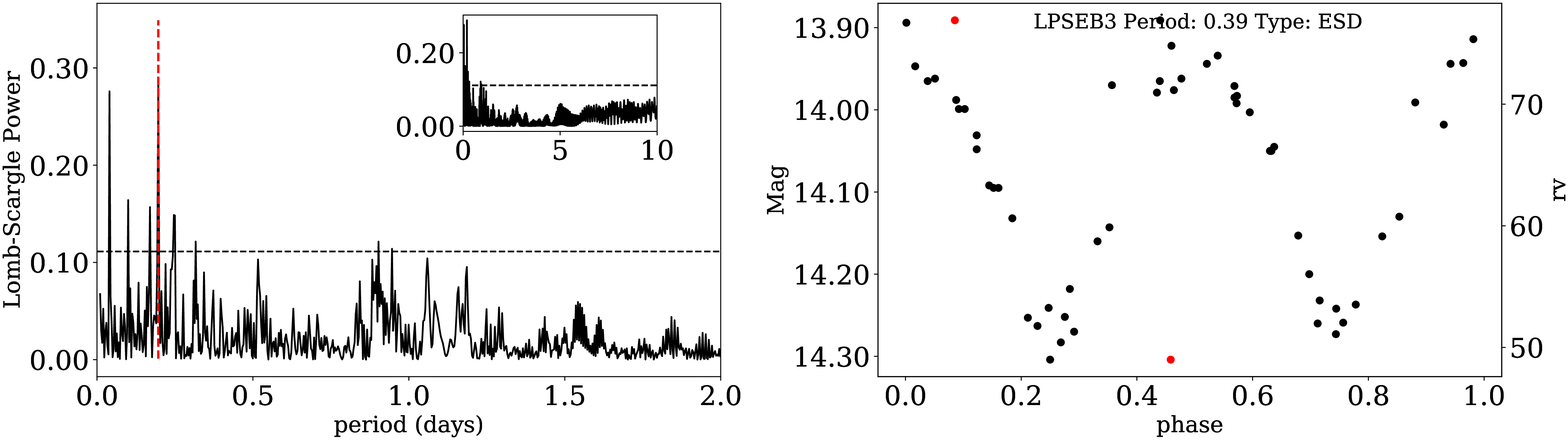}
  \caption{The periodograms (left column) and folded light curves (right column) of the three EB type binaries (LPSEB25, LPSEB38, LPSEB46) found in the literature and the SIMBAD database and three newly discovered binaries (LPSEB1, LPSEB2, LPSEB3). The inner subfigures show a broader period range. The black dashed lines in the periodograms indicate the Lomb-Scargle false positive probability which is 3$\sigma$ larger than the median power spectral density. The red vertical lines indicate the optimized peak which is half of the orbital period. The highest peak shown is possibly not the optimized peak. A solitary peak is treated as improbable if a clump of peaks appears when reducing the binning size. The light curves are shifted to have a major eclipse at $\theta$=0.25. The black point is the photometric brightness while red point presents the radial velocity with the same ephemeris as the light curve.}
  \label{ebex}
\end{figure*}

From examining the literature and the SIMBAD database, 75 of our sources have been previously identified as EBs. Among them, 3 sources are classified as $\beta$ type which is equivalent to our classification ESD type. The 3 $\beta$ type systems' Lomb-Scargle periodograms, their folded light curves, and the RV measurements with the same ephemeris as the light curve are shown in Figure \ref{ebex}, as an example. We also present 3 newly discovered EB in Figure \ref{ebex}. The full version of the plots is presented in supplementary material, Figure \ref{full}.

The folded light curves present the accuracy of the period finding method. The method derives the optimized peak rather than a solitary, highest peak, e.g., in the case of LPSEB2, LPSEB9, LPSEB69 (as shown in Fig. \ref{ebex}, \ref{full}). The intensity of the real peaks is smoothed to be smaller due to the binning in these cases. The real peak surrounded by a clump of peaks appears once the binning size is reduced. We test the precision of the period by checking the folded light curves with period offsets. The result implies period precision at the level of 1 percent for PTF light curves.

\section{Discussion}
\label{sect:data}

\subsection{Simulation Tests on Binary Detection Method}

To estimate the false positive probability of our software pipeline, we performed a Monte Carlo simulation. We firstly built a mock catalog with no binaries nor any pulsators. In order to be as close to the real PTF data as possible, the mock light curves come from shuffling the real PTF data. The luminosities were shuffled with a random time shift. Thus any structure in the light curves is removed but the observational features remain.

We then add some light curves of known variable sources. The basic shapes of light curve are taken from the General Catalogue of Variable Stars \citep{2017ARep...61...80S}, including Cepheids, RR Lyrae (including a, b, and c types). The time of the light curve is in the phase unit. The light curves are then extended in real-time based on period. The period is taken randomly according to the period distribution of the pulsators. The time baseline and sampling rate should be as close to PTF data as possible. We applied the baseline of 50-100 days and random sampling with an average rate of 1 day per data set, depending on the empirical knowledge of the PTF sampling from our data and descriptions from literature \citep{2009PASP..121.1395L,2009PASP..121.1334R}. A random photometric observational noise was added according to the relation shown in Figure \ref{accu}. We added 4000 pulsating light curves of variable stars into the mock catalog.

Following the above procedure, we created ten mock catalogs and ran them through our pipeline. Two parameters were monitored: the number of sources with the period from Lomb-Scargle and the number of sources accepted by the \textit{Flat Test}.

\begin{figure}[h]
  \centering
  \includegraphics[width=3.5in]{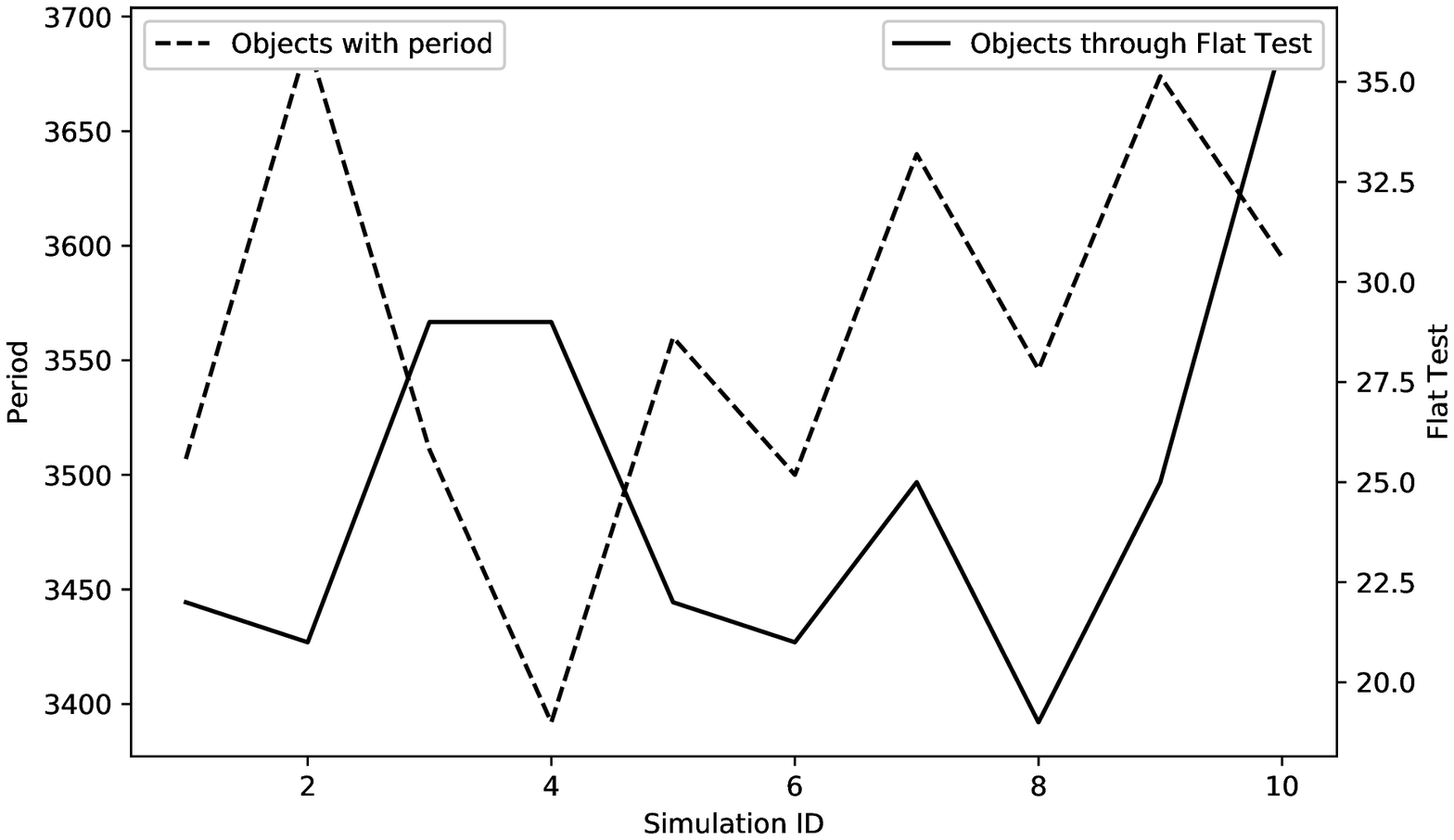}
  \caption{The results of the simulations. The solid line indicates the number of sources with a period from Lomb-Scargle. The dashed line indicates the number of sources accepted by the \textit{Flat Test}.}
  \label{shu}
\end{figure}

The ratio of the average number of sources accepted by \textit{Flat Test} in mock data and the sample size indicates the false detection rate of our pipeline. Combining the results of the simulations, 35614 sources are identified with the period from Lomb-Scargle. 249 of them got through the \textit{Flat Test}, the total number of the mock sources is 391790. The false detection rate is then 0.06$\%$ (249/(391790+4000$\times$10)). The parameters in every simulation were shown in Figure \ref{shu}.

Additionally, we applied the pipeline to 10000 stars from stripe 82 standard star catalog, among which 5818 stars were retrieved from PTF data. The result showed 11 sources were with a period while 2 of them were wrongly accepted by the \textit{Flat Test}. The false detection rate is $\sim$0.03$\%$(2/5818), similar to the simulation test.

The sample is not statistically complete due to the sampling strategies of LAMOST and PTF. In the detection procedure, the Lomb-Scargle false positive probability threshold we used was very strict. These strategies ensure the high purity of the catalog but reject some light curves with imperfect sampling. A credeble simulation test of completeness would be based on comprehensive simulated reproduction of PTF data, especially the estimation of sampling rate and photometry uncertainty. This is beyond the scope of this work. For a simple test, 1000 light curves of semi-detached and contact binaries were added into the mock catalog following the procedure above. The light curves come from the 10 basic shapes of binaries identified in \citet{2006MNRAS.368.1311P}, using the same simulated light curves creation method as described above.  82$\%$ of them are detected by our pipeline. Additionally, the detection rate significantly decreases when we try with detached binaries, and depends on the phase of transit duration.

\subsection{Orbital Period Cut-Off around 0.22 day}

\begin{figure}[!h]
  \centering
  \includegraphics[width=4in]{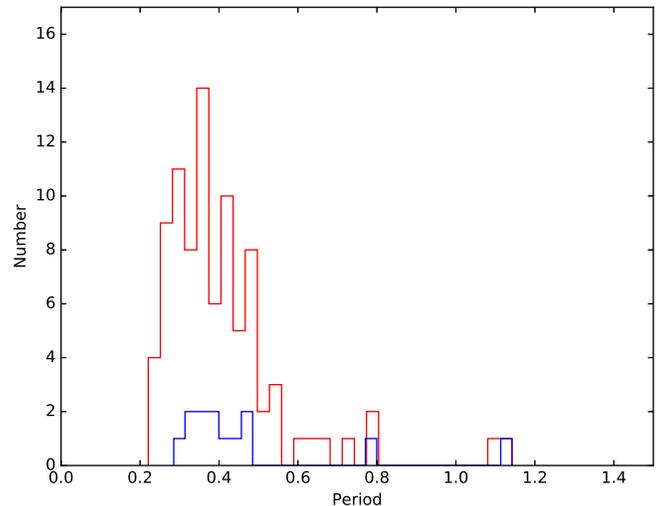}
  \caption{The binary system orbital period distribution. The red line indicates the distribution for the whole catalog while the blue line is for the newly discovered EBs only.}
  \label{ped}
\end{figure}

Our sample contains eclipsing binary stars with spectroscopic information. We recognize a steep cut-off at $\sim$0.22 day in the orbital period distribution (see in Figure \ref{ped}).

This cut-off has been known for at least the past two decades \citep{1992AJ....103..960R}. \citet{2006AcA....56..347S} argued that the cut-off is because low mass binary systems may not have enough time to fill the Roche Lobe. However, the existence of ultra-short period binary systems \citep{2015AcA....65...39S} challenges this model. \citet{2012MNRAS.421.2769J} proposed that the mass transfer for stars with initial primary masses lower than 0.63 M$_{\odot}$ is so fast that it quickly leads to the common envelope binary phase.

The theory of \citet{2012MNRAS.421.2769J} could be proved if we found very young binary systems with a short period such as 1 day or less. Such young binaries lead to short-period binaries with different periods at different phases of their evolution. The predicted period distribution of the young short-period binaries in the whole evolution phase without considering the duration of the evolution phase might not show the valley at 0.22 day. Then, the rareness of the binary with a period of 0.22 day should be due to the short-time existence in the certain evolution phase. Every source in our catalog has spectroscopic observations and so we could derive the age of the binaries. Such follow-up work may shed light on our understanding of low-mass binary evolution.

\section{summary}
\label{sect: summary}
Based on LAMOST and PTF data, we present our LPSEB catalog of eclipsing binary systems with both spectroscopic and photometric information. We build a software pipeline to find eclipsing binaries. The pipeline depends on the Lomb-Scargle periodogram to find the period and a new \textit{Flat Test} filter to recognize the key distinguishing feature of EB light curves. This pipeline is efficient in removing RR Lyrae contamination during the EB identification.

For binary systems in the catalog, we provide photometric information and spectral information. The false positive probability of the pipeline is estimated by Monte Carlo simulation as well as by real data from SDSS Stripe 82 Standard Catalog. Our sample shows the known cut-off at $\sim$0.22 day. Further studies can help constrain the evolution model of low-mass binaries.

\section{Appendix}

\subsection{RV observations for each system}

\setlength\LTcapwidth{3.1in}
\begin{longtable}{|c|c|c|}

\caption{RV observations. The Italic "jd" is taken from LAMOST data release 7 (\url{http://dr7.lamost.org/v1/search}) when not presented in LSS-GAC. The second observation of LPSEB21, LPSEB66 still misses the information of timestamp after the supplement that we abandon all the information in that epoch.}\label{Tab:publ-works}.\\
\hline

ID      &      Time(jd)  & RV (km s$^{-1}$)  \\

\endfirsthead
\multicolumn{3}{c}%
{\tablename\ \thetable\ -- \textit{Continued from previous column}} \\
\hline
ID      &      Time(jd)  & RV (km s$^{-1}$)  \\
\hline
\endhead
\hline \multicolumn{3}{c}{\textit{Continued on next column}} \\
\endfoot
\hline
\endlastfoot
\hline 
LPSEB1   &2456228.398750   &        1.3  \\
   &2456361.991065   &      -54.6  \\
   &2456650.207662   &        9.6  \\
LPSEB2   &2456611.015058   &      -59.5  \\
   &2456611.099259   &      -39.7  \\
LPSEB3   &2455903.234711   &       76.9  \\
   &2457010.222593   &       49.0  \\
LPSEB4   &2456608.325324   &       11.2  \\
   &2457005.220139   &       39.1  \\
LPSEB5   &2456321.062523   &      -75.8  \\
   &2456590.350637   &      -21.2  \\
   &\textit{2457464.993055}   &     -104.4  \\
LPSEB6   &2456608.325324   &       -3.5  \\
   &2457005.220139   &       80.4  \\
LPSEB7   &2456608.325324   &       54.8  \\
   &2457005.220139   &       88.2  \\
LPSEB8   &2457107.217894   &      -36.4  \\
   &2457146.139479   &        0.5  \\
LPSEB9   &2456984.952523   &      -34.0  \\
   &2456985.022176   &      -71.1  \\
LPSEB10   &2456978.208380   &      -40.6  \\
   &2457013.076609   &        0.0  \\
LPSEB11   &2456608.325324   &       18.1  \\
   &2456983.319468   &       32.8  \\
   &2457005.220139   &      -68.0  \\
LPSEB12   &2456608.325324   &      -14.2  \\
   &2457005.220139   &       32.5  \\
LPSEB13   &2456428.071887   &       26.3  \\
   &2457039.341748   &       38.5  \\
   &2457071.272130   &      -16.4  \\
LPSEB14   &2456952.193310   &      -21.2  \\
   &2457024.034537   &       94.2  \\
LPSEB15   &2456288.174248   &       62.7  \\
   &2456288.201586   &       64.9  \\
   &2456350.088762   &       48.0  \\
   &2456350.115926   &       74.5  \\
   &2456983.319468   &       24.6  \\
   &2457005.220139   &       54.3  \\
LPSEB16   &2456608.325324   &       28.0  \\
   &2456966.337755   &       59.5  \\
   &2456966.403588   &       79.2  \\
LPSEB17   &2456266.073102   &      -87.4  \\
   &2456652.053924   &      -35.9  \\
LPSEB18   &2456317.305613   &        4.5  \\
   &2456317.386944   &       22.7  \\
   &2456321.305914   &      -29.1  \\
LPSEB19   &2455939.010255   &      -40.9  \\
   &2456308.961007   &       18.2  \\
   &2456309.971285   &       84.0  \\
LPSEB20   &2457138.190498   &       11.4  \\
   &2457167.146748   &      -43.7  \\
LPSEB21   &2457004.290000   &      -91.5  \\
LPSEB22   &2456230.391192   &       60.2  \\
   &2456288.174248   &       16.4  \\
   &2456600.361042   &        3.1  \\
   &2456608.325324   &        3.4  \\
   &2457005.286389   &       41.4  \\
LPSEB23   &2457053.122269   &      -42.4  \\
   &2457055.126053   &       14.7  \\
LPSEB24   &2456412.328519   &       53.9  \\
   &2456412.233947   &      -77.5  \\
LPSEB25   &2456306.373970   &      -20.9  \\
   &2456346.250243   &      -15.1  \\
   &2457033.415694   &       -0.7  \\
   &2457038.354433   &       36.7  \\
LPSEB26   &2456570.246123   &      -82.2  \\
   &2457327.152731   &      -22.5  \\
LPSEB27   &2456561.125995   &     -103.5  \\
   &2456568.106019   &      -43.4  \\
   &\textit{2457367.960416}   &      -55.1  \\
LPSEB28   &2456231.967870   &      -18.6  \\
   &2456231.999363   &       -3.0  \\
   &2456243.974630   &       42.4  \\
   &2456246.014225   &      -12.7  \\
   &2457300.026227   &       -3.9  \\
LPSEB29   &2456611.015058   &      -39.6  \\
   &2456611.099259   &       -6.0  \\
   &2457285.151759   &      -70.2  \\
   &2457285.182639   &      -51.9  \\
   &2457285.197245   &      -58.9  \\
   &2457285.166944   &      -65.9  \\
   &2457285.261377   &      -29.1  \\
   &2457285.211829   &      -51.7  \\
   &2457285.246817   &      -21.2  \\
   &2457285.232338   &      -35.3  \\
LPSEB30   &2456608.379560   &       28.4  \\
   &2457005.220139   &       22.6  \\
   &2457005.286389   &       14.7  \\
LPSEB31   &2456973.037049   &      -13.2  \\
   &2457008.966817   &      -27.3  \\
LPSEB32   &2456266.335799   &       16.9  \\
   &2456608.379560   &       80.3  \\
   &2457005.220139   &       83.0  \\
   &2457005.286389   &       30.1  \\
LPSEB33   &2456625.340856   &       48.5  \\
   &2457090.055000   &      -19.0  \\
LPSEB34   &2456362.343507   &      -56.2  \\
   &2457096.301887   &      -39.5  \\
LPSEB35   &2456423.203299   &      -60.5  \\
   &2456423.277188   &        7.5  \\
LPSEB36   &2457005.220139   &       -9.8  \\
   &2457005.286389   &        7.6  \\
LPSEB37   &2456350.088762   &       74.0  \\
   &2456350.115926   &       91.8  \\
   &2456983.319468   &      -62.1  \\
   &2457005.220139   &      -69.5  \\
LPSEB38   &2456326.160185   &       92.2  \\
   &2456369.134132   &       -2.4  \\
   &2456656.276146   &      -41.0  \\
   &2456656.318993   &      -69.5  \\
LPSEB39   &2457003.302812   &      -69.5  \\
   &\textit{2457391.266666}   &      -48.1  \\
LPSEB40   &2455920.306134   &        9.4  \\
   &2456287.293738   &       32.0  \\
LPSEB41   &2456700.394873   &      -14.7  \\
   &2456780.213206   &      -37.5  \\
LPSEB42   &2457004.290000   &       71.8  \\
   &2457025.185914   &      123.6  \\
   &\textit{2457400.209722}   &      -47.5  \\
LPSEB43   &2455840.221806   &      -39.3  \\
   &2455863.102581   &      -39.6  \\
   &2456267.991539   &       35.5  \\
LPSEB44   &2456239.385961   &       24.4  \\
   &2456622.307581   &       57.4  \\
   &2456687.145347   &       15.7  \\
LPSEB45   &2456983.319468   &        9.5  \\
   &2457062.106655   &       35.0  \\
LPSEB46   &2456683.202292   &      -91.6  \\
   &\textit{2457400.295833}   &      -14.2  \\
LPSEB47   &2456317.109711   &      -56.2  \\
   &2456317.136551   &      -50.6  \\
   &2456321.088773   &       21.8  \\
   &2456590.389039   &       -6.4  \\
   &2456700.065810   &      -11.4  \\
   &2456700.103750   &        4.7  \\
LPSEB48   &2456608.325324   &       43.4  \\
   &2457005.220139   &       32.5  \\
   &2457005.286389   &       60.0  \\
LPSEB49   &2456550.241447   &       23.2  \\
   &2457280.262396   &       50.9  \\
LPSEB50   &2457005.220139   &       67.5  \\
   &2457005.286389   &       95.8  \\
LPSEB51   &2456608.379560   &      108.0  \\
   &2457005.220139   &       28.8  \\
   &2457005.286389   &       37.4  \\
   &2457042.183356   &       48.8  \\
LPSEB52   &2456980.341725   &       16.7  \\
   &2457032.209016   &       96.2  \\
LPSEB53   &2456267.062245   &       20.8  \\
   &2456956.183970   &        0.4  \\
   &2456973.037049   &       -8.3  \\
LPSEB54   &2457025.946516   &      -24.6  \\
   &\textit{2457360.079861}   &        4.6  \\
LPSEB55   &2457020.380984   &       15.2  \\
   &2457021.389699   &       44.7  \\
LPSEB56   &2456367.333484   &      -19.8  \\
   &\textit{2457478.297916}   &       61.4  \\
LPSEB57   &2456424.148125   &       -2.3  \\
   &2456424.190313   &      -32.6  \\
LPSEB58   &2456321.351979   &        2.9  \\
   &2456380.171863   &     -198.3  \\
   &2456740.235671   &      -33.4  \\
LPSEB59   &2456266.335799   &       36.2  \\
   &2456288.174248   &       23.7  \\
   &2456385.024676   &       54.5  \\
LPSEB60   &2456966.337755   &      -35.9  \\
   &2456966.403588   &      -27.4  \\
&\textit{2457378.243750}   &      -58.2  \\
&\textit{2457378.309722}   &      -46.2  \\
LPSEB61   &2456625.275637   &     -114.3  \\
   &2457033.197188   &      -36.3  \\
   &2457090.055000   &     -120.7  \\
LPSEB62   &2456316.331782   &      -40.6  \\
   &2456752.096204   &       -8.0  \\
LPSEB63   &2456249.382593   &       14.5  \\
   &2457063.147905   &       47.3  \\
LPSEB64   &2456952.193310   &       -1.1  \\
   &2456975.125914   &      -34.0  \\
LPSEB65   &2456608.325324   &        4.6  \\
   &2456983.319468   &        3.5  \\
   &2457005.220139   &       -9.2  \\
   &2457005.286389   &       24.1  \\
LPSEB66   &2456980.341725   &      -22.9  \\
LPSEB67   &2456637.387731   &       39.3  \\
   &2456665.343380   &        4.6  \\
LPSEB68   &2456984.952523   &      -91.6  \\
   &2456985.022176   &      -54.8  \\
LPSEB69   &2456361.073252   &       87.5  \\
   &2456692.069433   &       50.6  \\
   &2457010.222593   &       87.8  \\
LPSEB70   &2456980.341725   &       19.2  \\
   &2456990.327315   &       56.7  \\
LPSEB71   &2456984.952523   &     -118.2  \\
   &2456985.022176   &     -156.1  \\
LPSEB72   &2455923.410914   &      -89.2  \\
   &2457062.269606   &      -50.8  \\
LPSEB73   &2456384.998368   &       67.6  \\
   &2456743.027083   &      107.0  \\
LPSEB74   &2456966.337755   &       48.5  \\
   &2456966.403588   &       96.2  \\
   &\textit{2457378.243750}   &       -2.4  \\
   &\textit{2457378.309722}   &       41.6  \\
LPSEB75   &2456611.099259   &     -100.6  \\
   &2456641.949792   &      -93.2  \\
   &2457285.182639   &     -117.4  \\
   &2457285.197245   &      -86.5  \\
   &2457285.261377   &      -89.6  \\
   &2457285.246817   &      -92.0  \\
   &2457286.143044   &      -84.7  \\
   &2457286.157789   &      -93.0  \\
   &2457286.230972   &     -106.7  \\
   &2457286.172431   &      -81.5  \\
   &2457286.245822   &     -115.6  \\
   &2457286.186956   &      -81.7  \\
   &2457286.201493   &      -89.8  \\
   &2457286.260486   &     -121.9  \\
   &2457286.216088   &     -103.1  \\
LPSEB76   &2455953.336609   &       35.8  \\
   &2456785.045231   &       -5.2  \\
LPSEB77   &2456966.337755   &      -53.0  \\
   &2456990.327315   &      -11.3  \\
LPSEB78   &2456385.228322   &       19.9  \\
   &2456385.308495   &      -21.9  \\
LPSEB79   &\textit{2457378.243750}   &      109.0  \\
   & \textit{2457378.309722}        &       59.7 \\
LPSEB80   &2456640.320035   &        5.9  \\
   &2456980.341725   &       50.3  \\
LPSEB81   &2456350.088762   &       31.5  \\
   &2457005.220139   &      -13.2  \\
LPSEB82   &2457285.342755   &       51.6  \\
   &2457286.332789   &       96.3  \\
   &2457286.362176   &       87.8  \\
   &2457286.347442   &       82.1  \\
   &2457286.318148   &       84.9  \\
LPSEB83   &2457330.110463   &      -96.4  \\
   &2457304.150313   &      -51.5  \\
LPSEB84   &2456326.160185   &       30.6  \\
   &2456656.276146   &      -16.1  \\
   &2456656.318993   &       -3.5  \\
LPSEB85   &2457005.220139   &       -7.7  \\
   &2457005.286389   &       39.1  \\
LPSEB86   &2456392.183611   &      -25.1  \\
   &\textit{2457439.365277}   &      -72.8  \\
LPSEB87   &2456224.202720   &     -100.0  \\
   &2456224.154884   &     -108.6  \\
LPSEB88   &2457015.367141   &      -69.5  \\
   &\textit{2457389.402777}   &      -21.5  \\
\hline

\end{longtable}

\subsection{Full Light Curves of Binaries}

\newpage
\begin{figure*}[!htb]%
  \centering
  \includegraphics[width=4.5in]{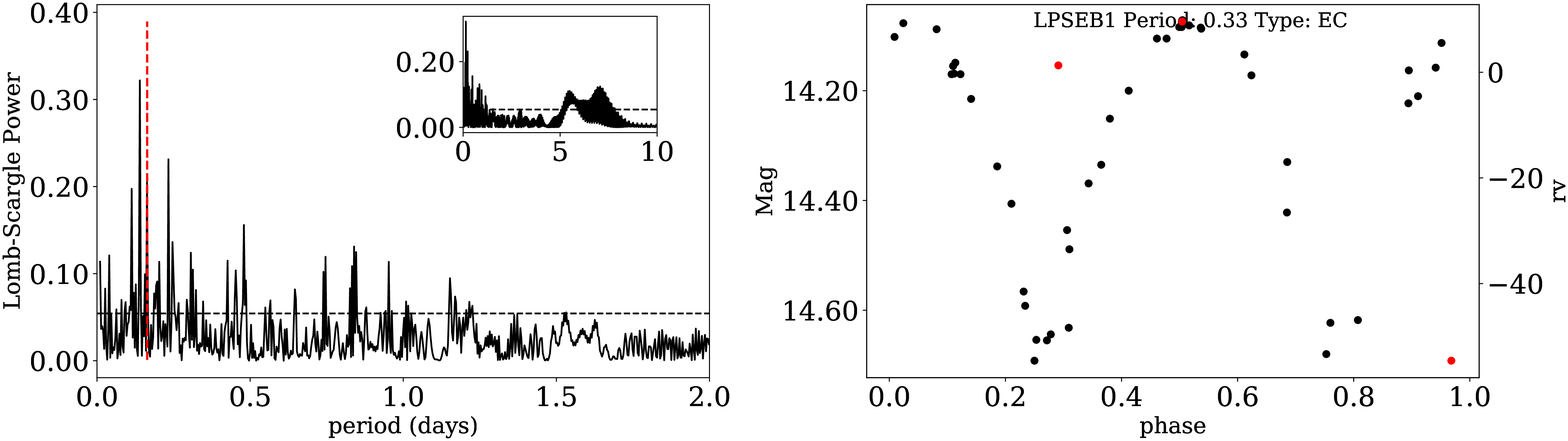}
  \includegraphics[width=4.5in]{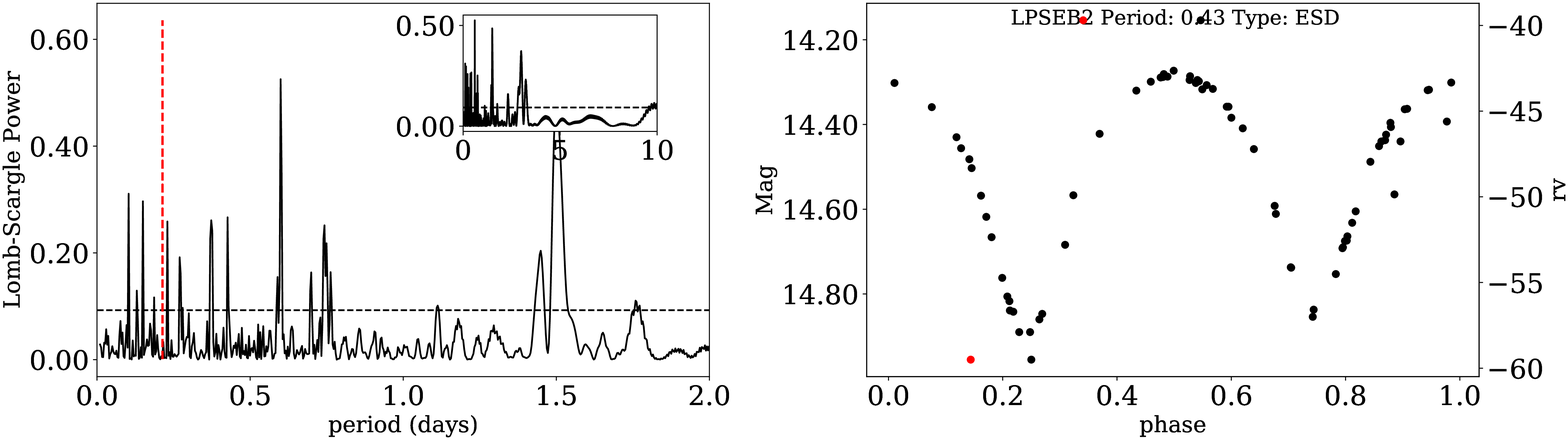}
  \includegraphics[width=4.5in]{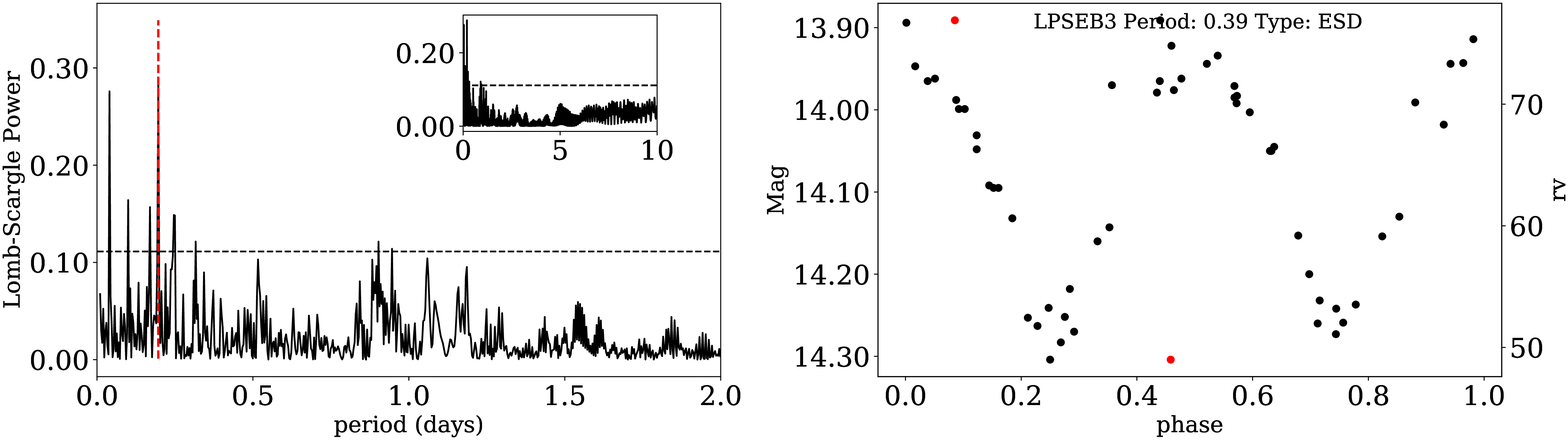}
  \includegraphics[width=4.5in]{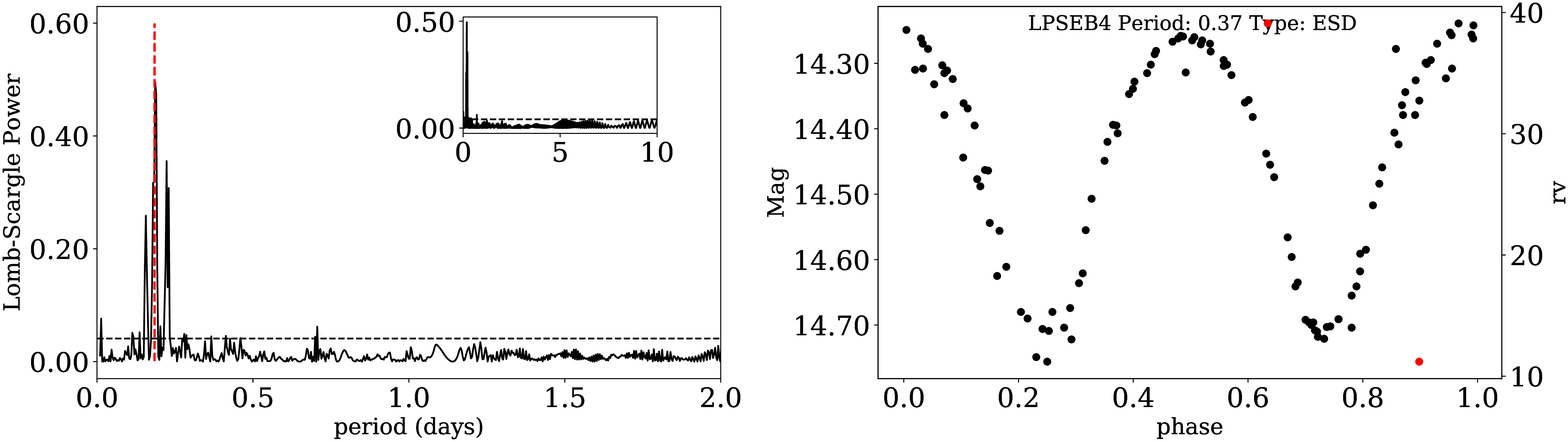}
  \includegraphics[width=4.5in]{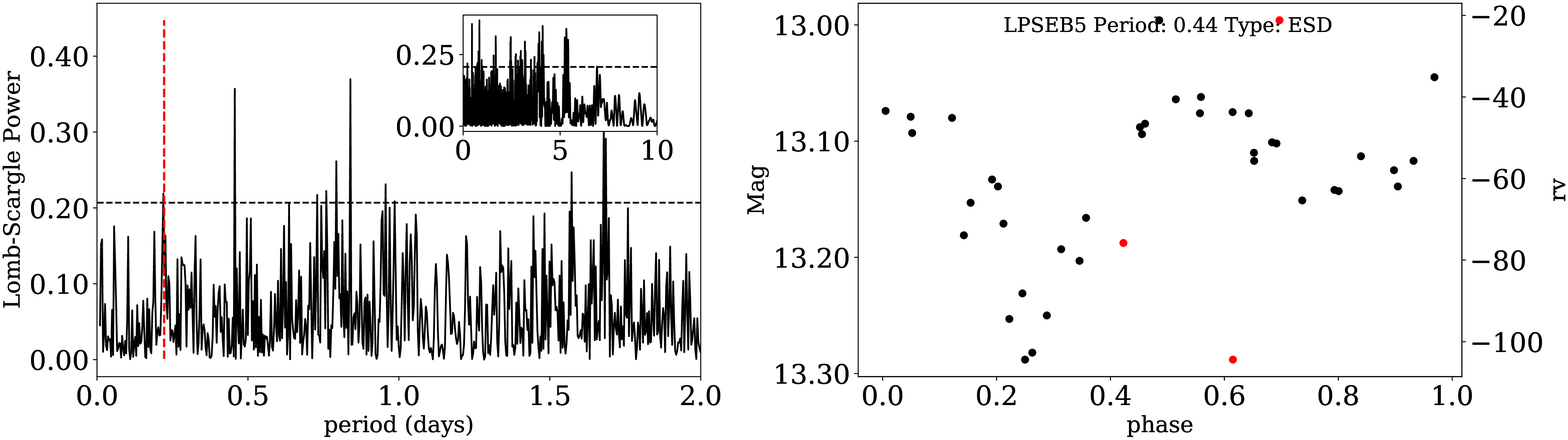}
  \includegraphics[width=4.5in]{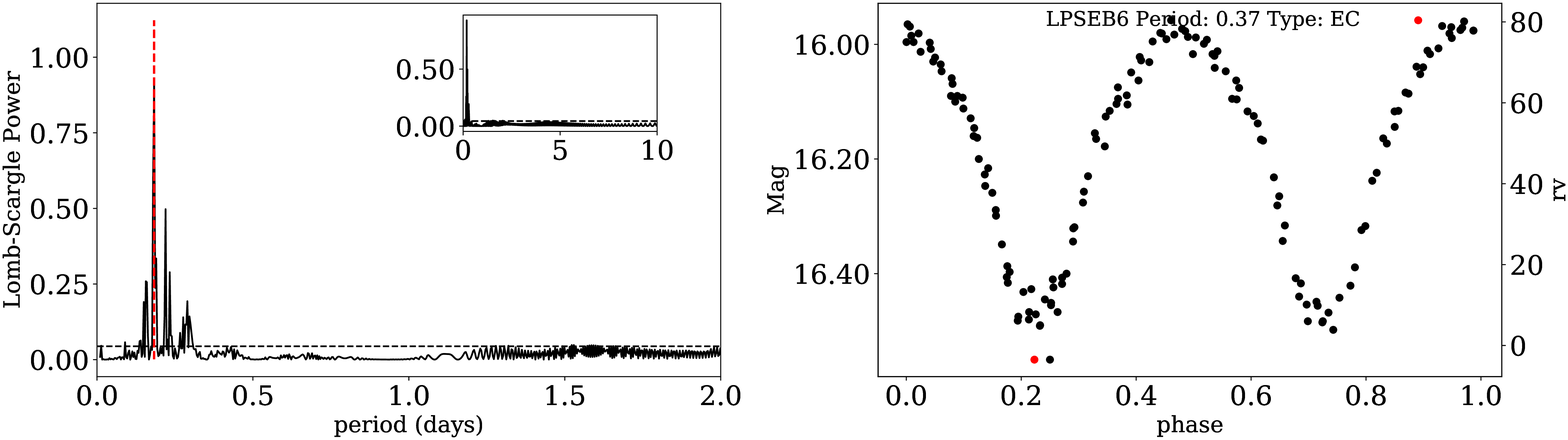}

  \caption{The periodograms (left column) and folded light curves (right column) of all EB. The rows from top to bottom are LPSEB1 to LPSEB88. The inner subfigure inside the left column shows a broader period range. The dashed line in the periodogram indicates the Lomb-Scargle false positive probability. The red vertical lines indicate the optimized peak which is half of the orbital period. The highest peak shown is possibly not the optimized peak. A solitary peak is treated as improbable if a clump of peaks appears when reducing the binning size. The light curves are shifted to have a major eclipse at $\theta$=0.25. The black points are the photometric brightness while red points present the radial velocity with the same ephemeris as the light curve.
}
  \label{full}
\end{figure*}

\newpage
\addtocounter{figure}{-1}
\begin{figure*}[!htb]%
  
  \centering
 
  \includegraphics[width=4.5in]{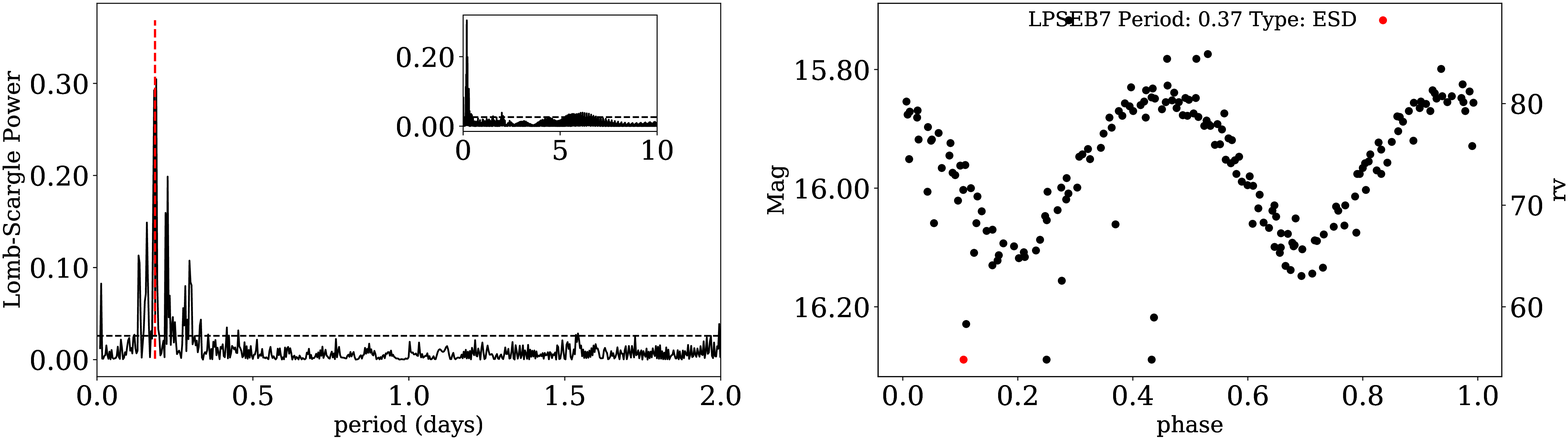}
   \includegraphics[width=4.5in]{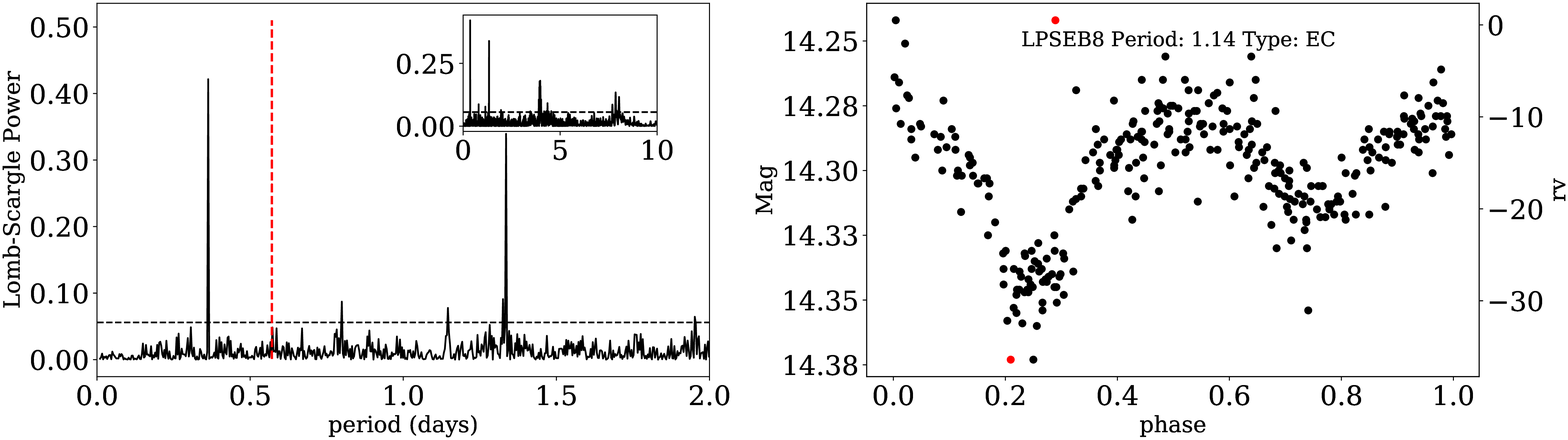}
  \includegraphics[width=4.5in]{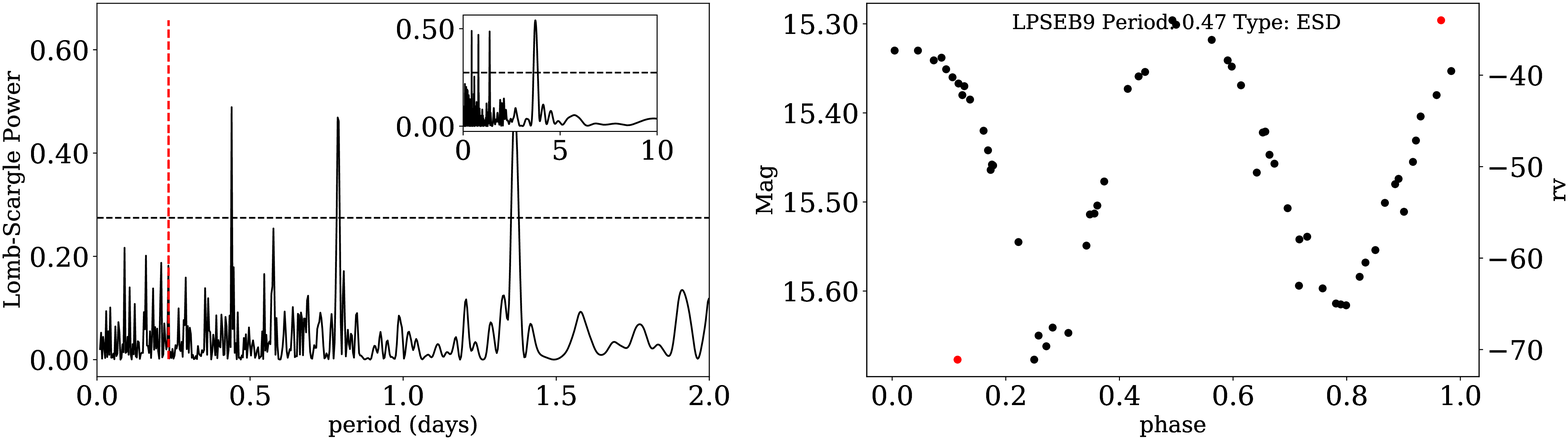}
  \includegraphics[width=4.5in]{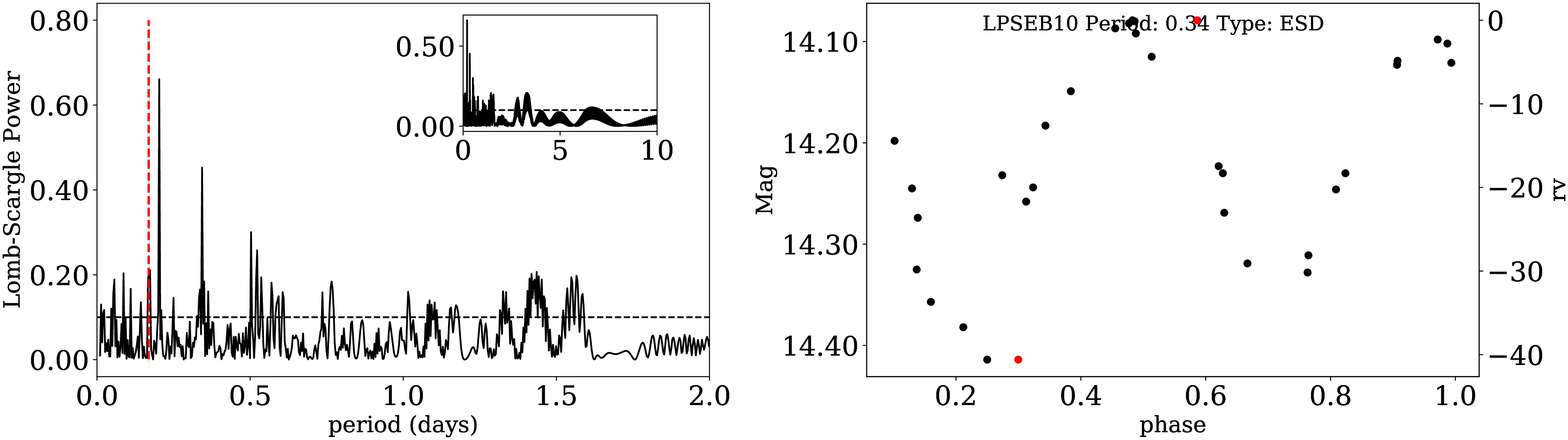}
  \includegraphics[width=4.5in]{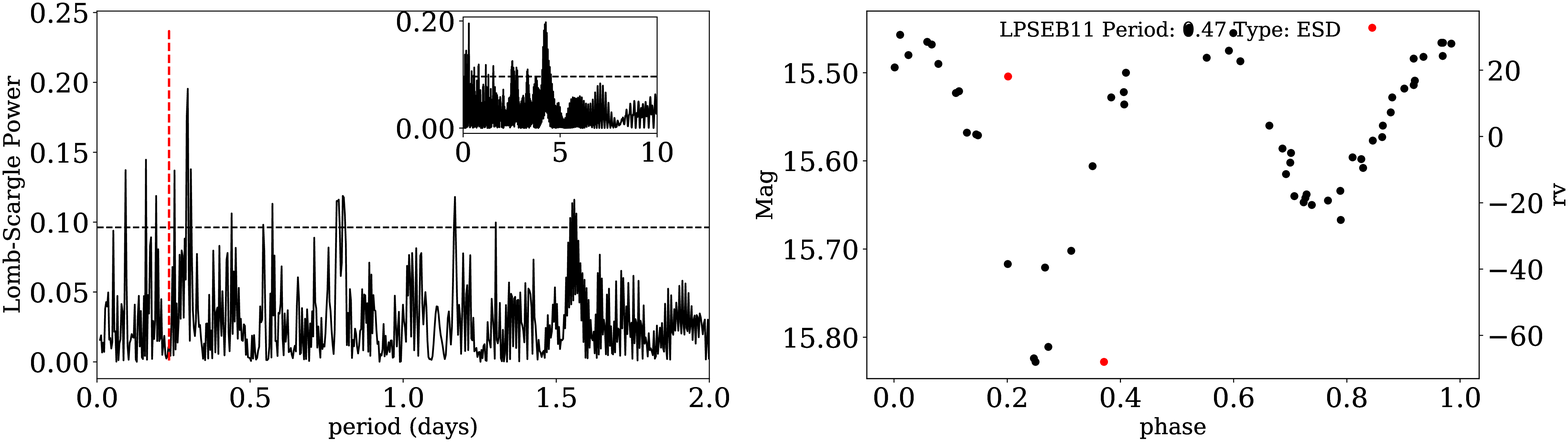}
  \includegraphics[width=4.5in]{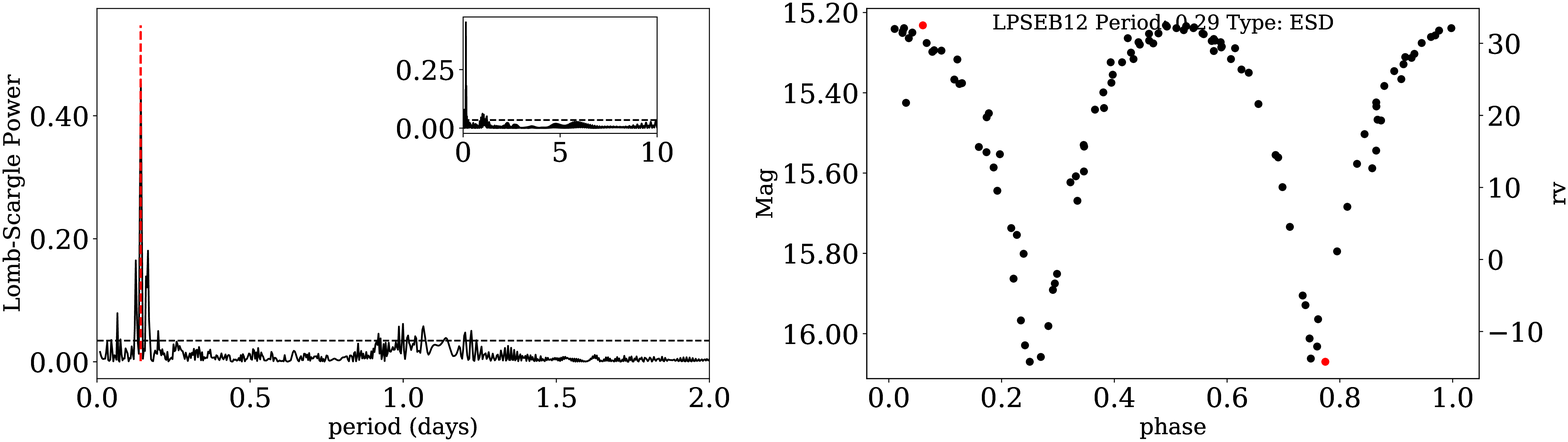}
  
\caption{(Continued) }
  \label{fig1}
\end{figure*}

\newpage
\addtocounter{figure}{-1}
\begin{figure*}[!htb]%
  
  \centering
 \includegraphics[width=4.5in]{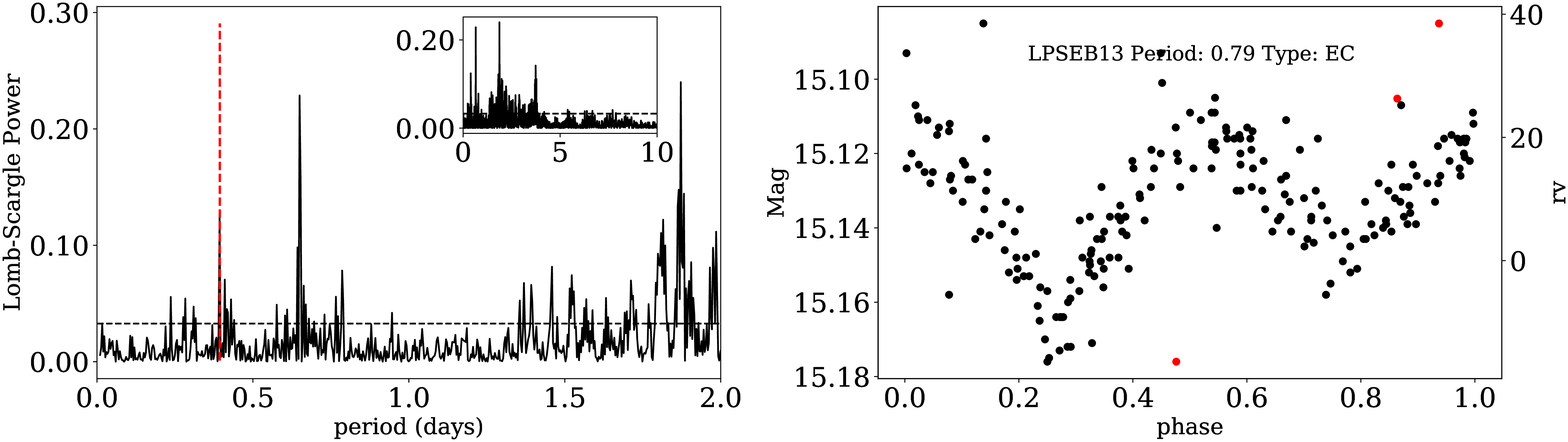}
   \includegraphics[width=4.5in]{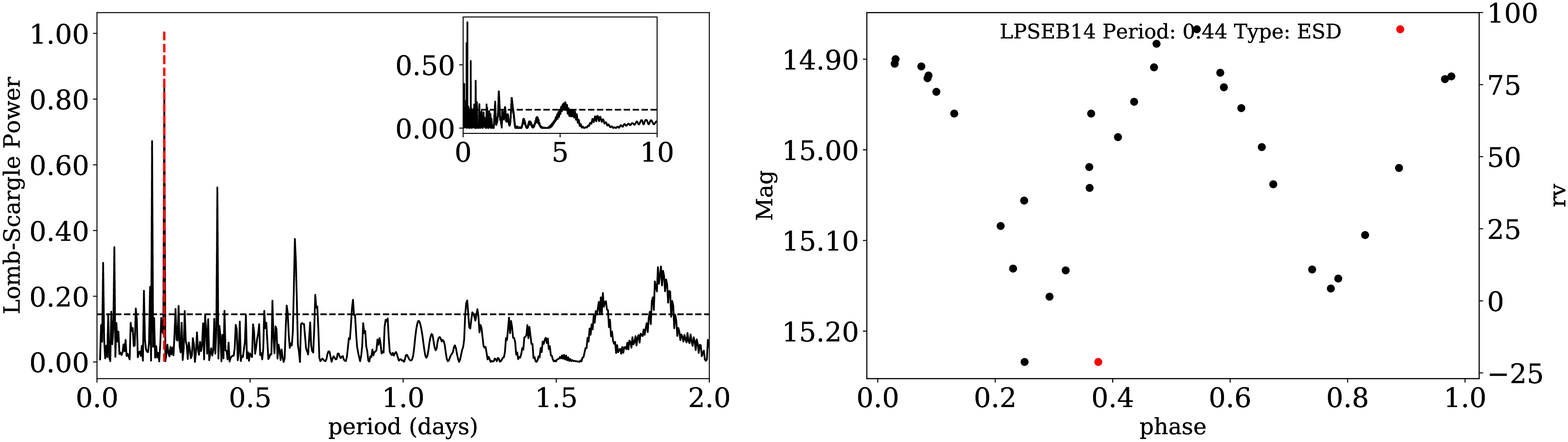}
  \includegraphics[width=4.5in]{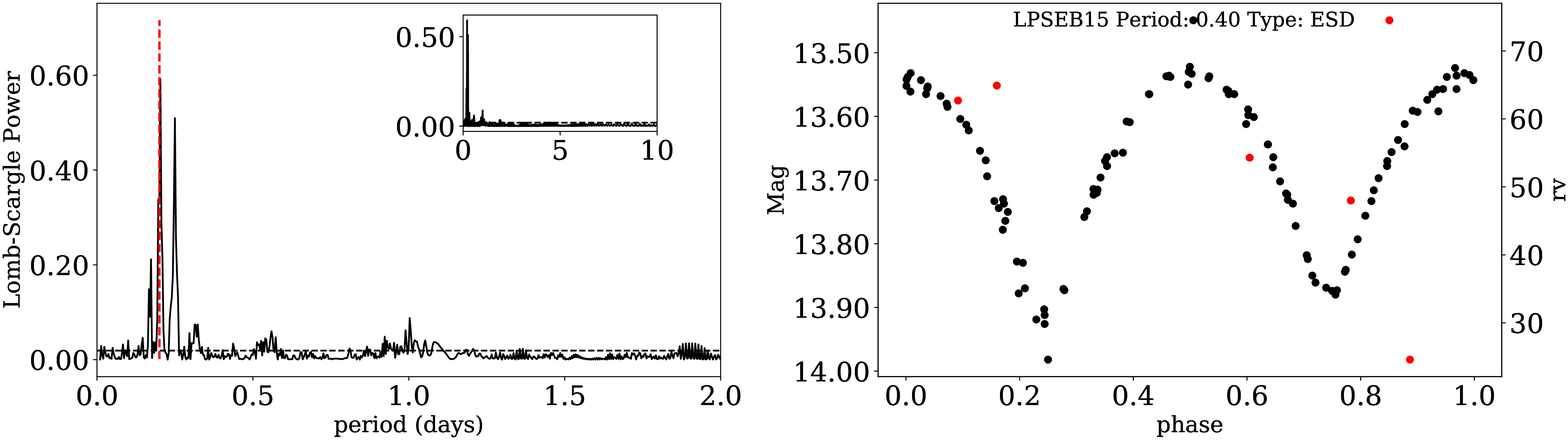}
  \includegraphics[width=4.5in]{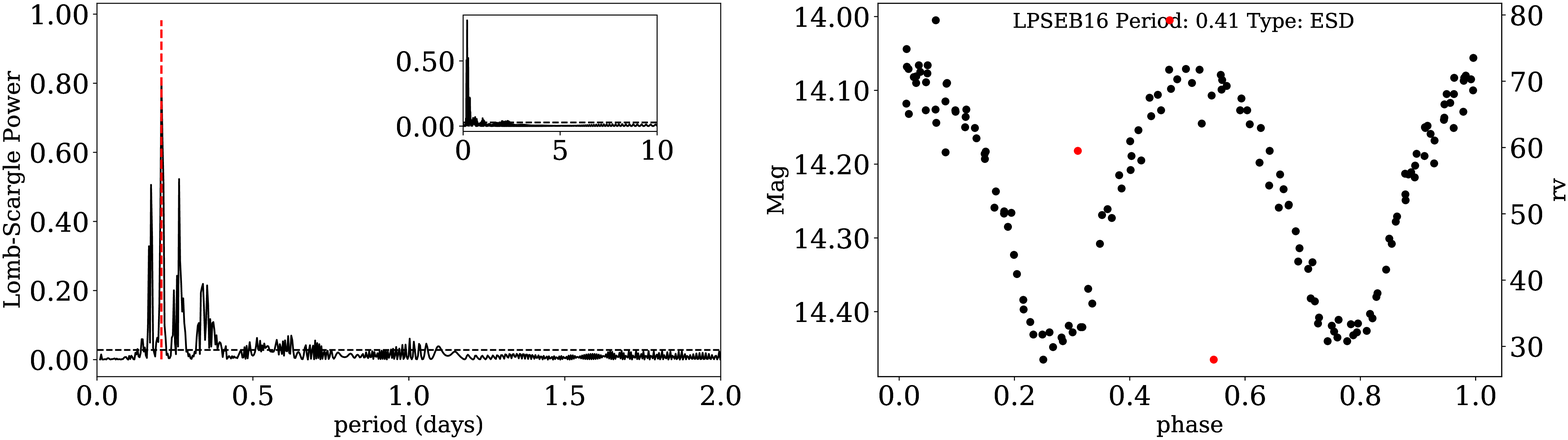}
  \includegraphics[width=4.5in]{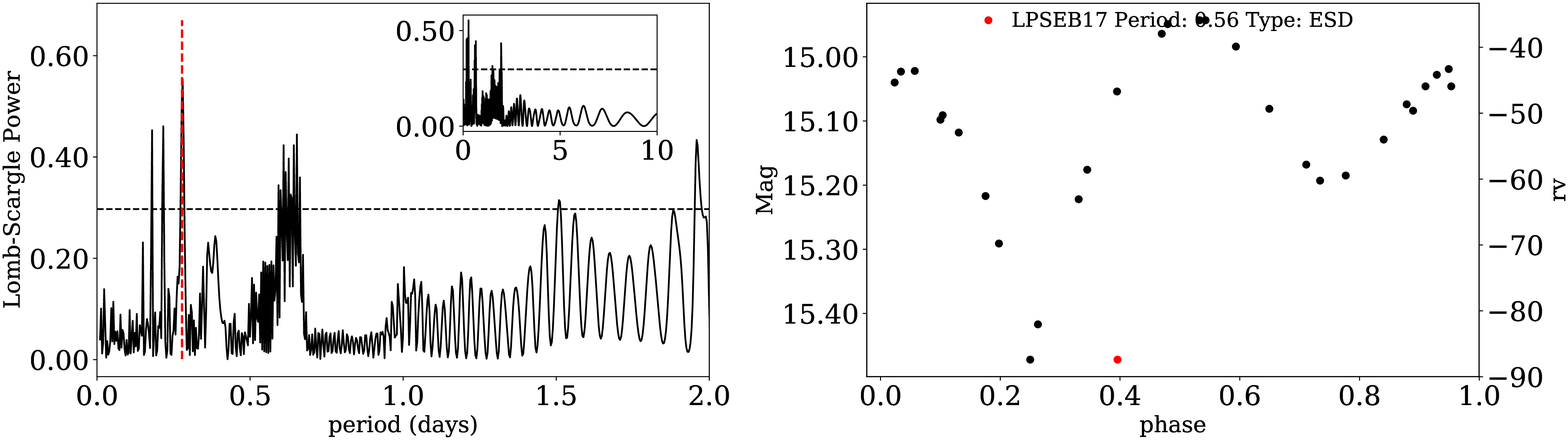}
  \includegraphics[width=4.5in]{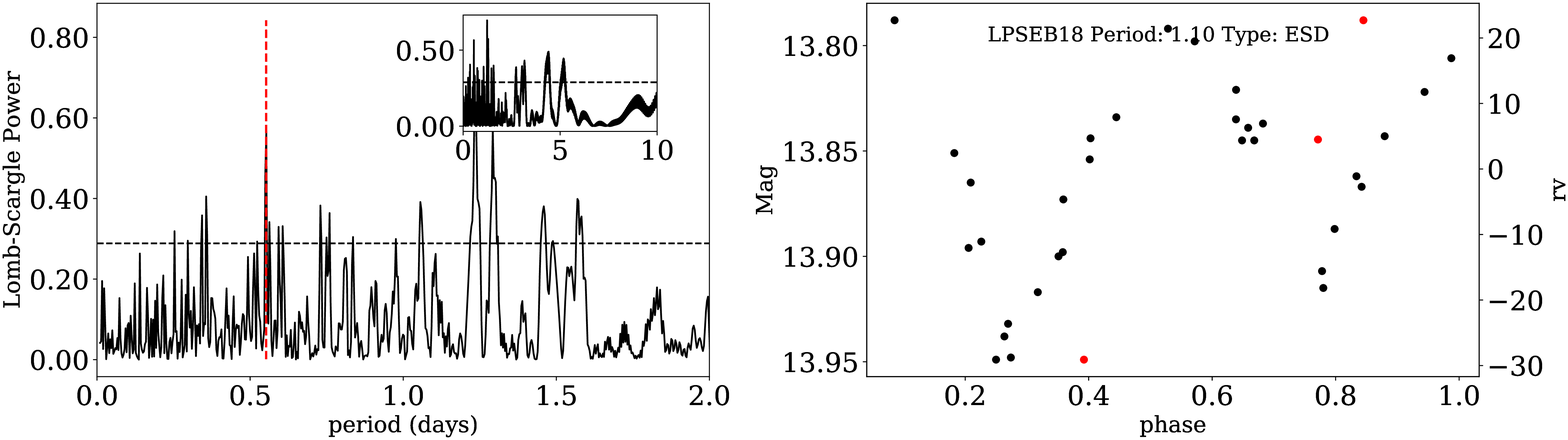}
    
\caption{(Continued) }
  \label{full}
\end{figure*}

\newpage
\addtocounter{figure}{-1}
\begin{figure*}[!htb]%
  
  \centering
 \includegraphics[width=4.5in]{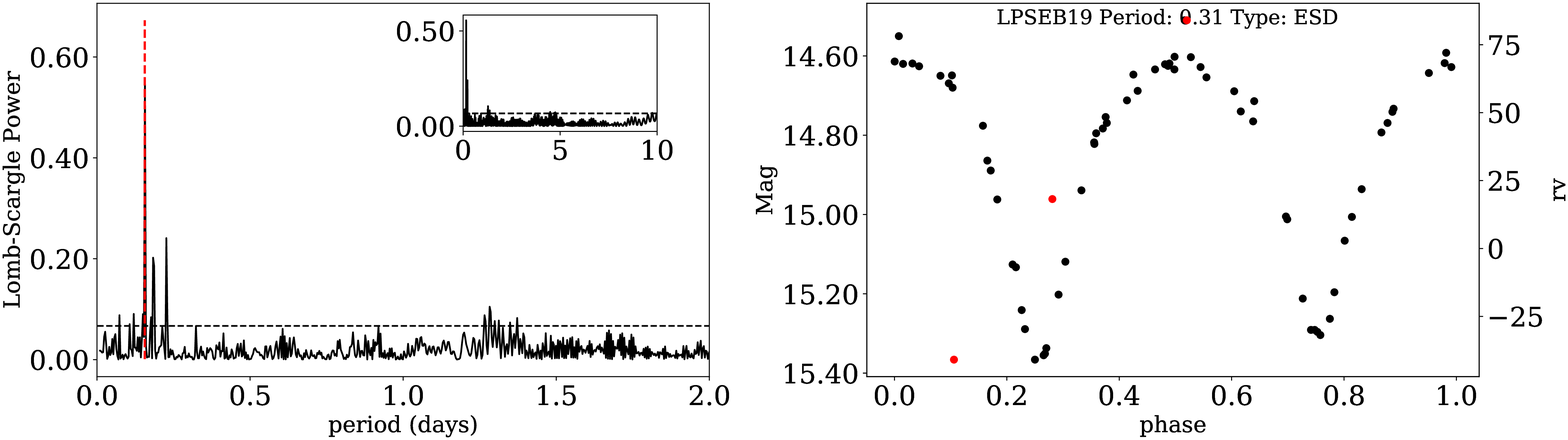}
   \includegraphics[width=4.5in]{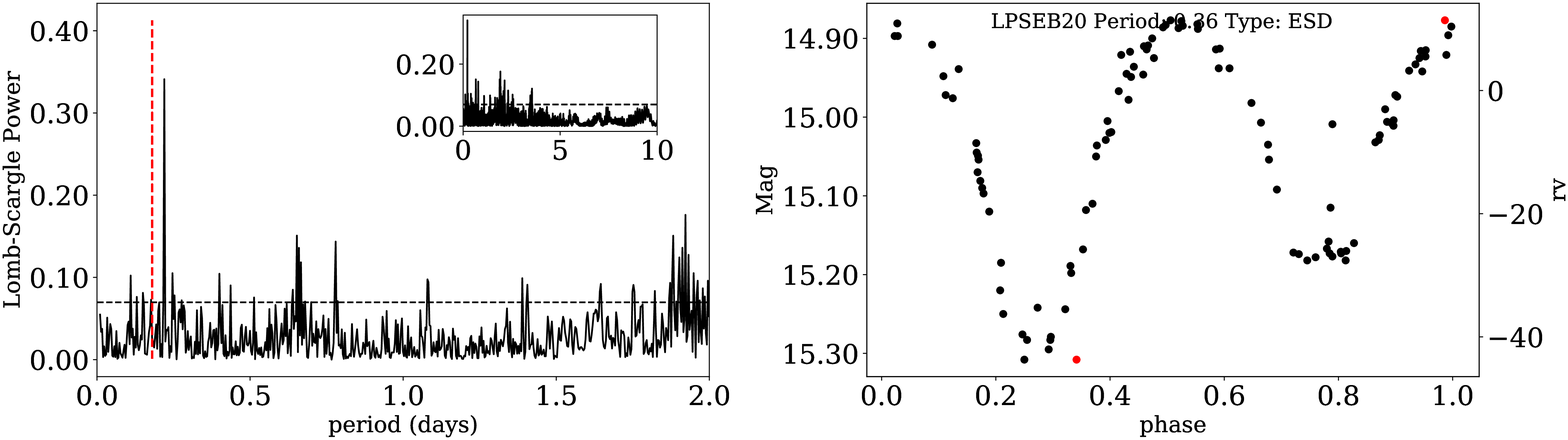}
  \includegraphics[width=4.5in]{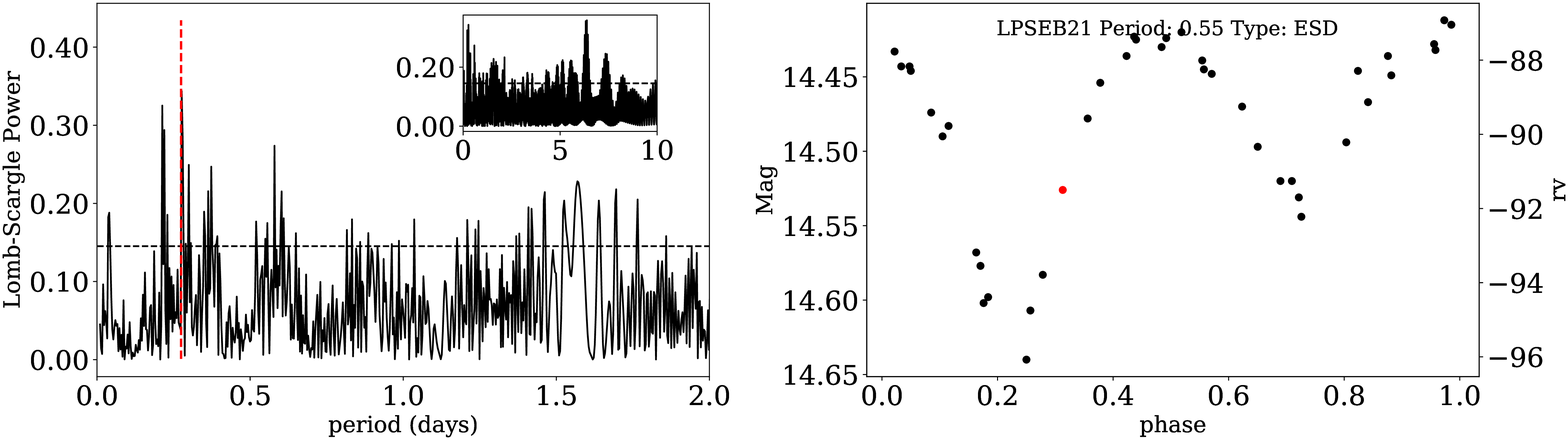}
  \includegraphics[width=4.5in]{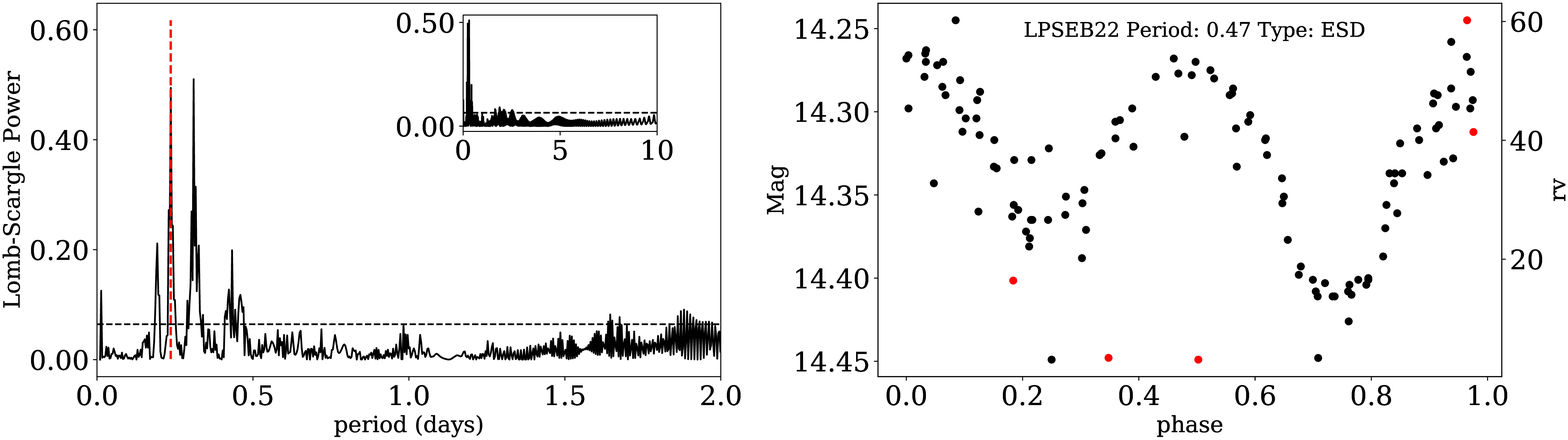}
  \includegraphics[width=4.5in]{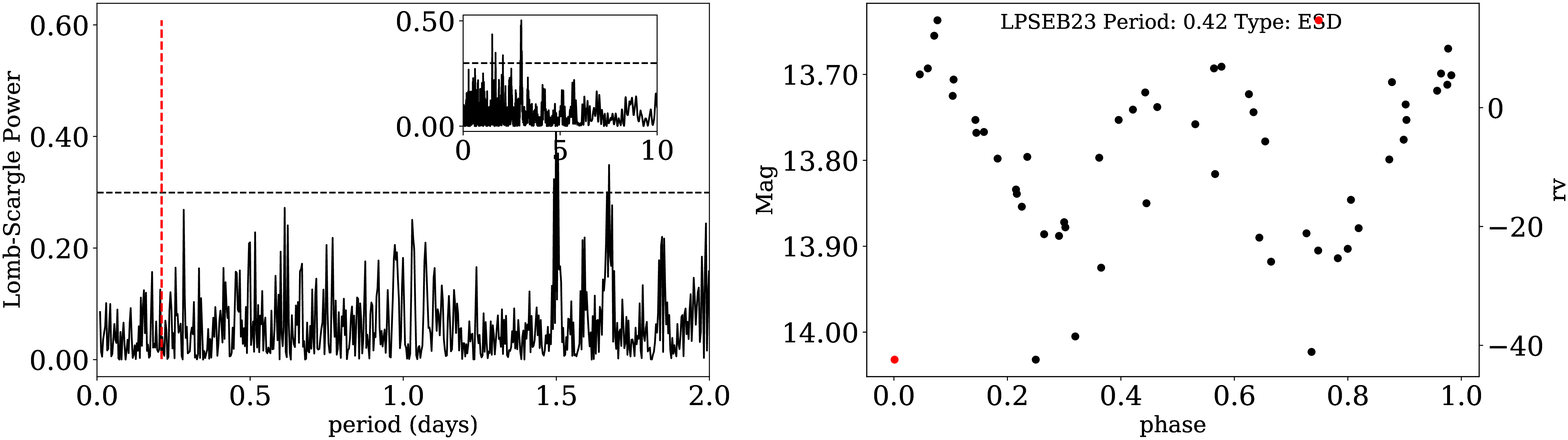}
  \includegraphics[width=4.5in]{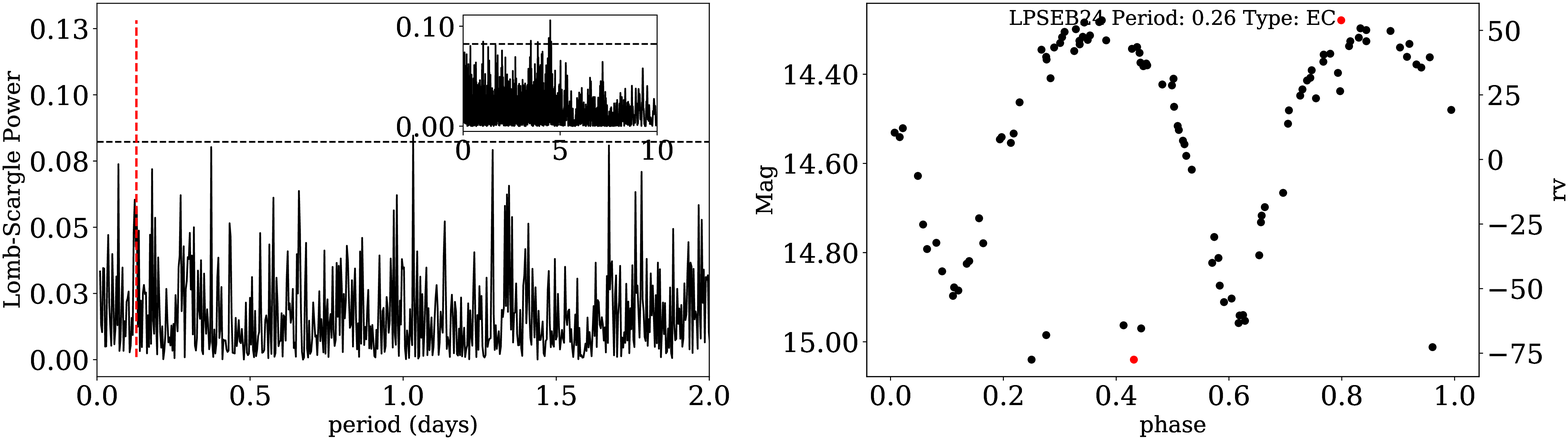}
    
\caption{(Continued) }
  \label{full}
\end{figure*}

\newpage
\addtocounter{figure}{-1}
\begin{figure*}[!htb]%
  
  \centering
 \includegraphics[width=4.5in]{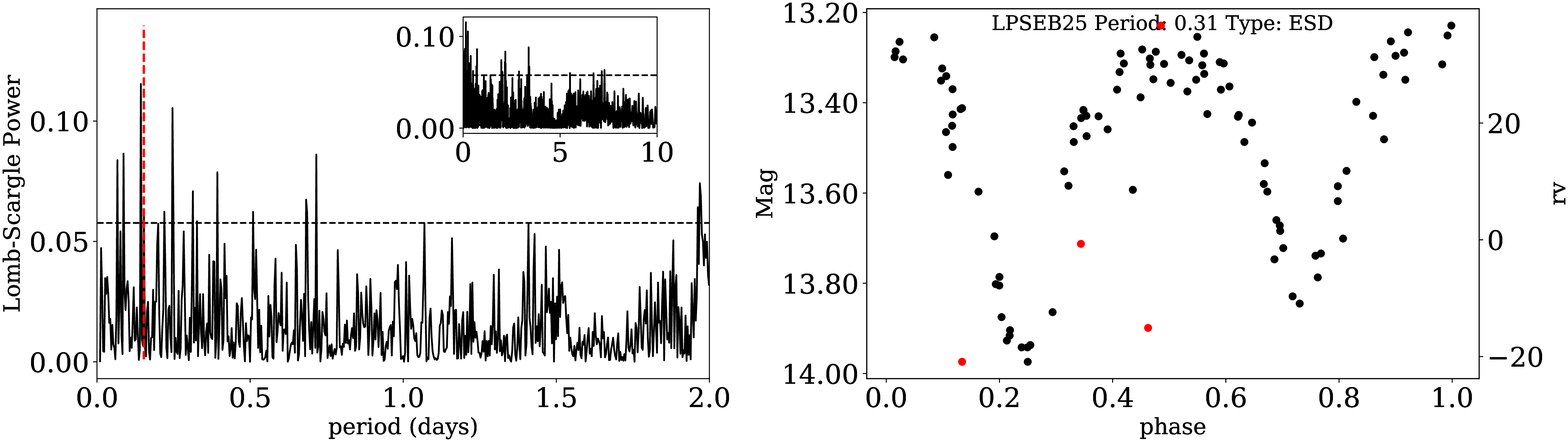}
   \includegraphics[width=4.5in]{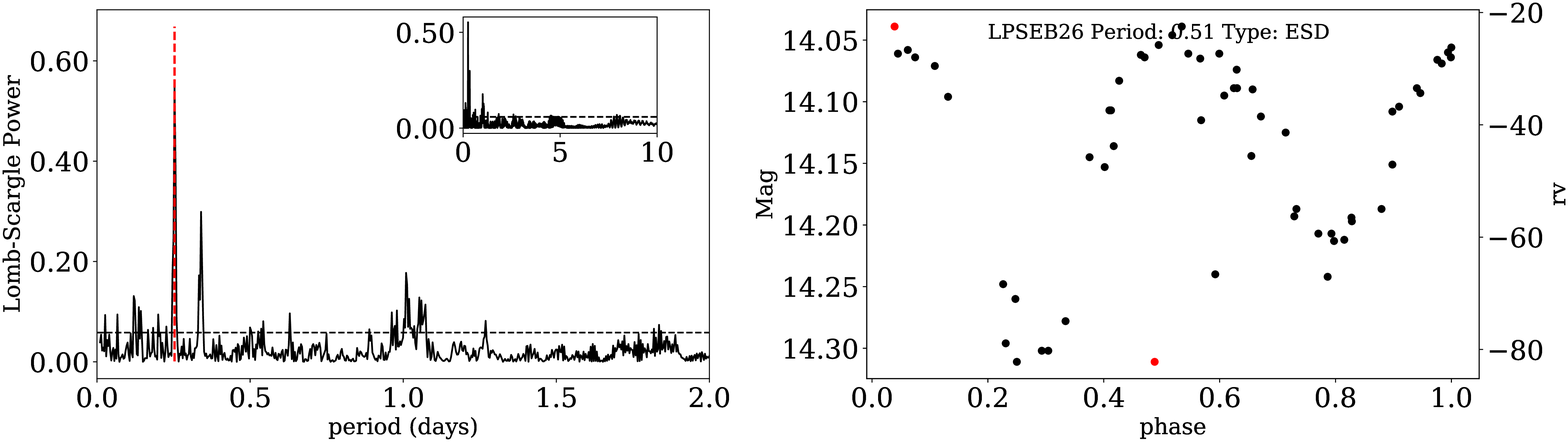}
  \includegraphics[width=4.5in]{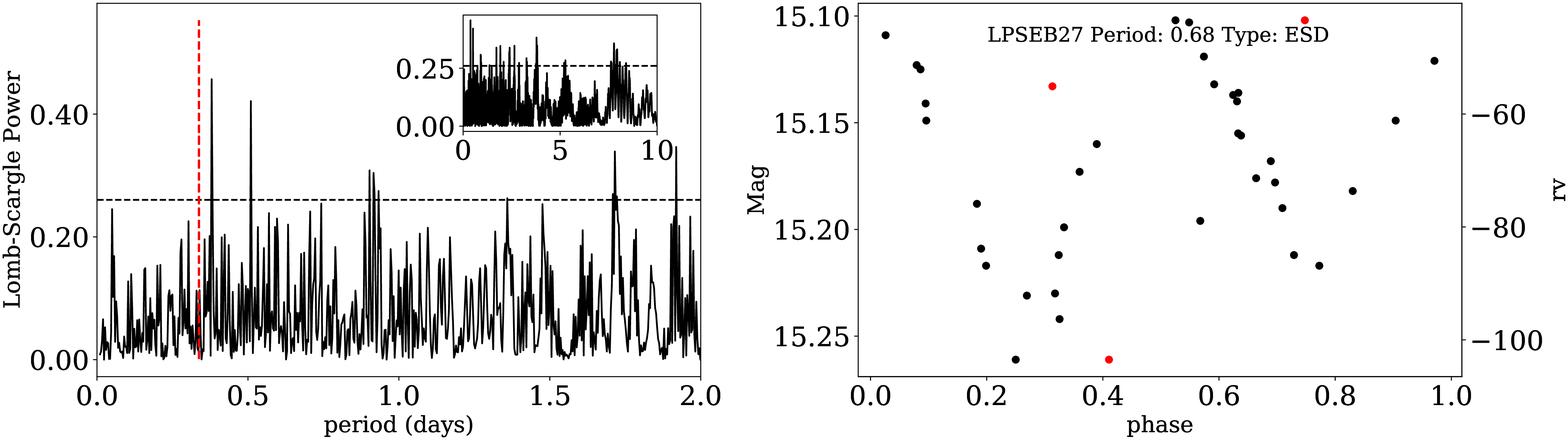}
  \includegraphics[width=4.5in]{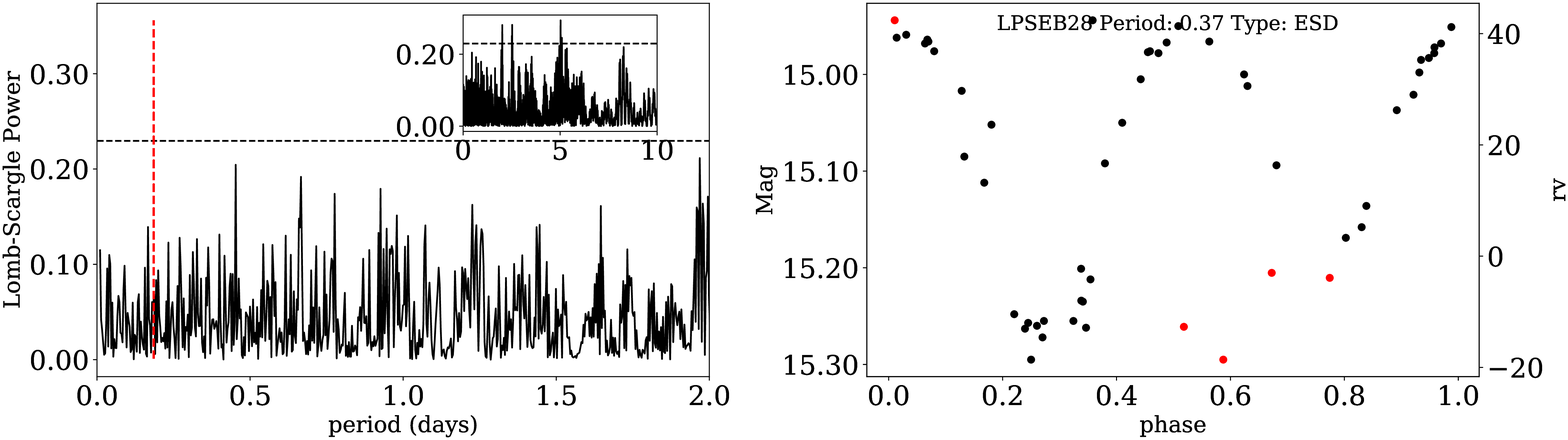}
  \includegraphics[width=4.5in]{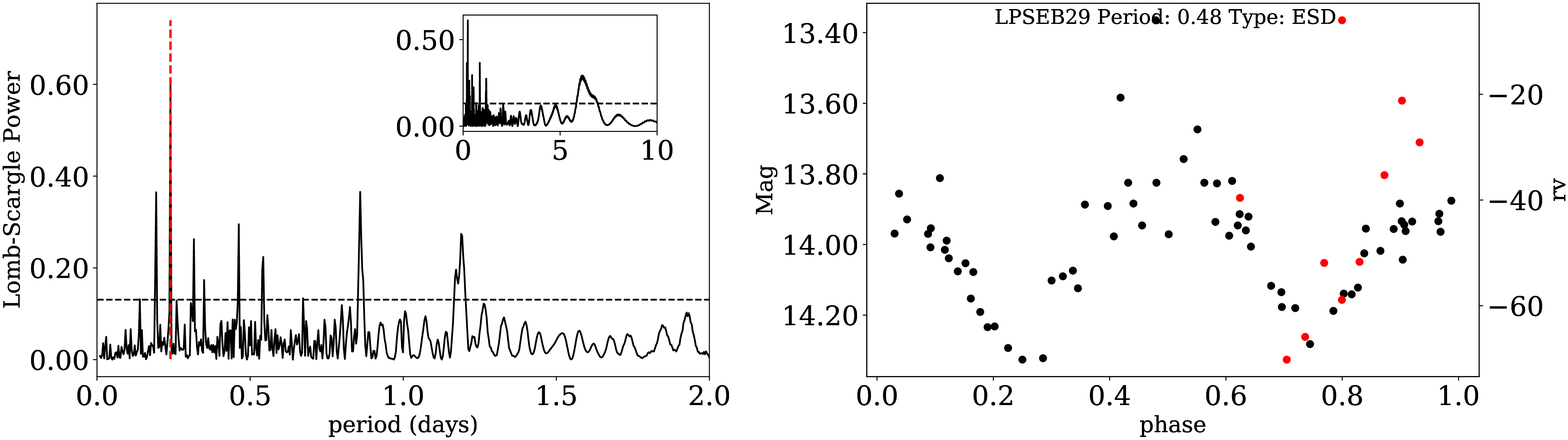}
  \includegraphics[width=4.5in]{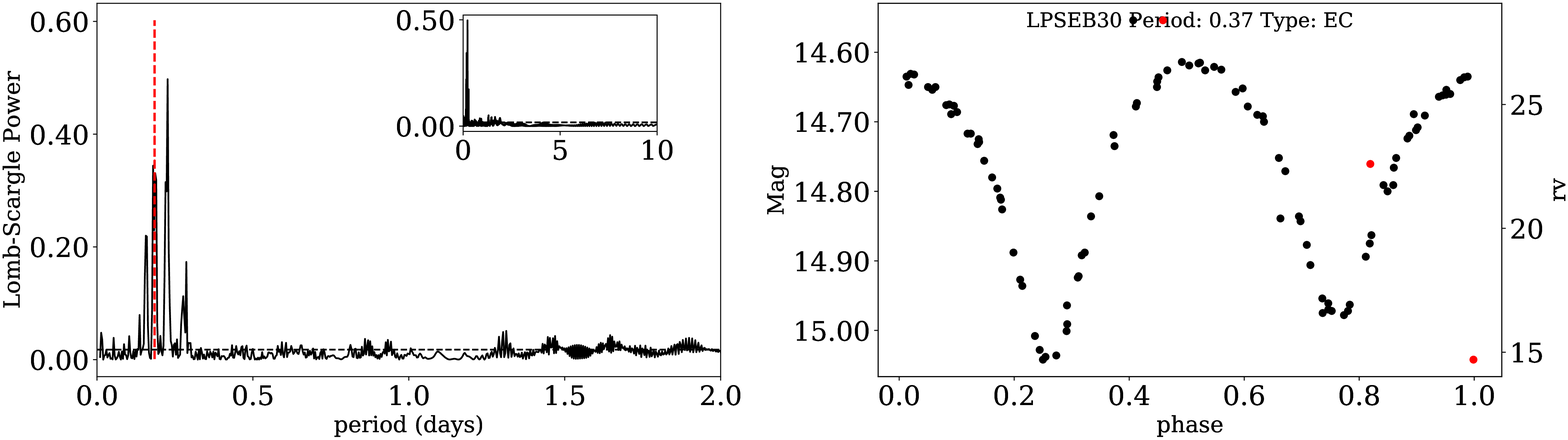}
    
\caption{(Continued) }
  \label{fig1}
\end{figure*}

\newpage
\addtocounter{figure}{-1}
\begin{figure*}[!htb]%
  
  \centering
 \includegraphics[width=4.5in]{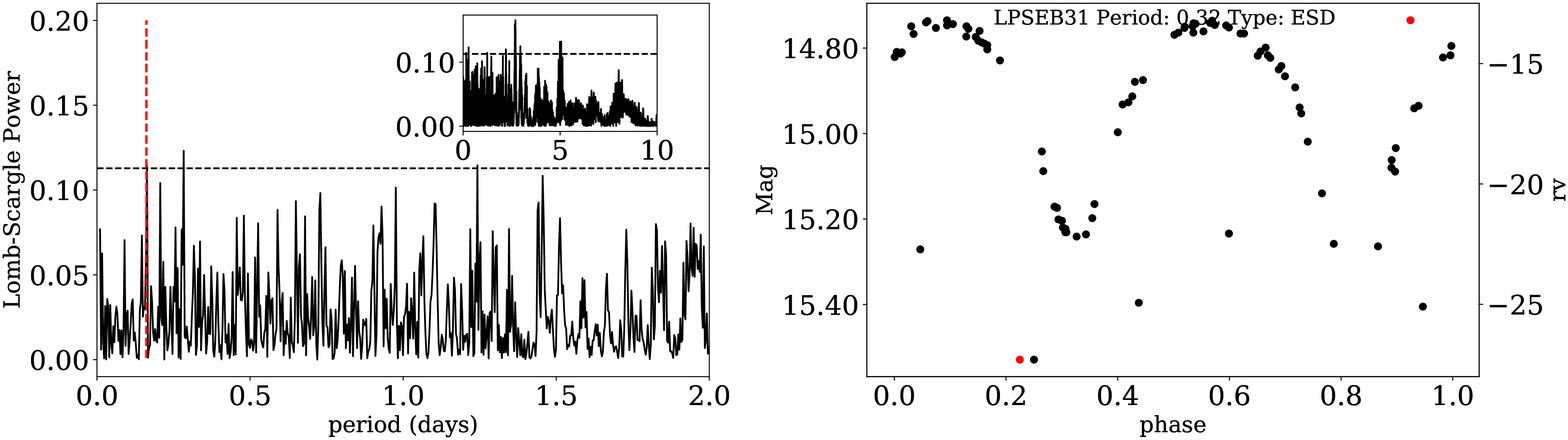}
   \includegraphics[width=4.5in]{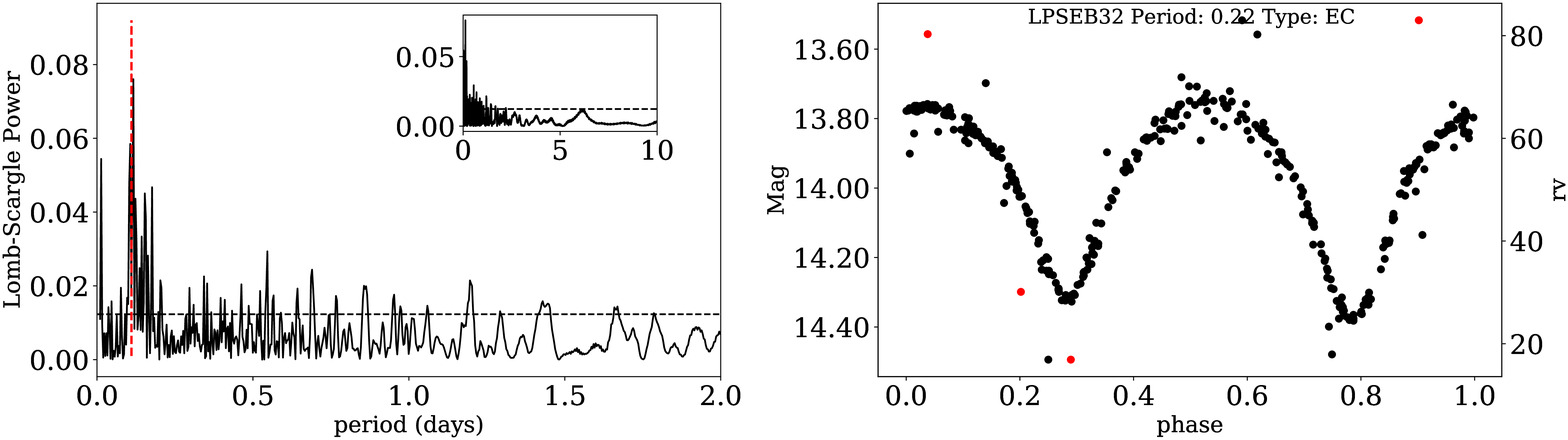}
  \includegraphics[width=4.5in]{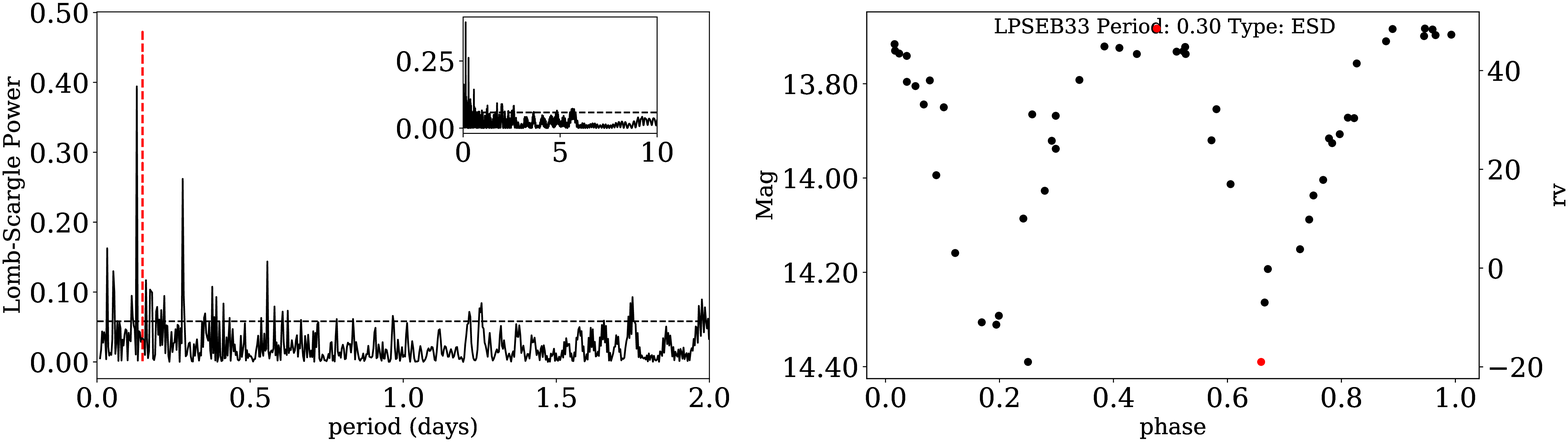}
  \includegraphics[width=4.5in]{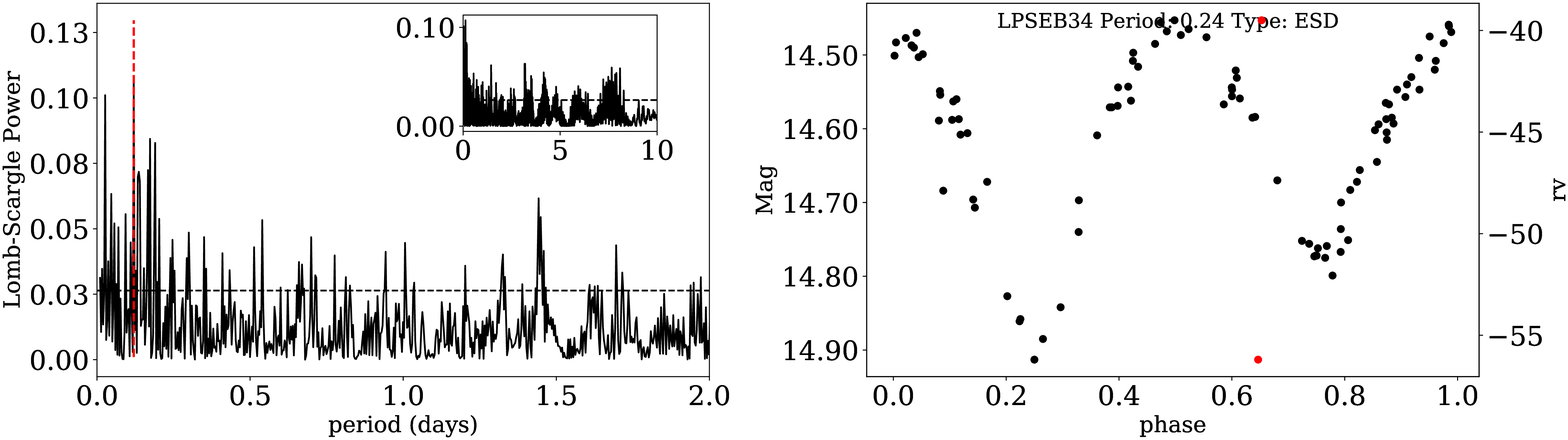}
  \includegraphics[width=4.5in]{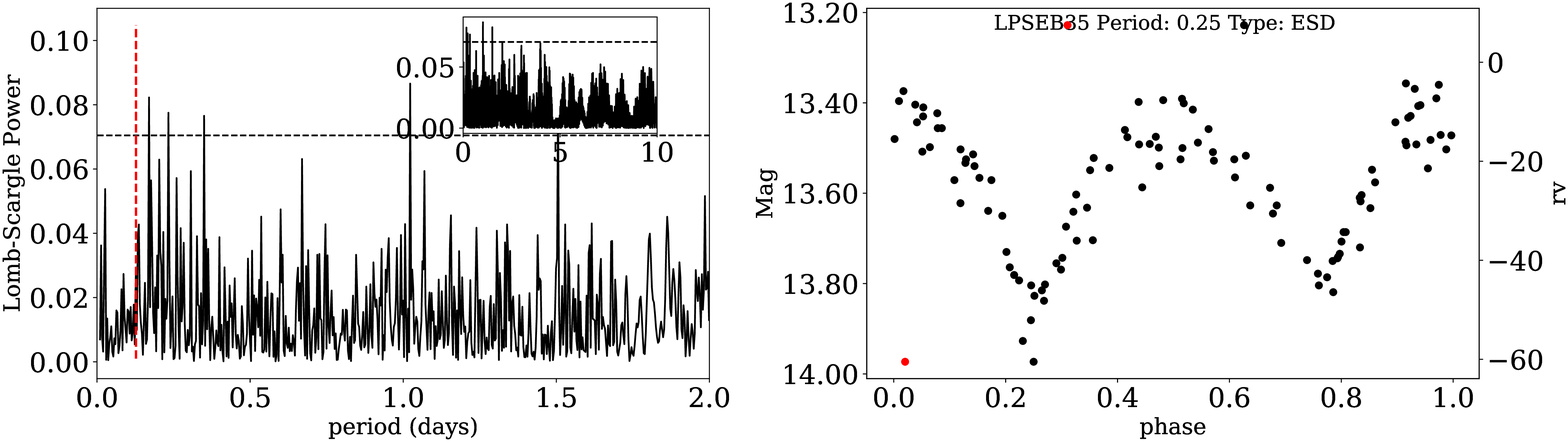}
  \includegraphics[width=4.5in]{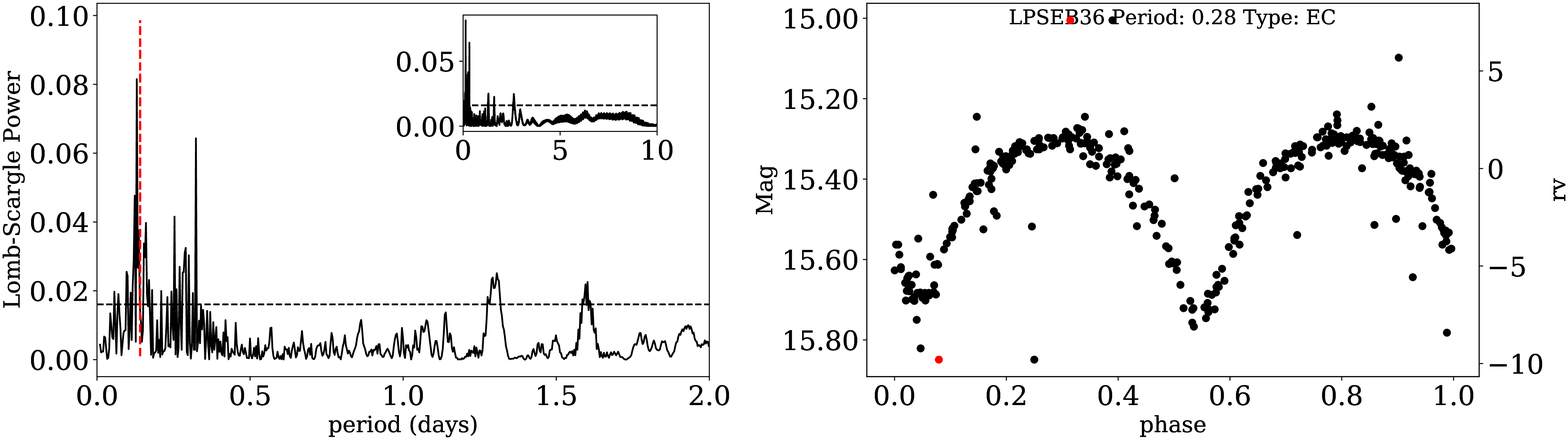}
    
\caption{(Continued) }
  \label{full}
\end{figure*}

\newpage
\addtocounter{figure}{-1}
\begin{figure*}[!htb]%
  
  \centering
 \includegraphics[width=4.5in]{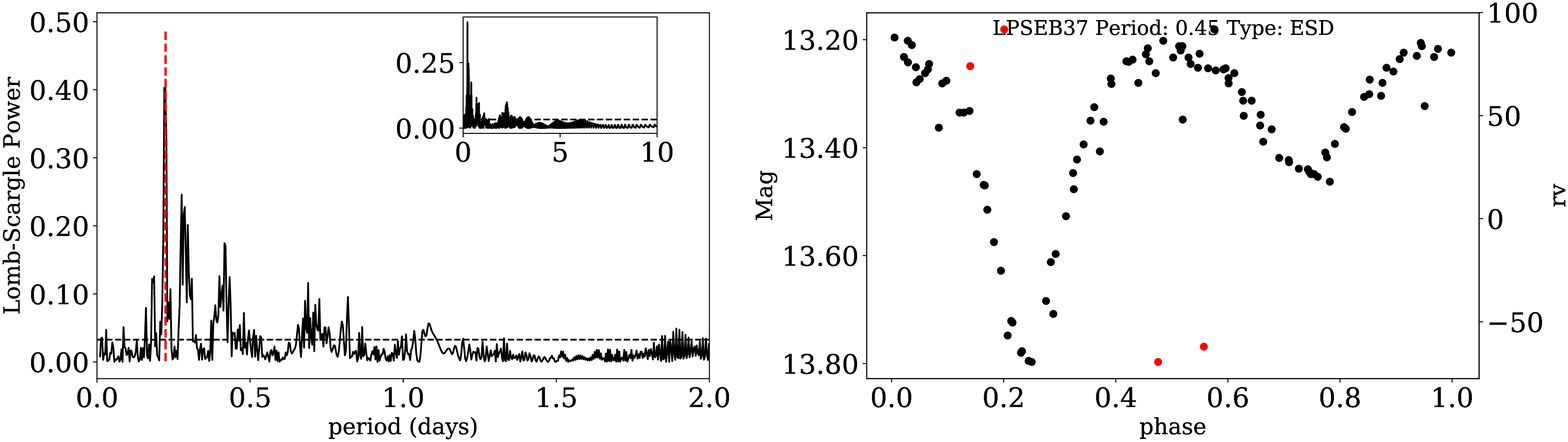}
   \includegraphics[width=4.5in]{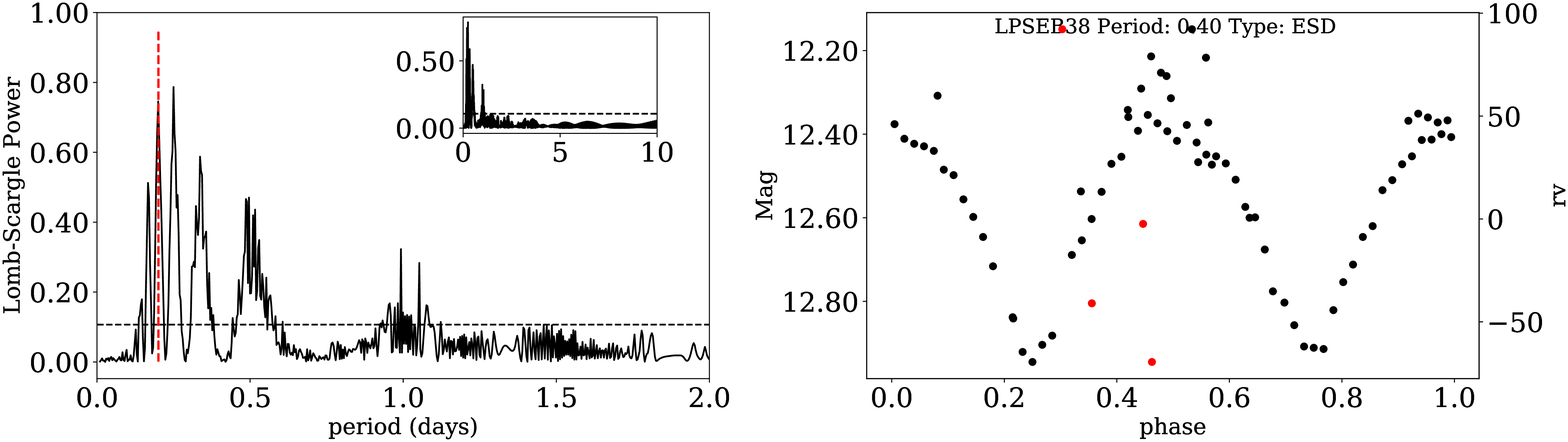}
  \includegraphics[width=4.5in]{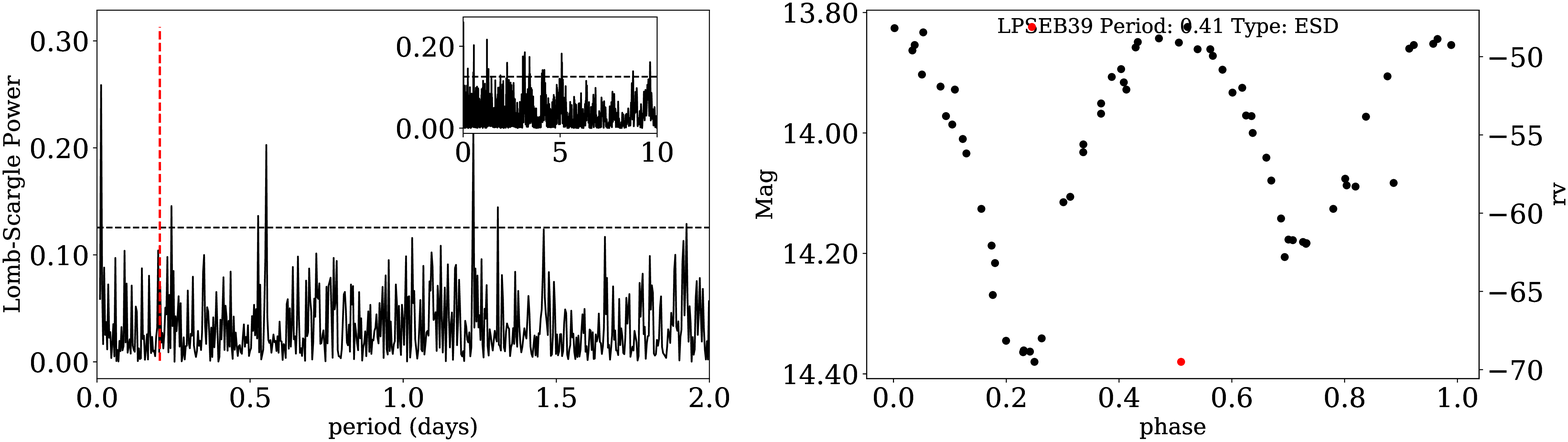}
  \includegraphics[width=4.5in]{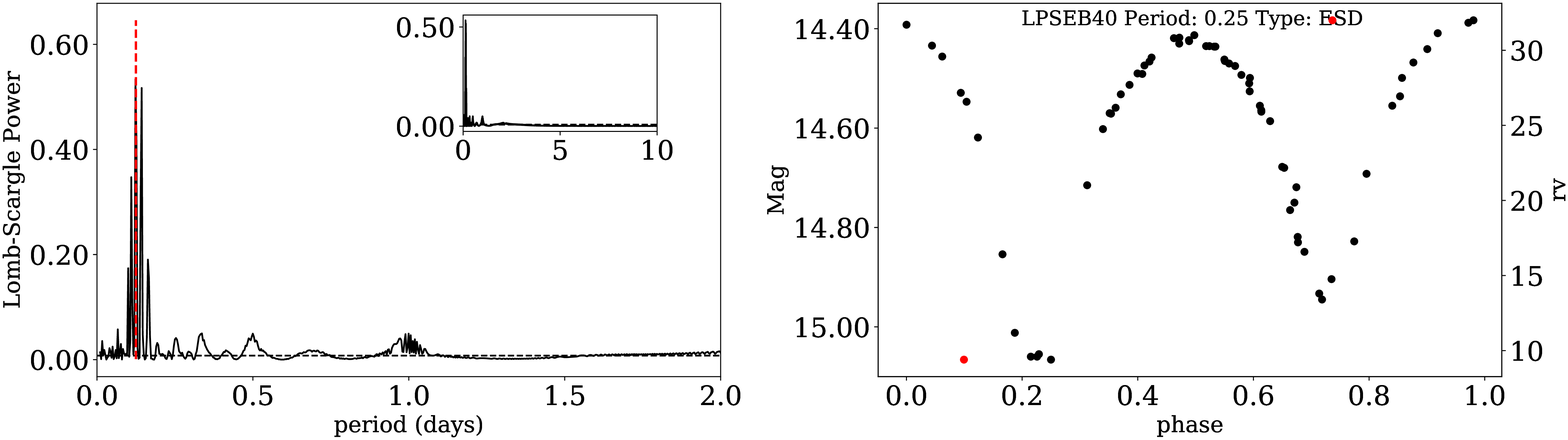}
  \includegraphics[width=4.5in]{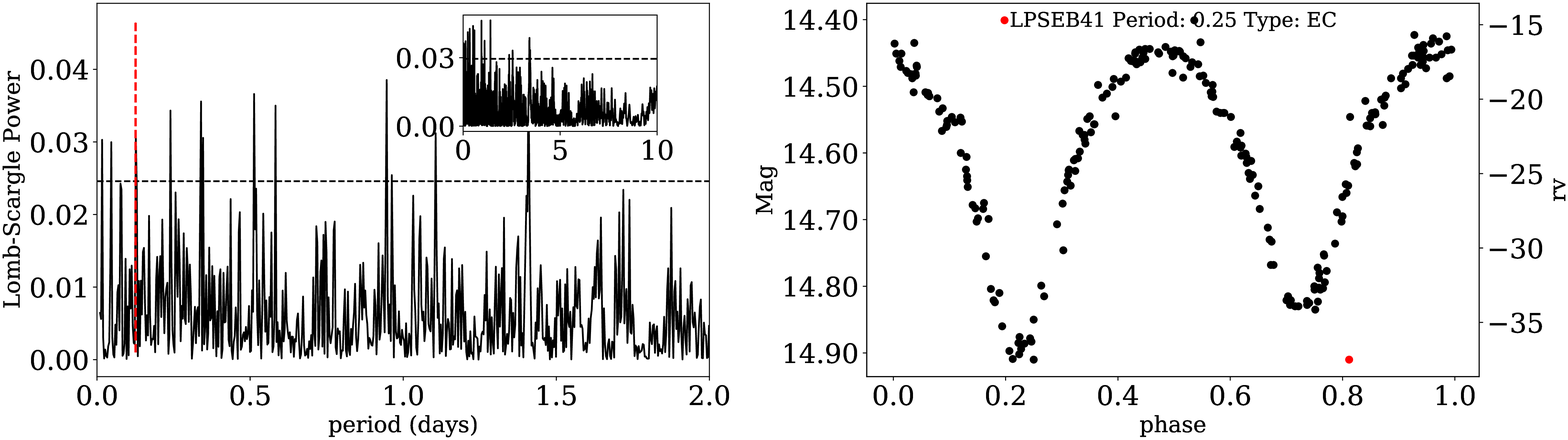}
  \includegraphics[width=4.5in]{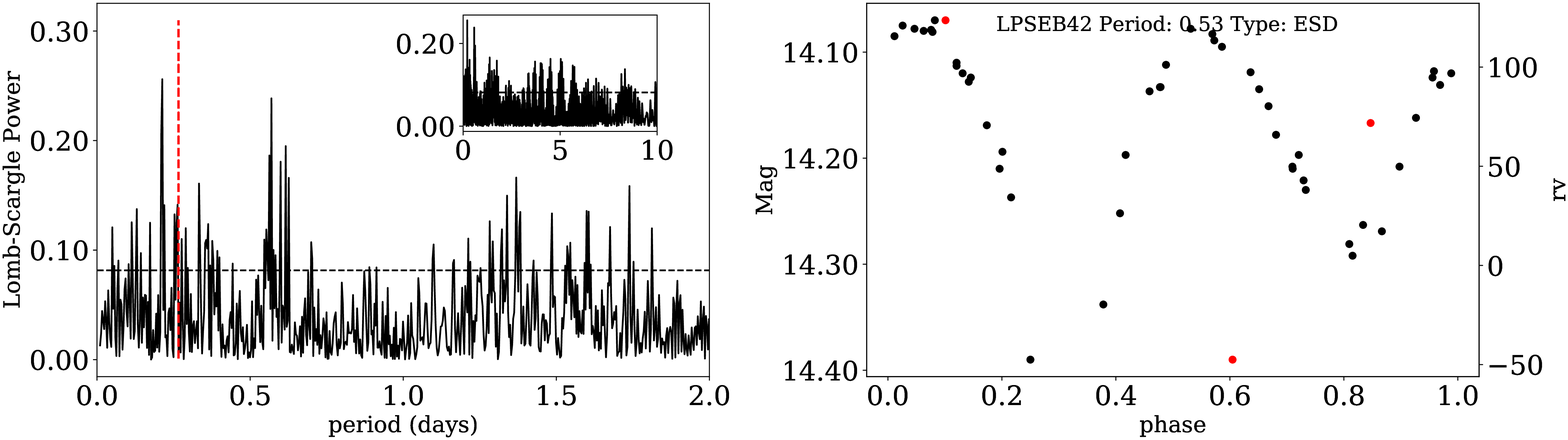}
    
\caption{(Continued) }
  \label{fig1}
\end{figure*}

\newpage
\addtocounter{figure}{-1}
\begin{figure*}[!htb]%
  
  \centering
 \includegraphics[width=4.5in]{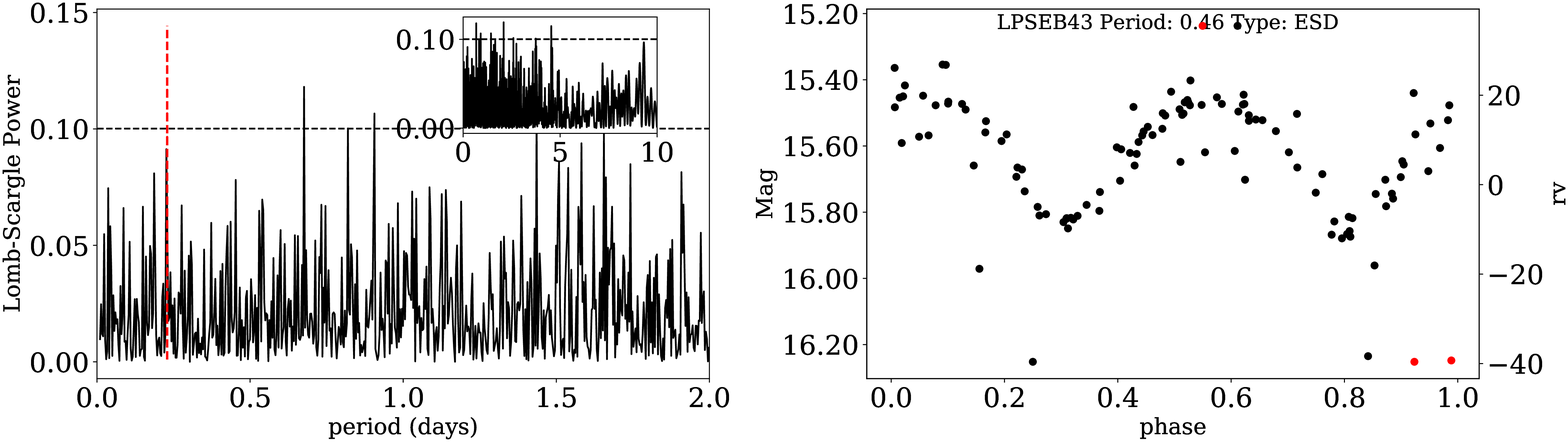}
   \includegraphics[width=4.5in]{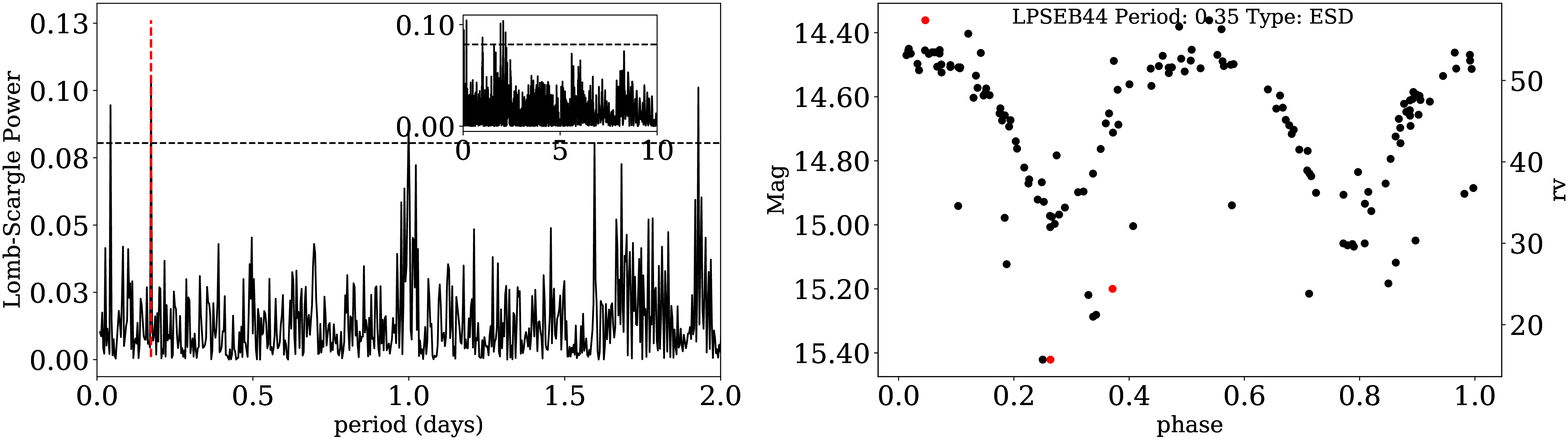}
  \includegraphics[width=4.5in]{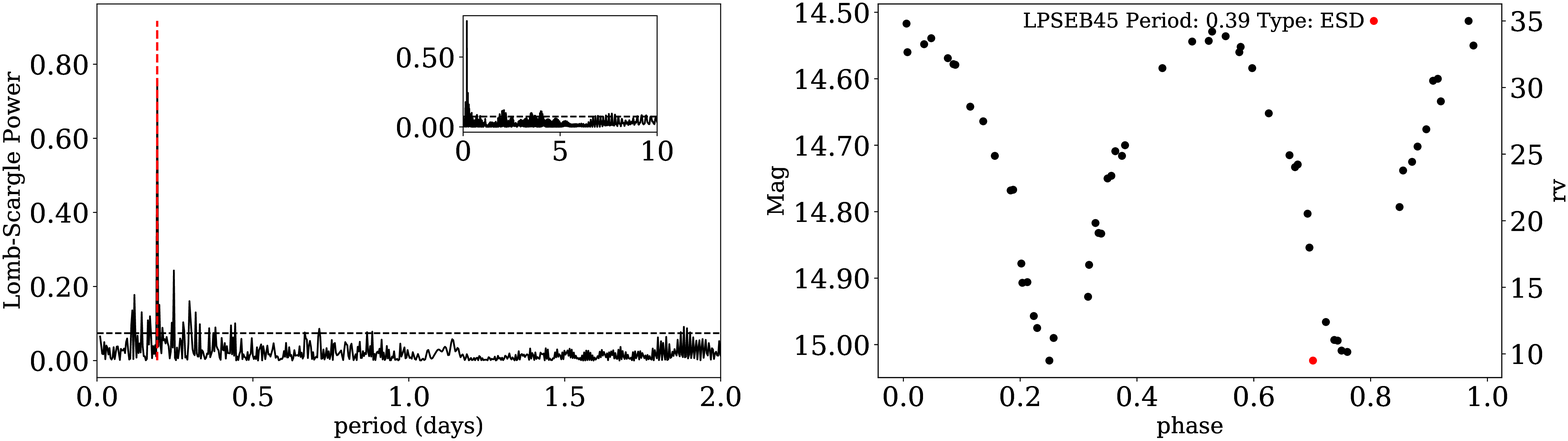}
  \includegraphics[width=4.5in]{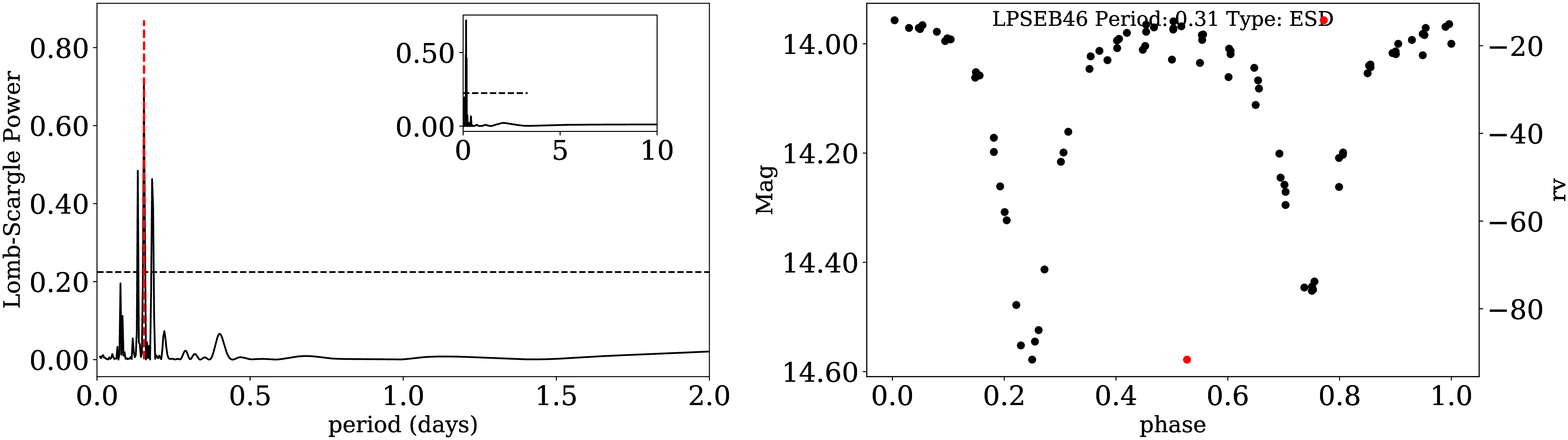}
  \includegraphics[width=4.5in]{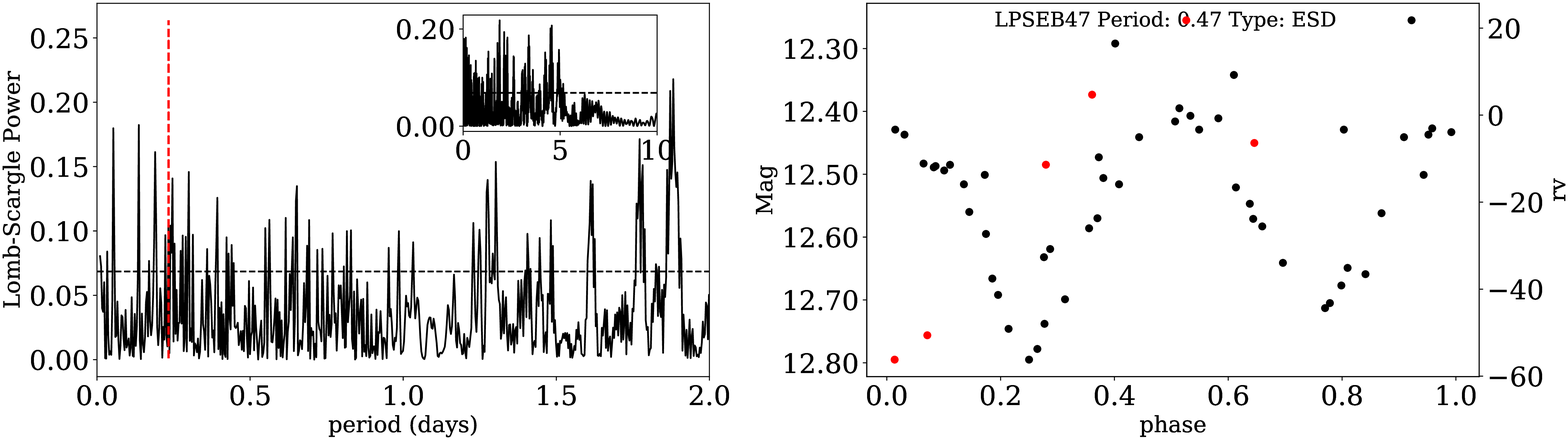}
  \includegraphics[width=4.5in]{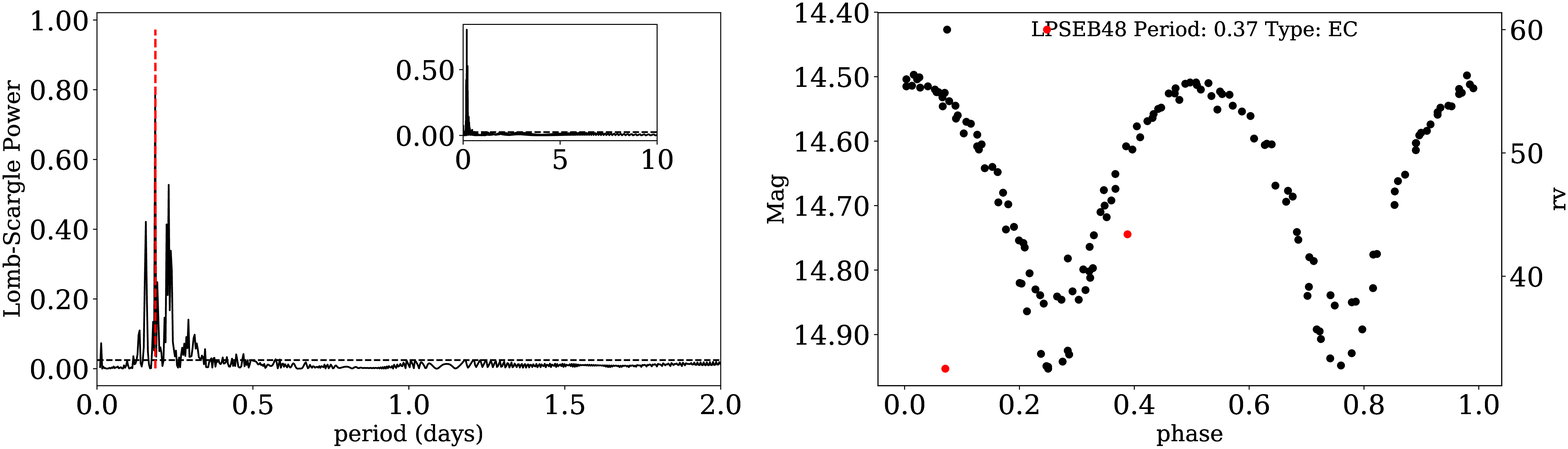}
    
\caption{(Continued) }
  \label{full}
\end{figure*}

\newpage
\addtocounter{figure}{-1}
\begin{figure*}[!htb]%
  
  \centering
 \includegraphics[width=4.5in]{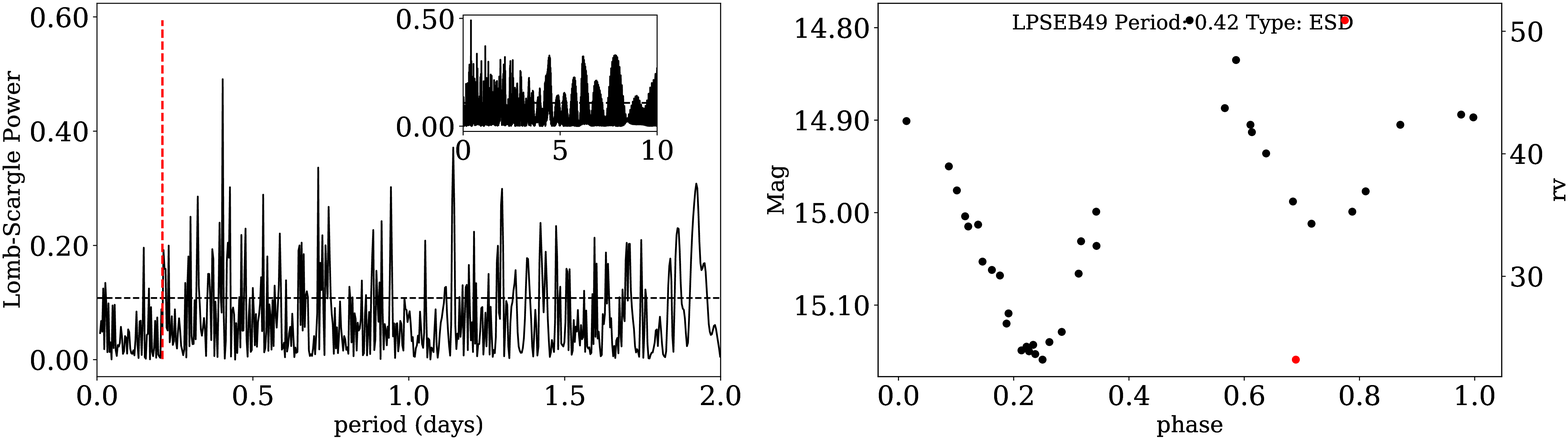}
   \includegraphics[width=4.5in]{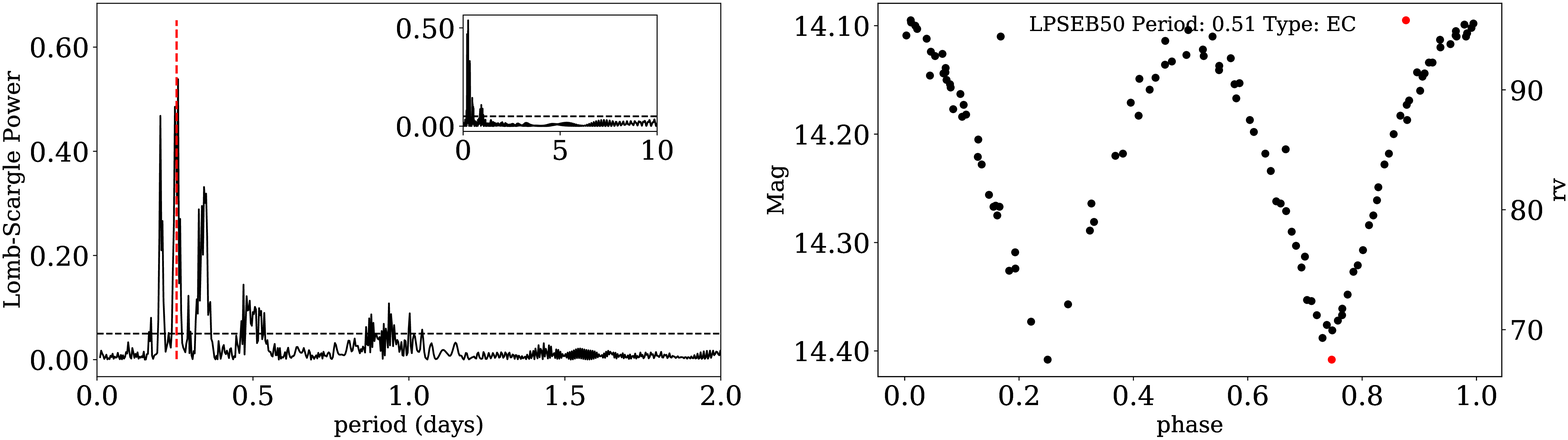}
  \includegraphics[width=4.5in]{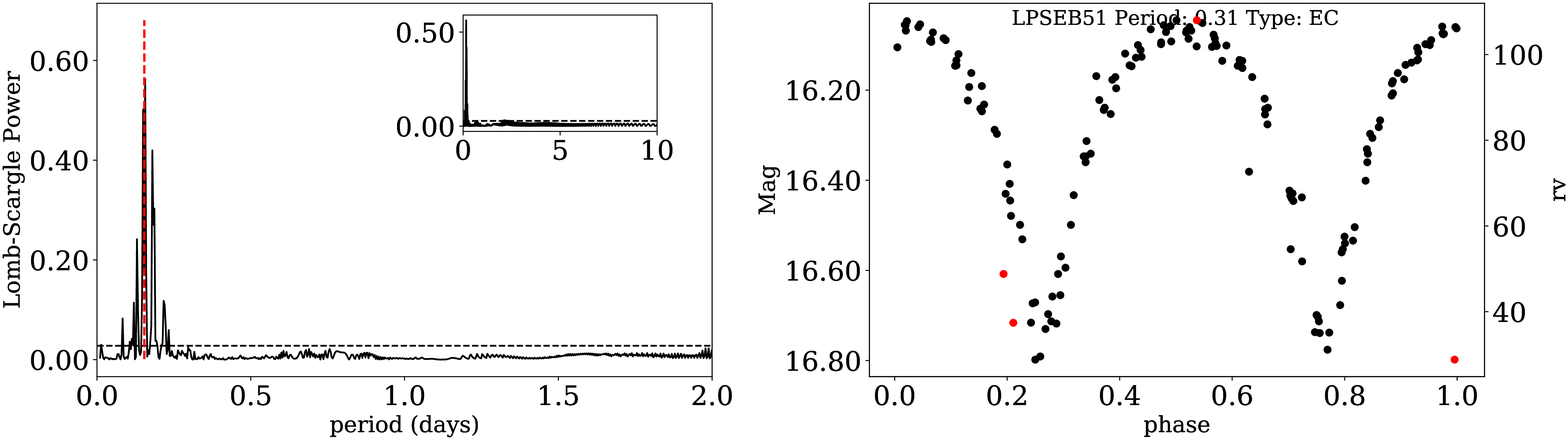}
  \includegraphics[width=4.5in]{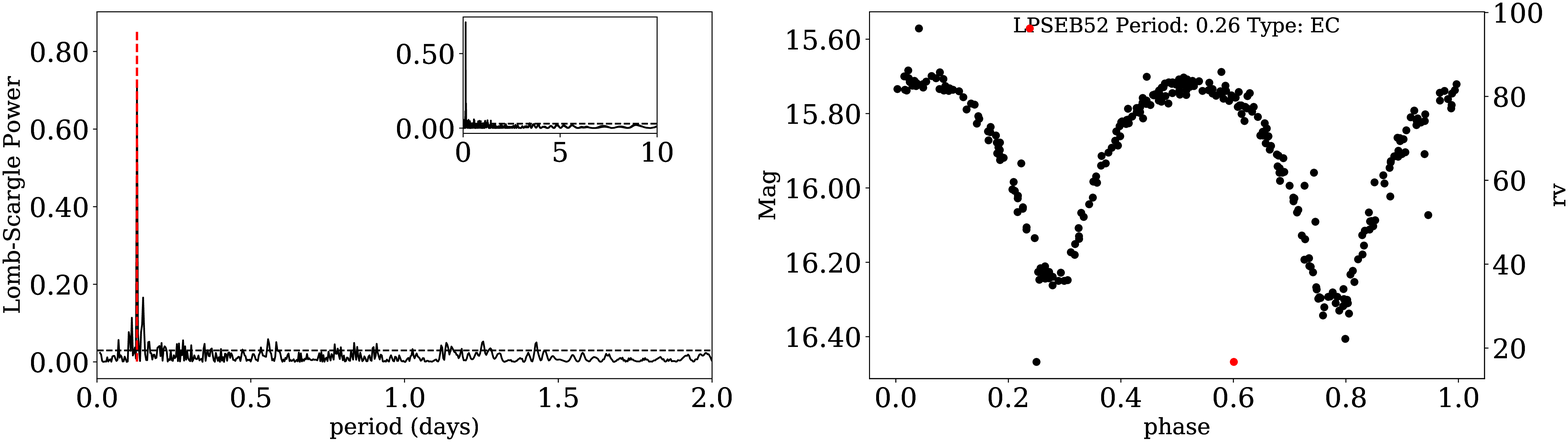}
  \includegraphics[width=4.5in]{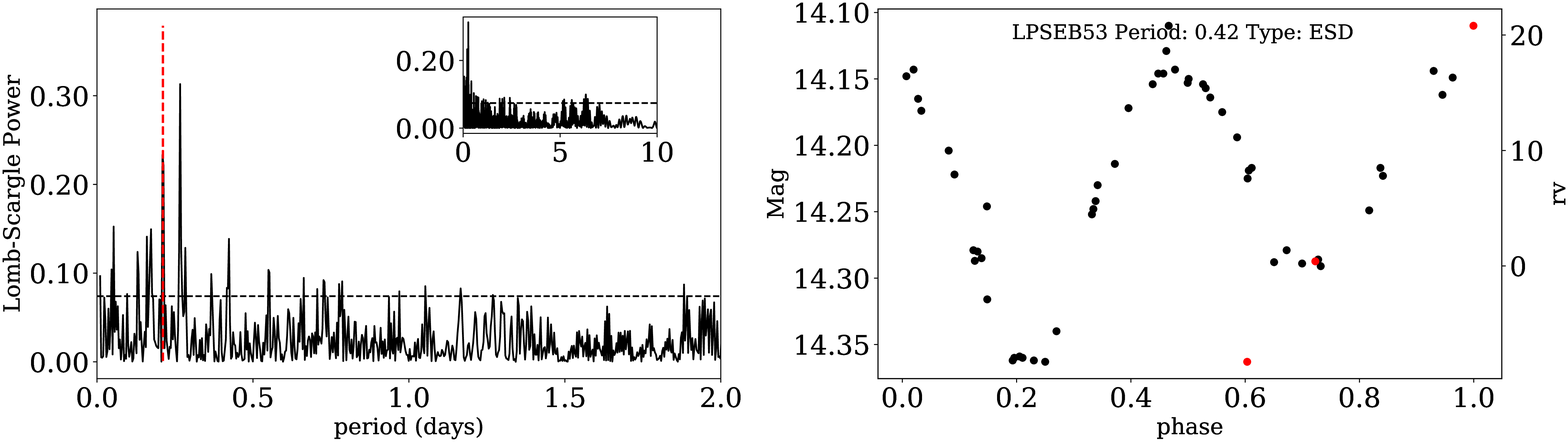}
  \includegraphics[width=4.5in]{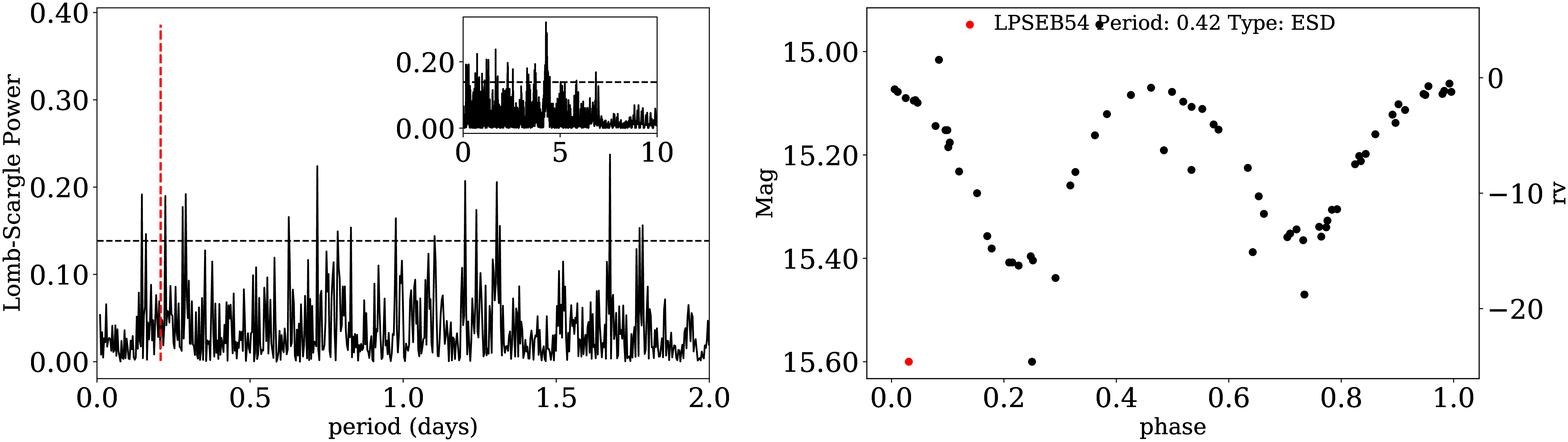}
    
\caption{(Continued) }
  \label{full}
\end{figure*}

\newpage
\addtocounter{figure}{-1}
\begin{figure*}[!htb]%
  
  \centering
 \includegraphics[width=4.5in]{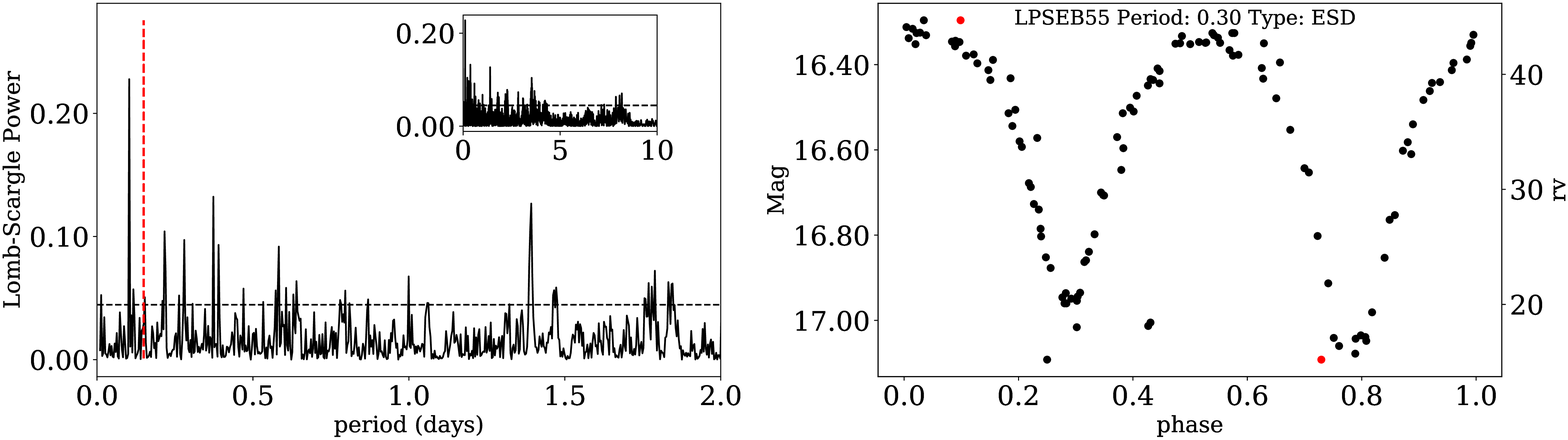}
   \includegraphics[width=4.5in]{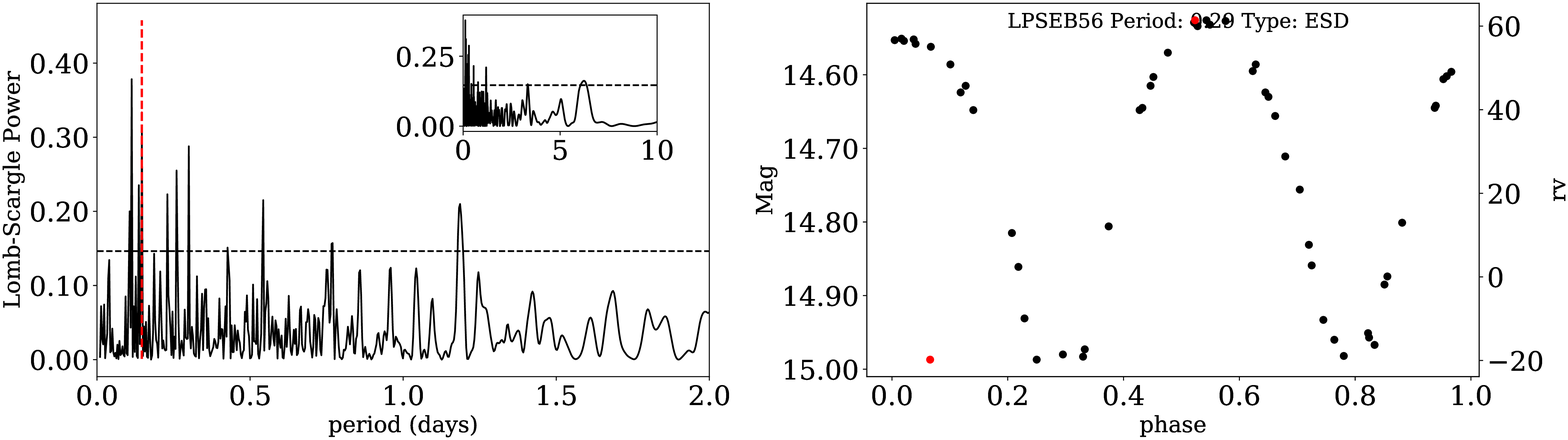}
  \includegraphics[width=4.5in]{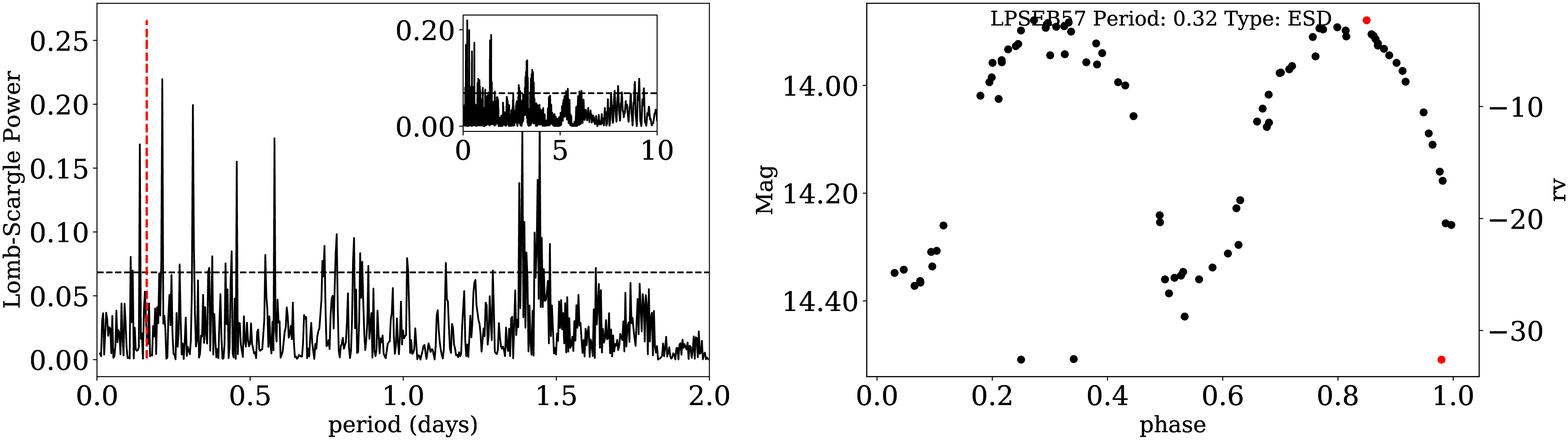}
  \includegraphics[width=4.5in]{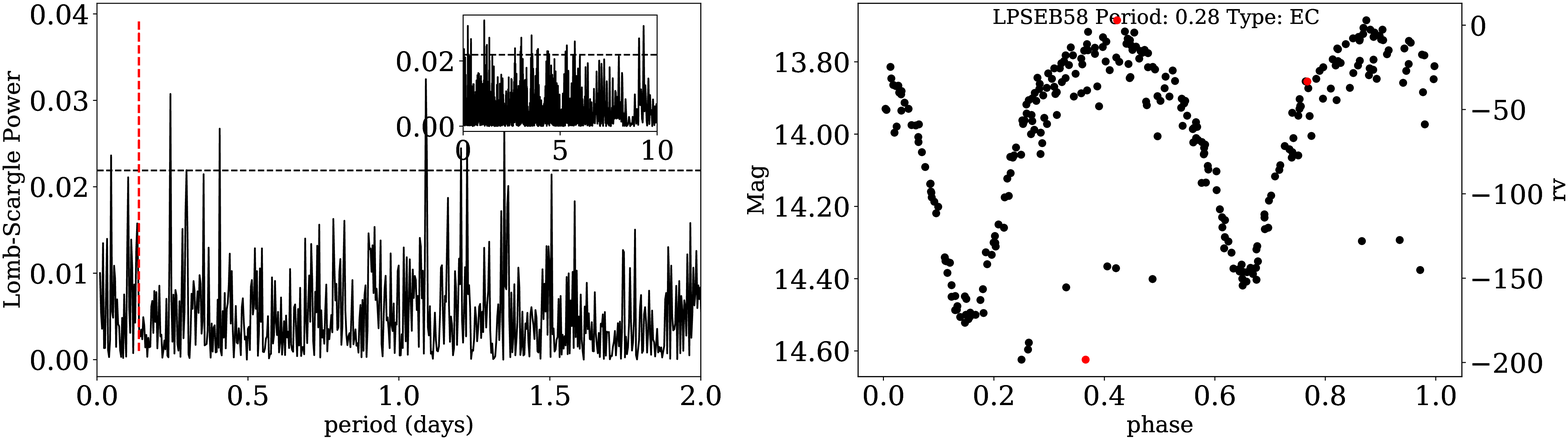}
  \includegraphics[width=4.5in]{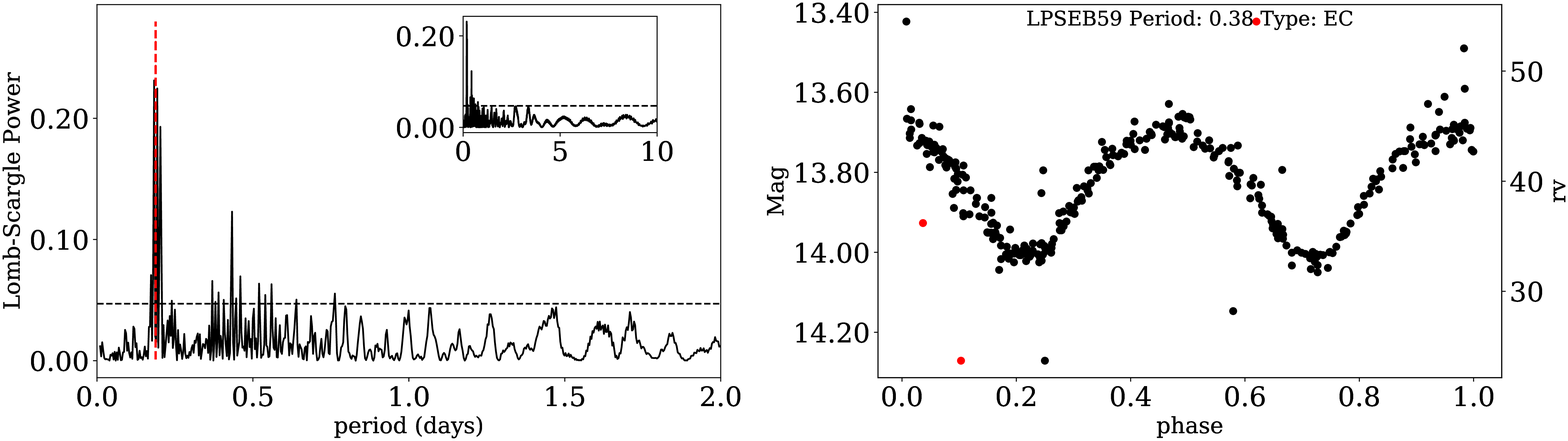}
  \includegraphics[width=4.5in]{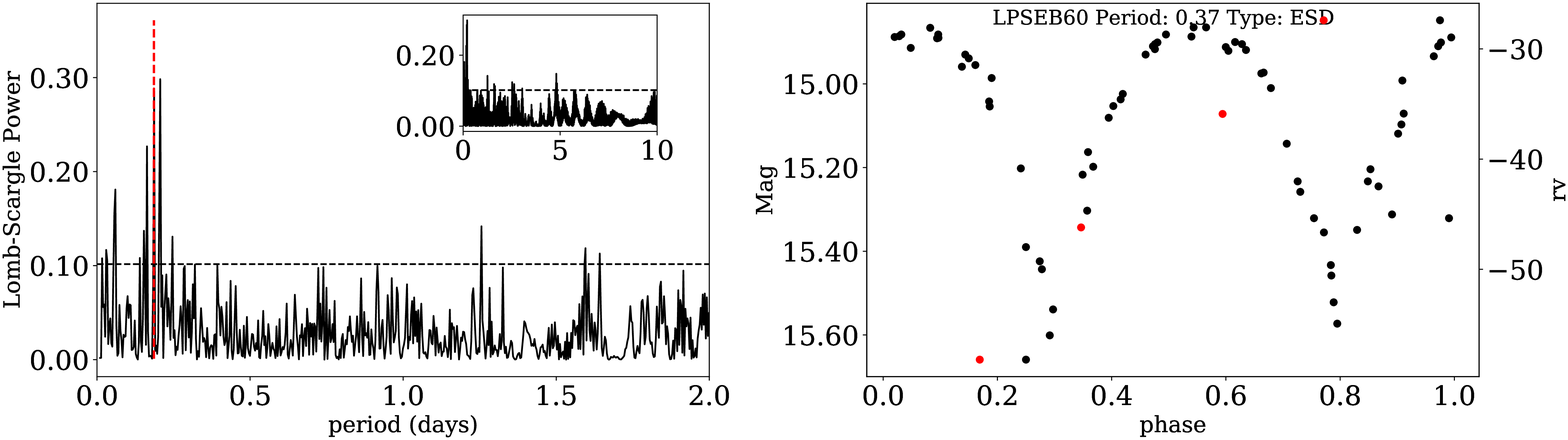}
    
\caption{(Continued) }
  \label{full}
\end{figure*}

\newpage
\addtocounter{figure}{-1}
\begin{figure*}[!htb]%
  
  \centering
 \includegraphics[width=4.5in]{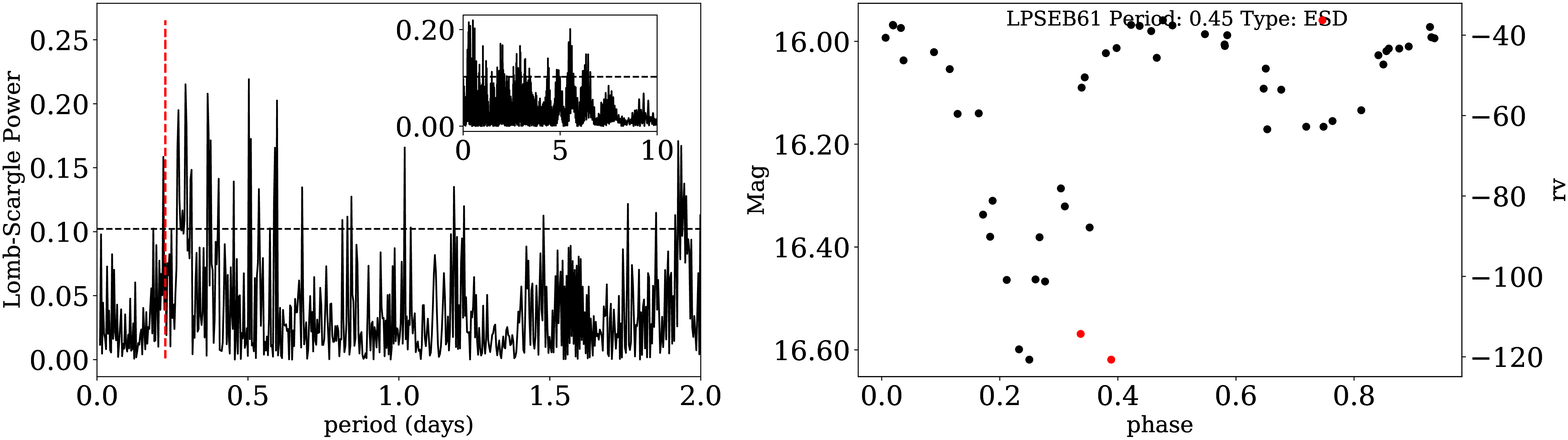}
   \includegraphics[width=4.5in]{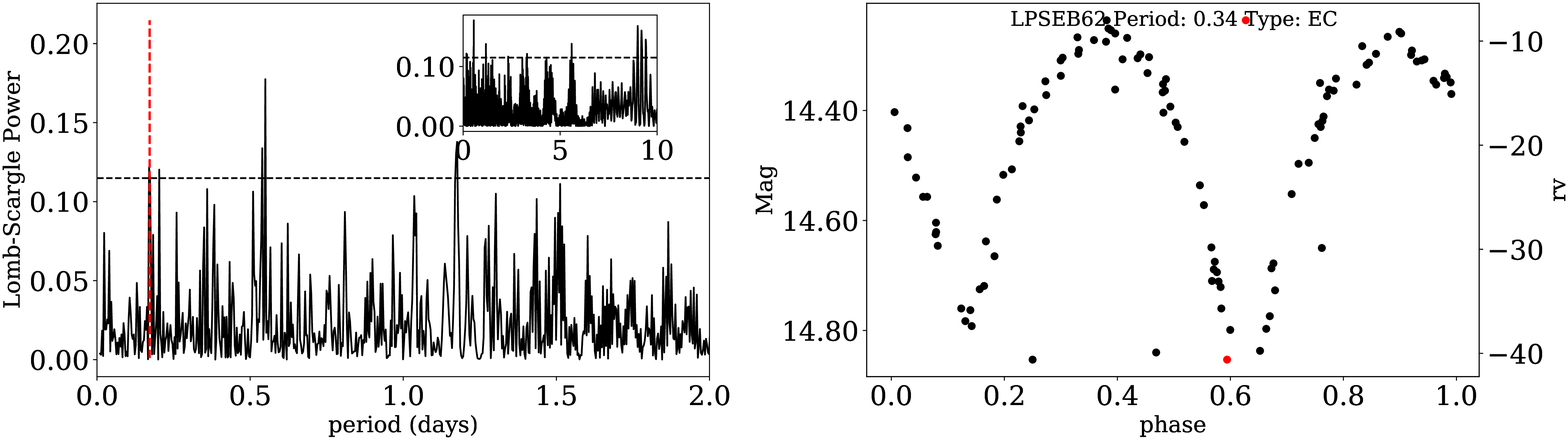}
  \includegraphics[width=4.5in]{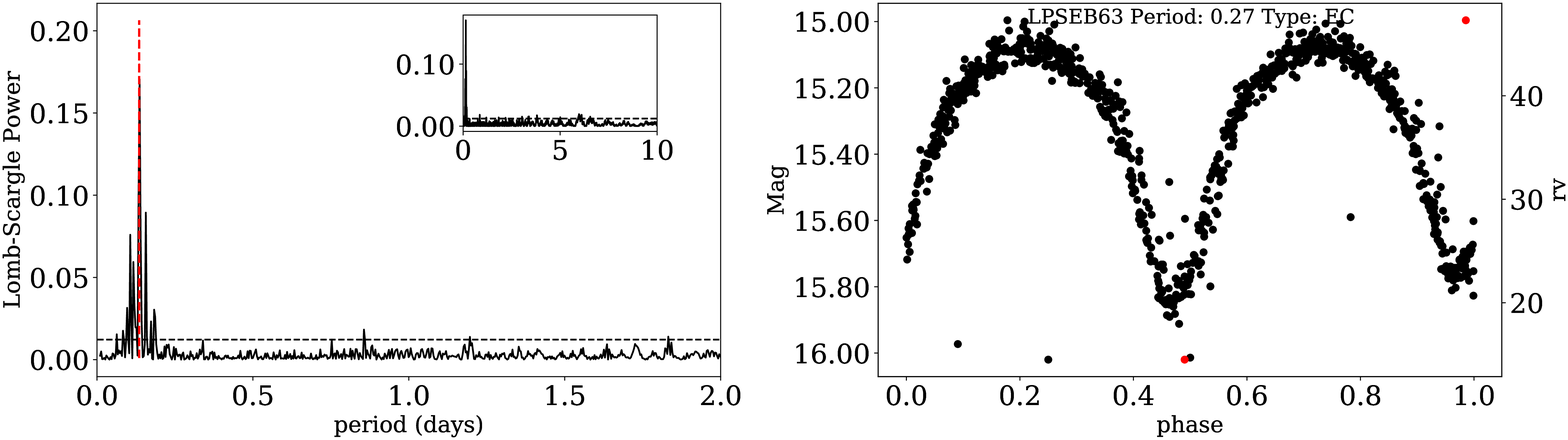}
  \includegraphics[width=4.5in]{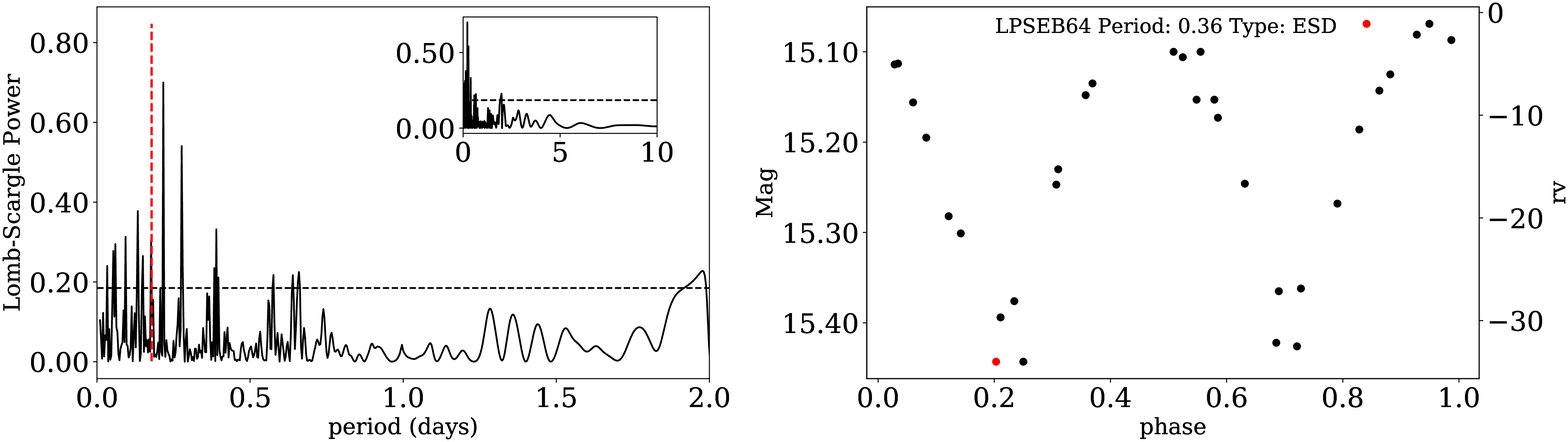}
  \includegraphics[width=4.5in]{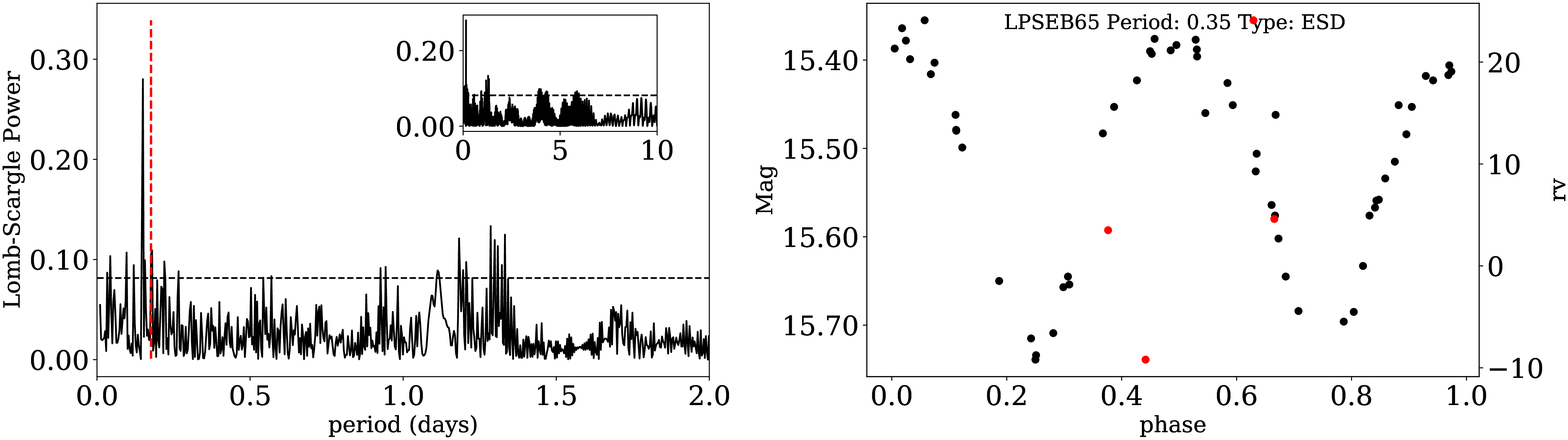}
  \includegraphics[width=4.5in]{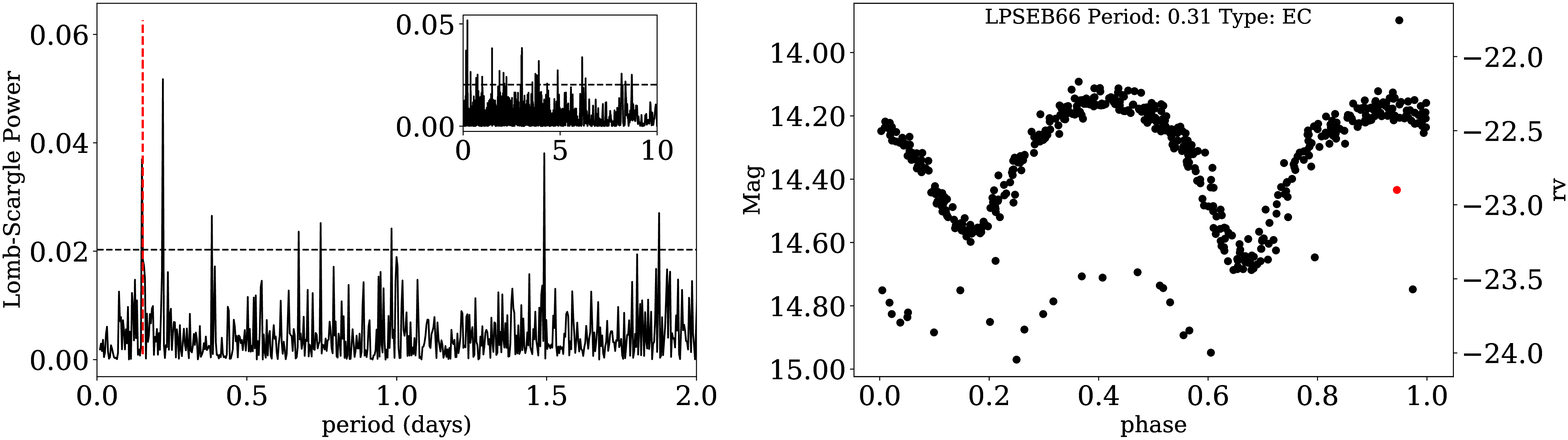}
    
\caption{(Continued) }
  \label{full}
\end{figure*}

\newpage
\addtocounter{figure}{-1}
\begin{figure*}[!htb]%
  
  \centering
 \includegraphics[width=4.5in]{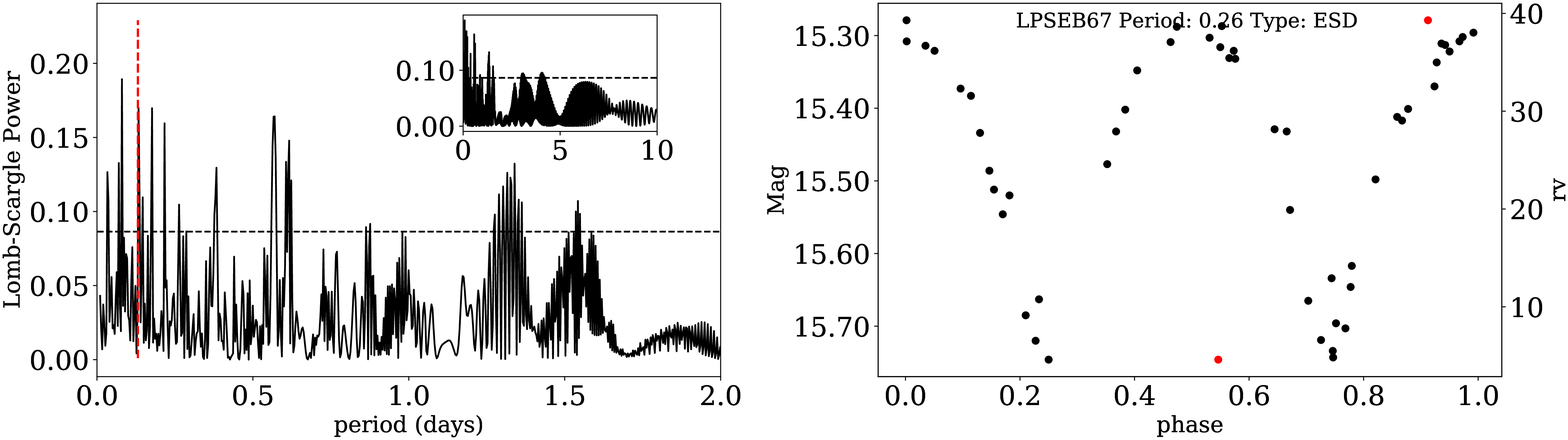}
   \includegraphics[width=4.5in]{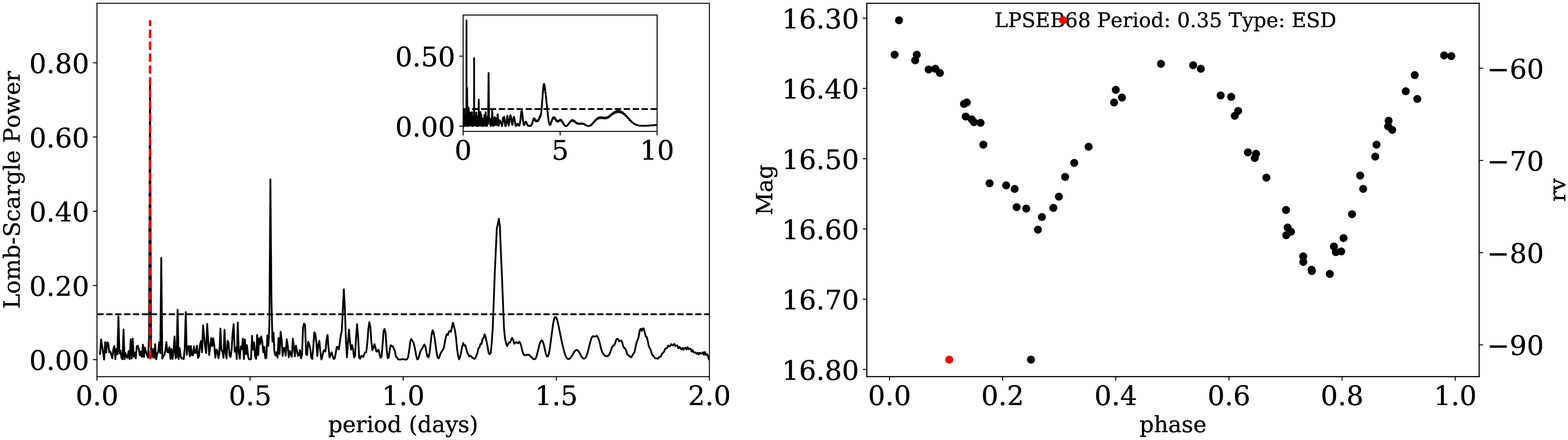}
  \includegraphics[width=4.5in]{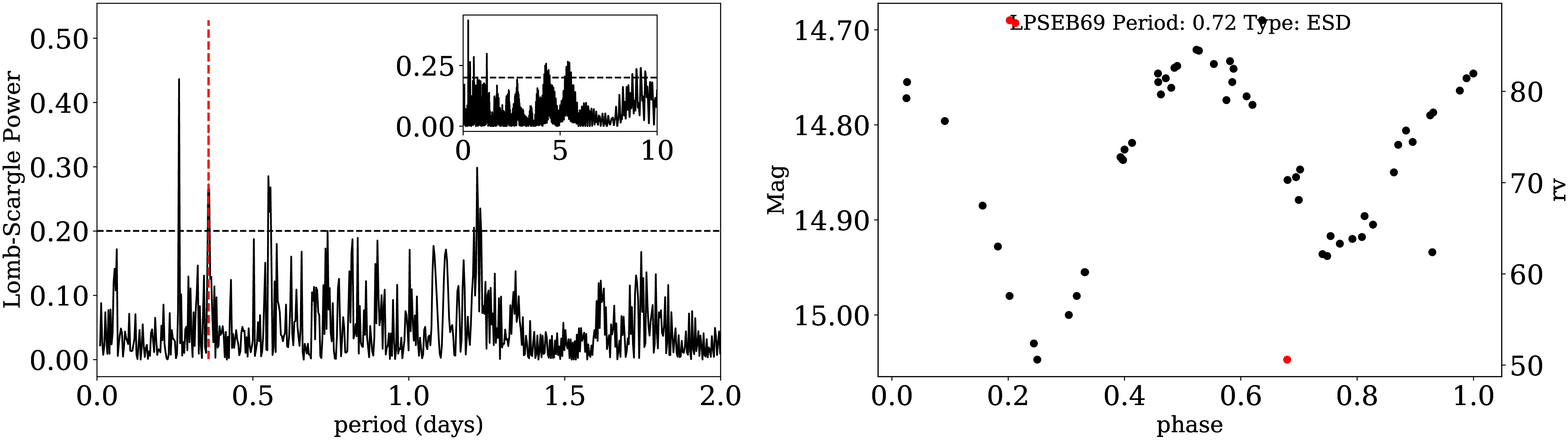}
  \includegraphics[width=4.5in]{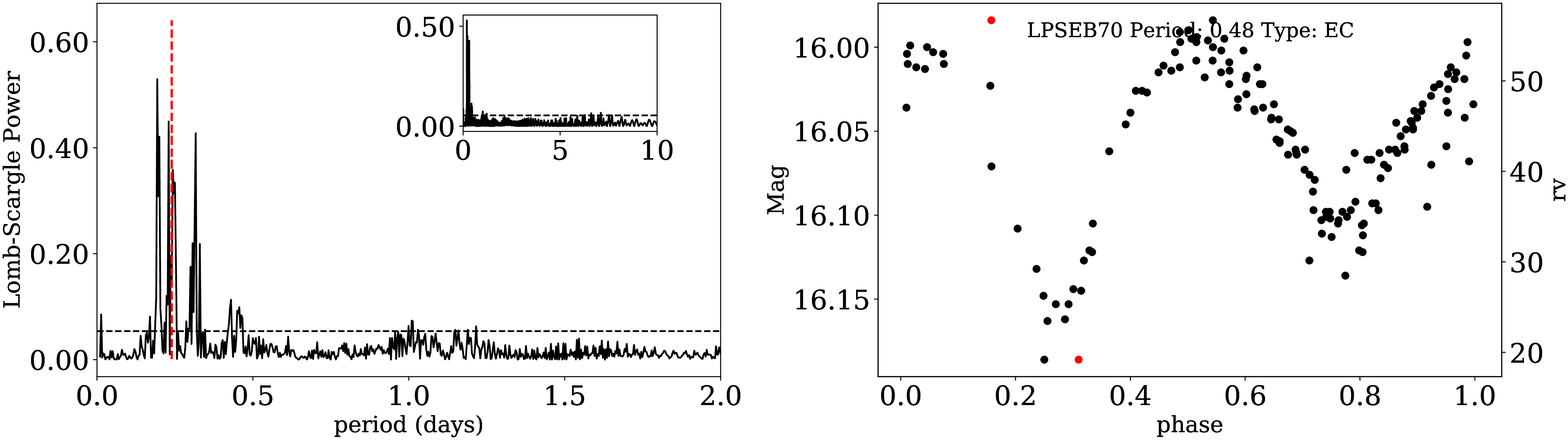}
  \includegraphics[width=4.5in]{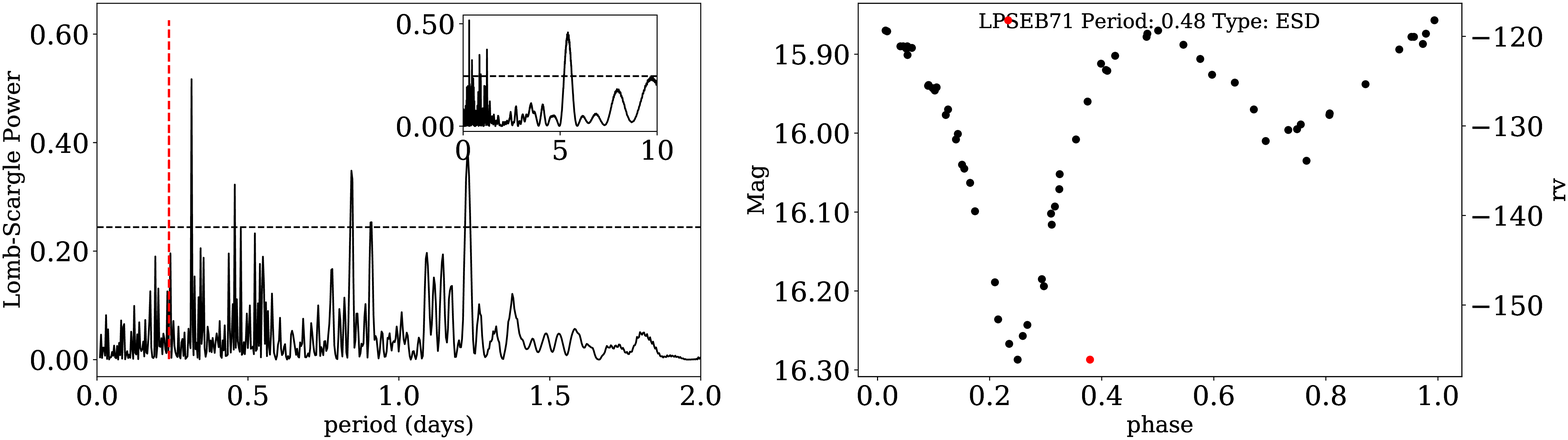}
  \includegraphics[width=4.5in]{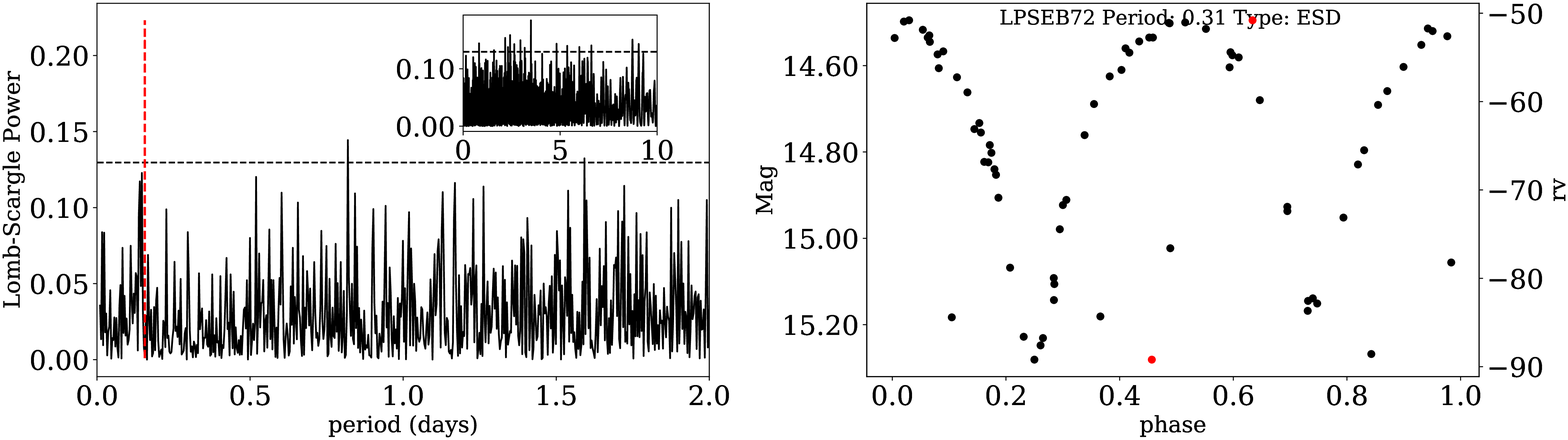}
    
\caption{(Continued) }
  \label{fig1}
\end{figure*}

\newpage
\addtocounter{figure}{-1}
\begin{figure*}[!htb]%
  
  \centering
 
   \includegraphics[width=4.5in]{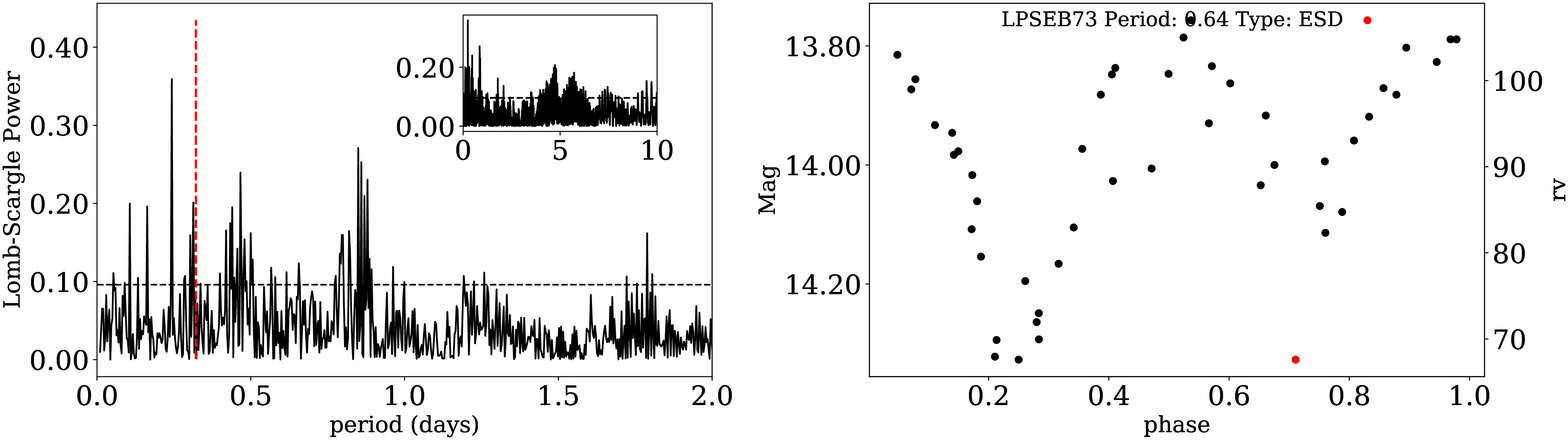}
  \includegraphics[width=4.5in]{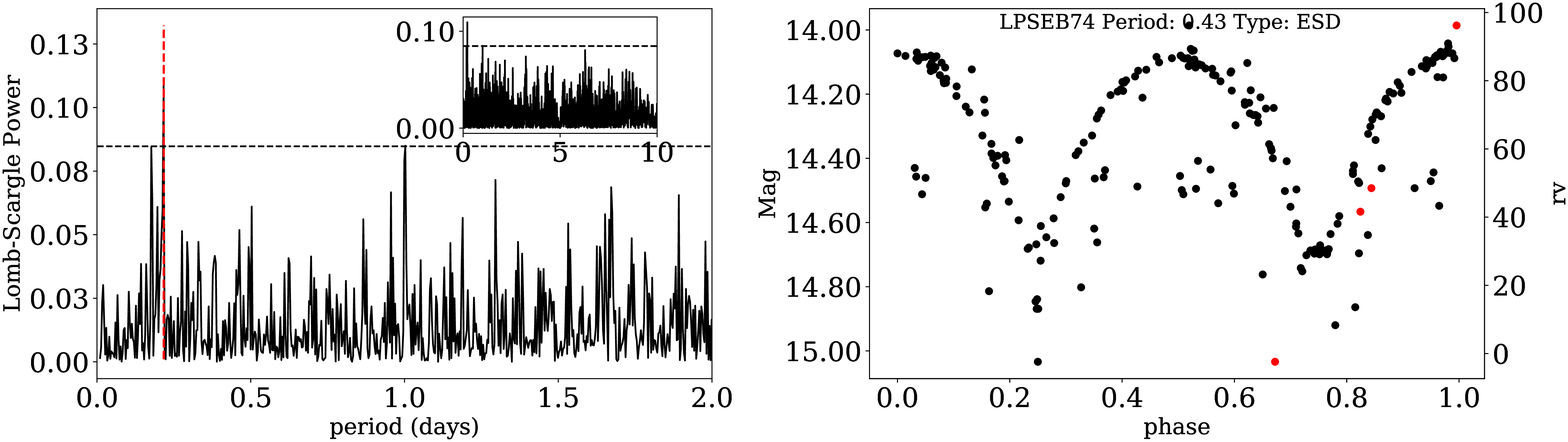}
  \includegraphics[width=4.5in]{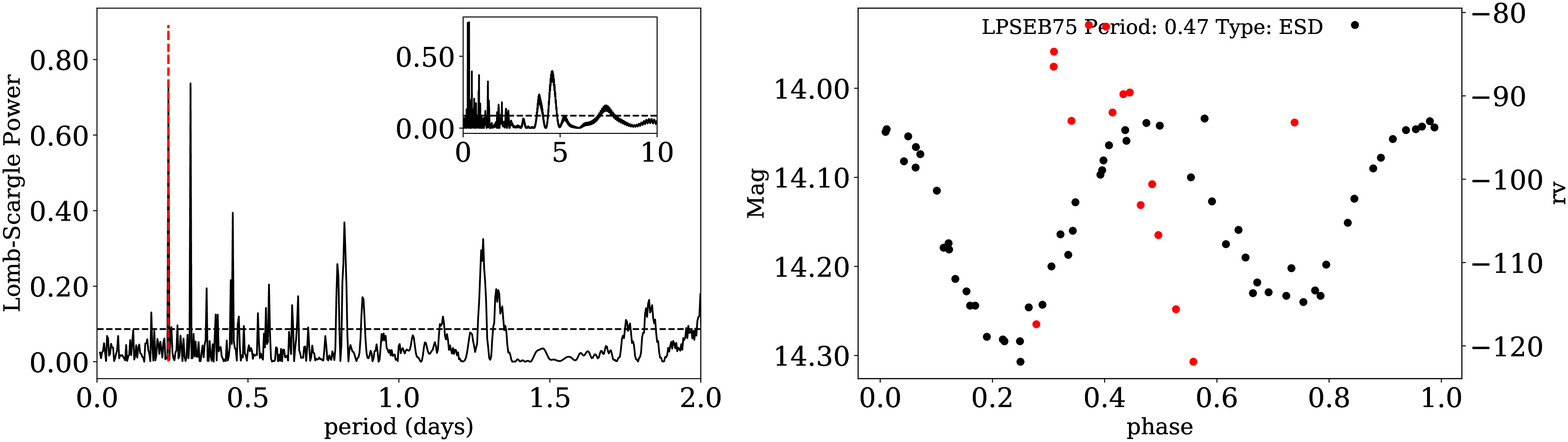}
  \includegraphics[width=4.5in]{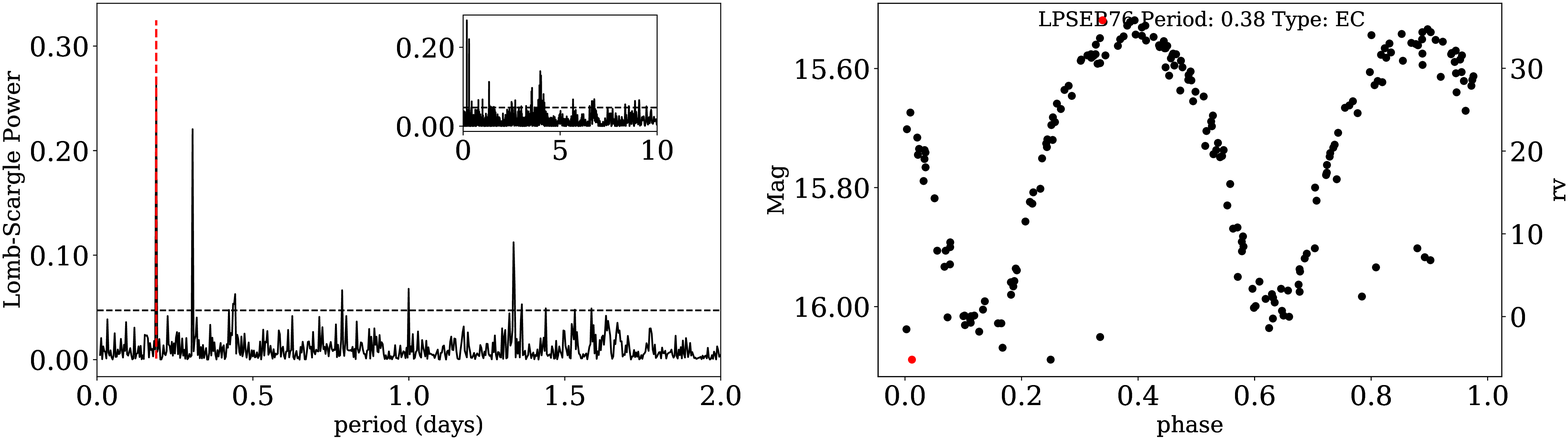}
  \includegraphics[width=4.5in]{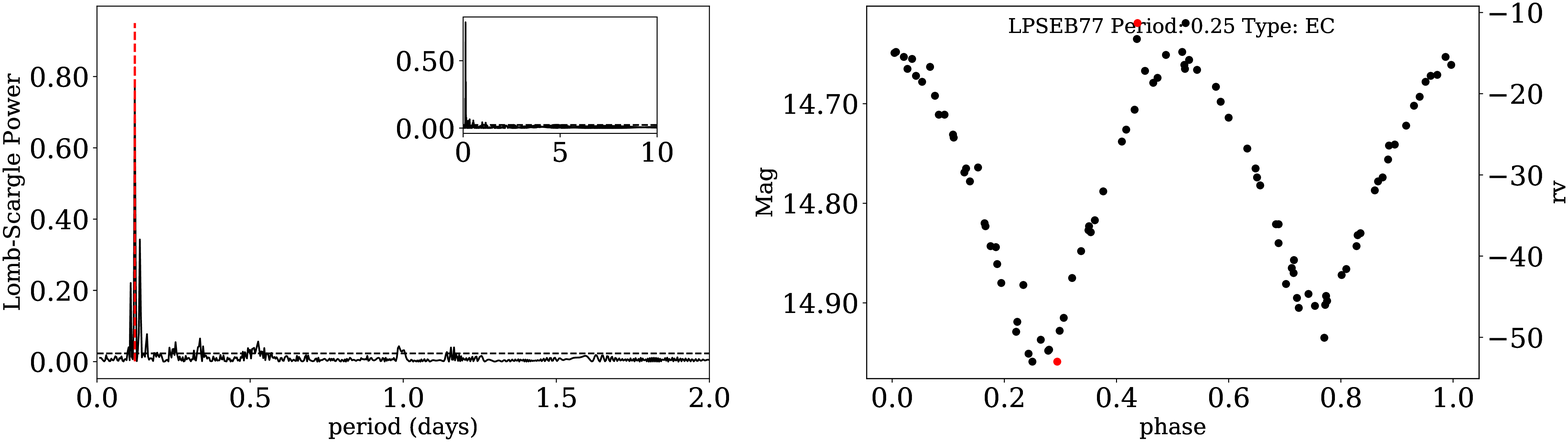}
  \includegraphics[width=4.5in]{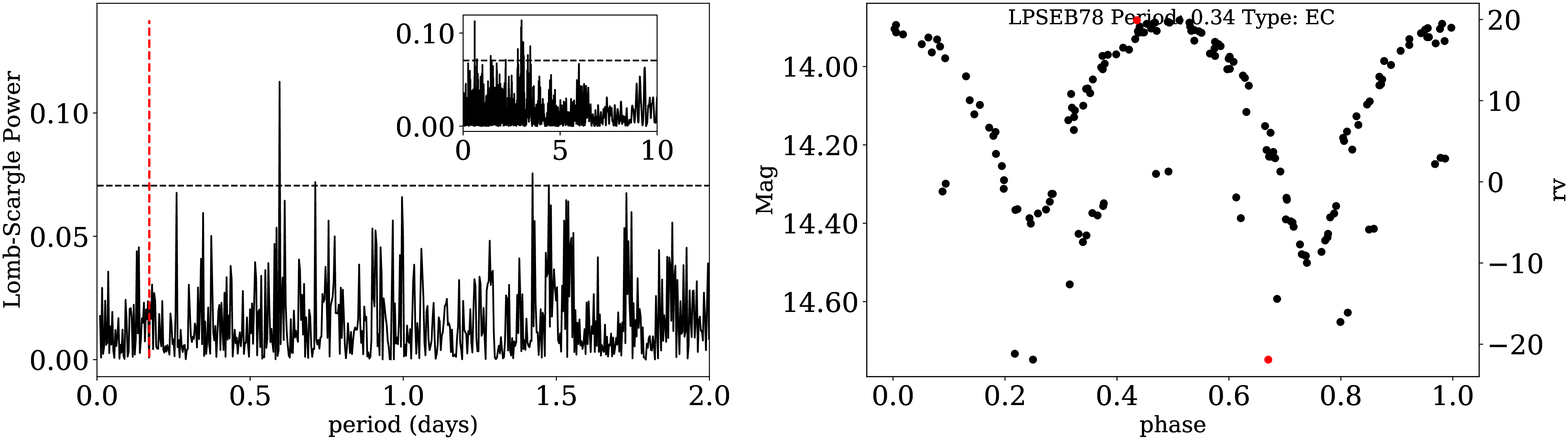}
\caption{(Continued) }
  \label{full}
\end{figure*}

\newpage
\addtocounter{figure}{-1}
\begin{figure*}[!htb]%
  
  \centering
 
   \includegraphics[width=4.5in]{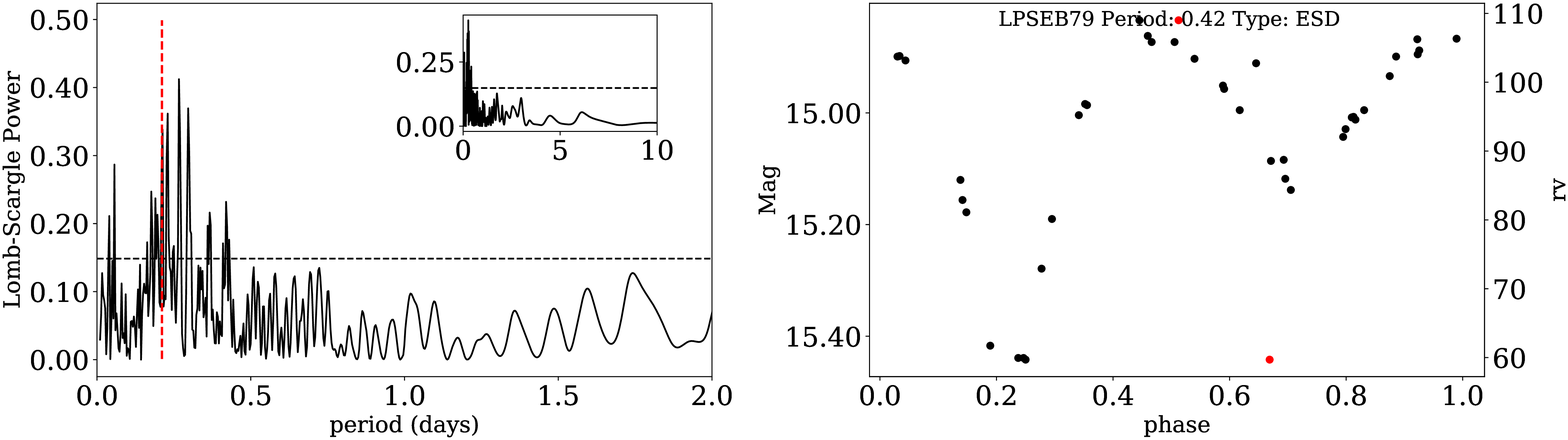}
  \includegraphics[width=4.5in]{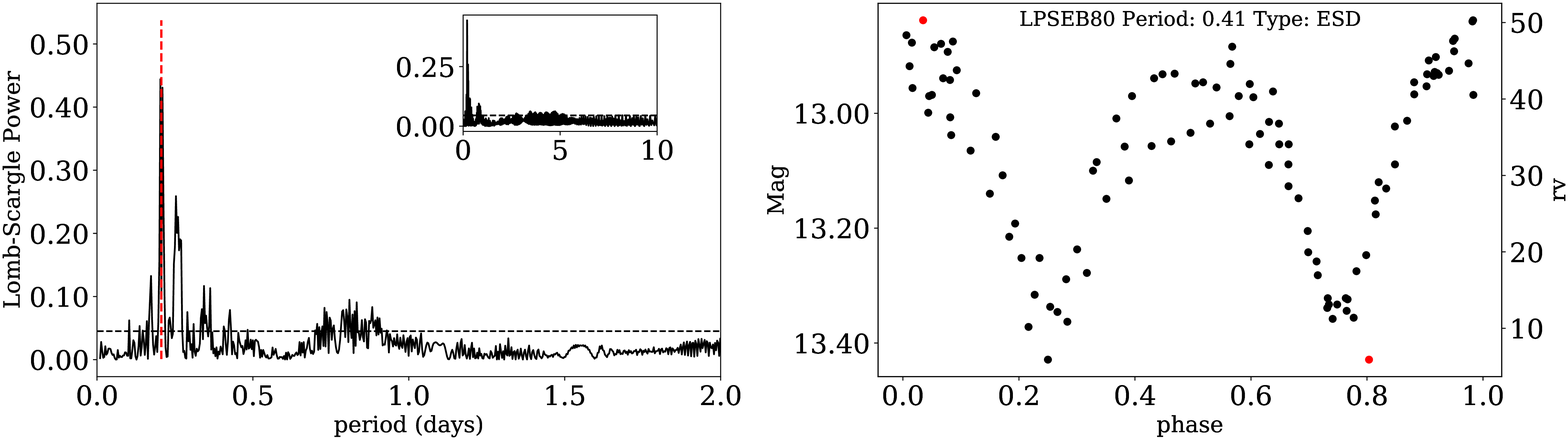}
  \includegraphics[width=4.5in]{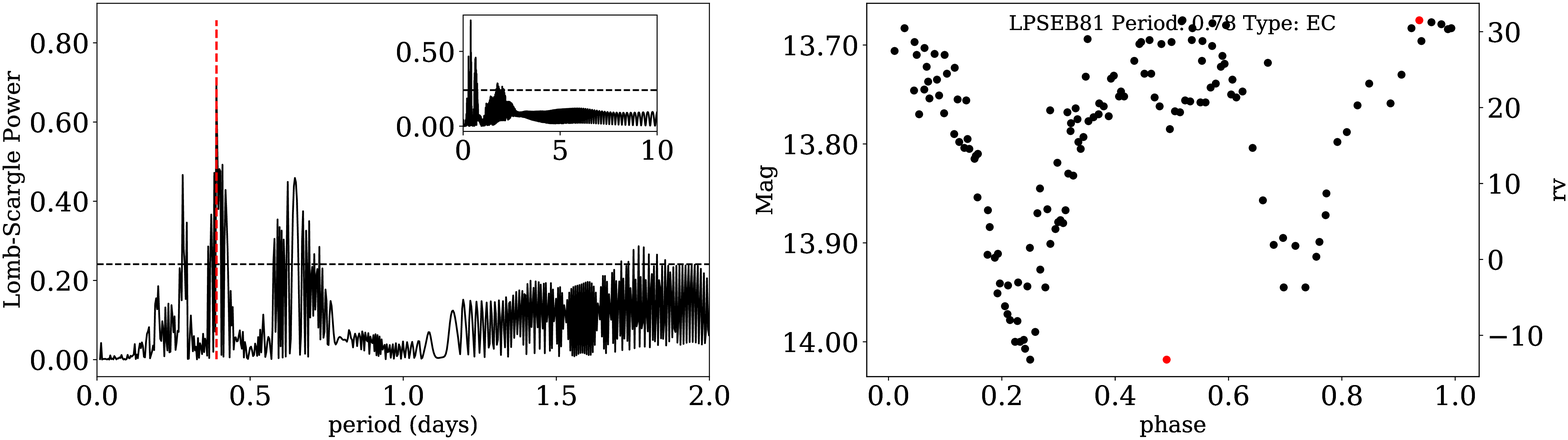}
  \includegraphics[width=4.5in]{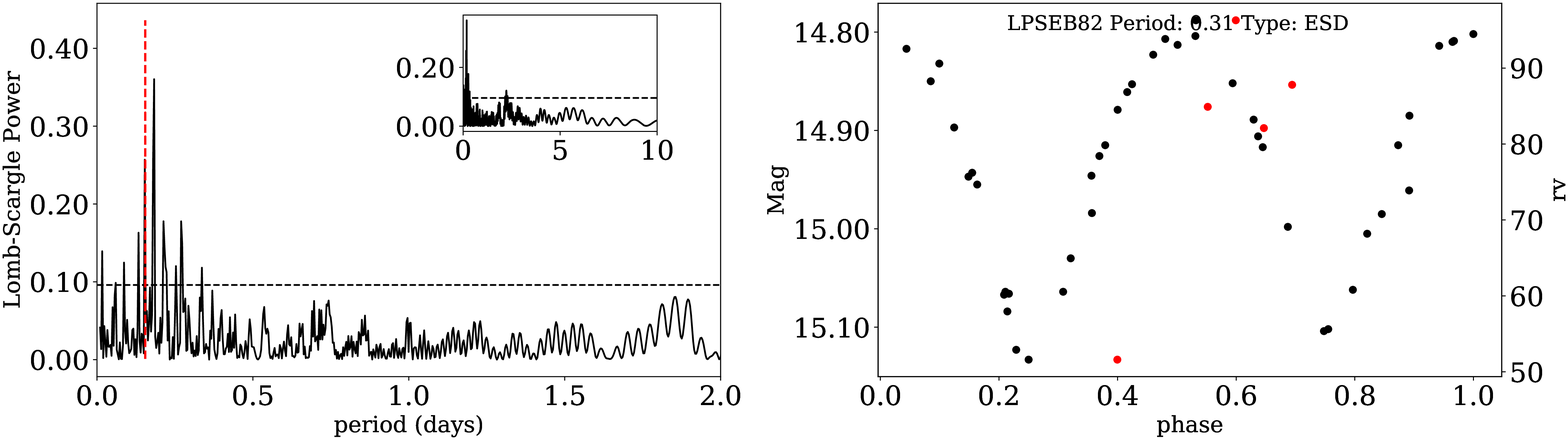}
  \includegraphics[width=4.5in]{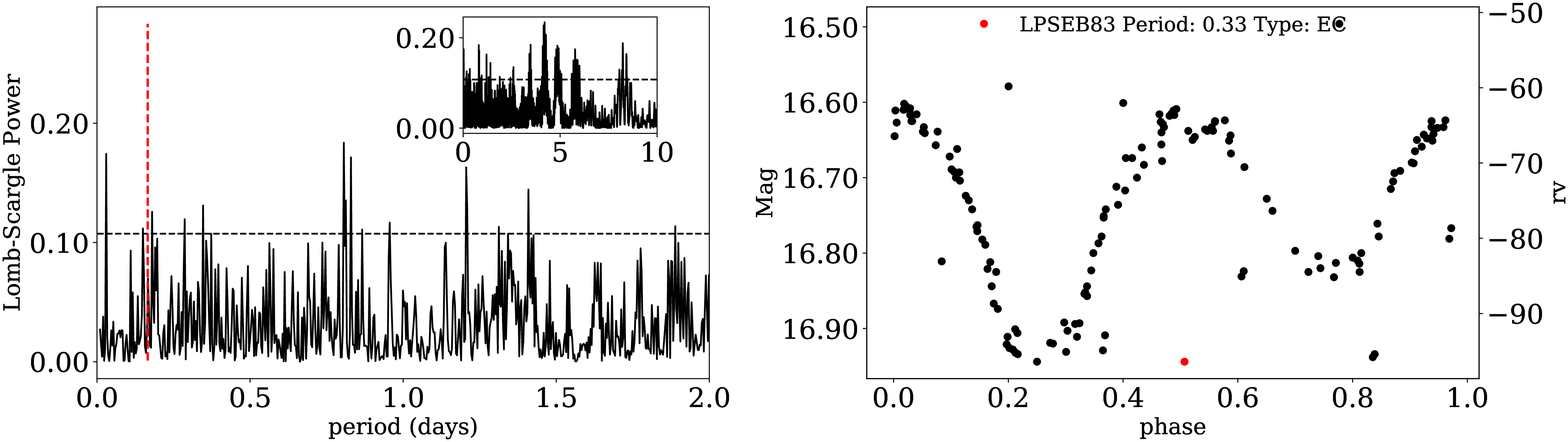}
  \includegraphics[width=4.5in]{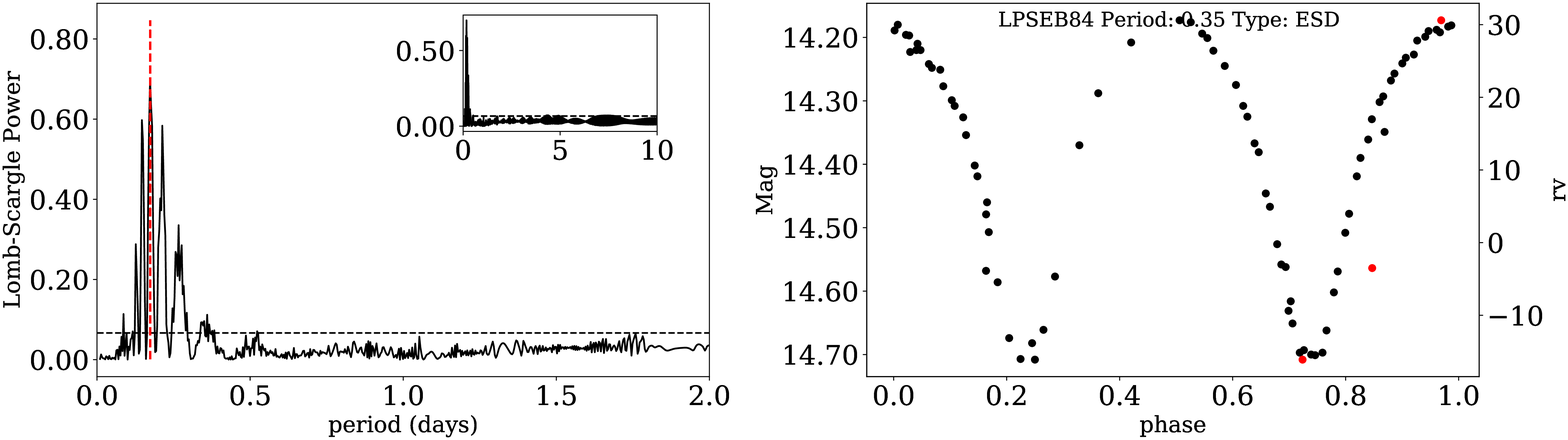}
\caption{(Continued) }
  \label{full}
\end{figure*}

\newpage
\addtocounter{figure}{-1}
\begin{figure*}[!htb]%
  
  \centering

  \includegraphics[width=4.5in]{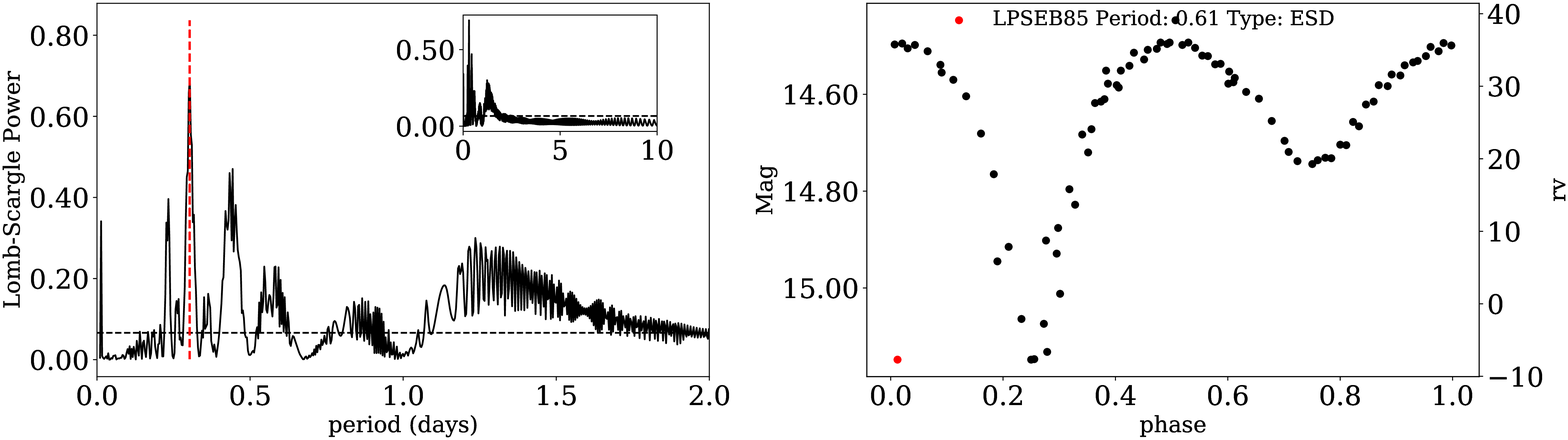}
  \includegraphics[width=4.5in]{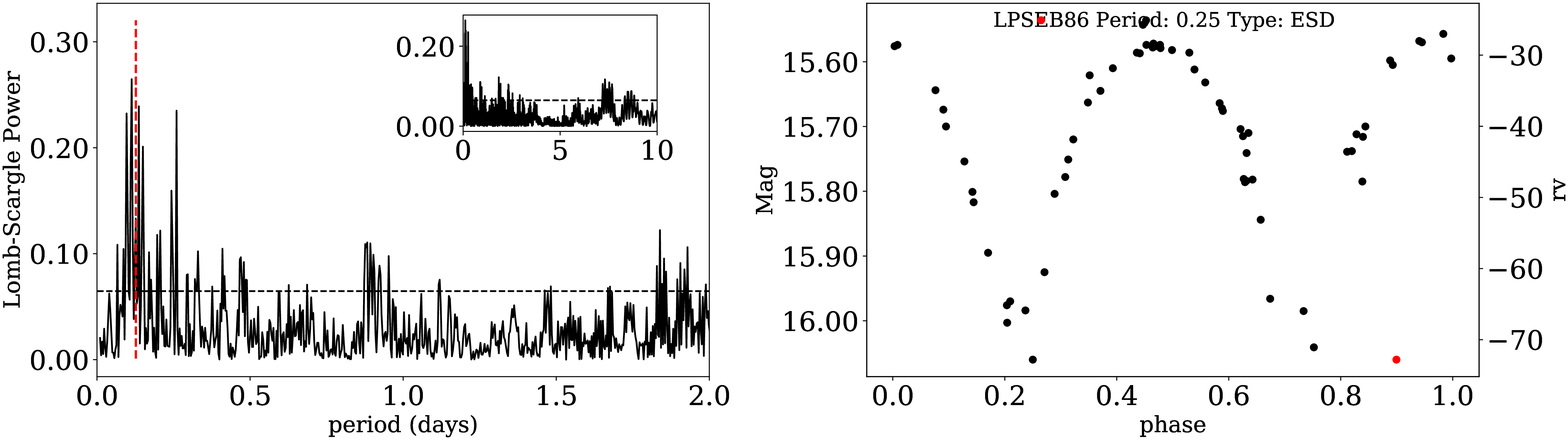}
  \includegraphics[width=4.5in]{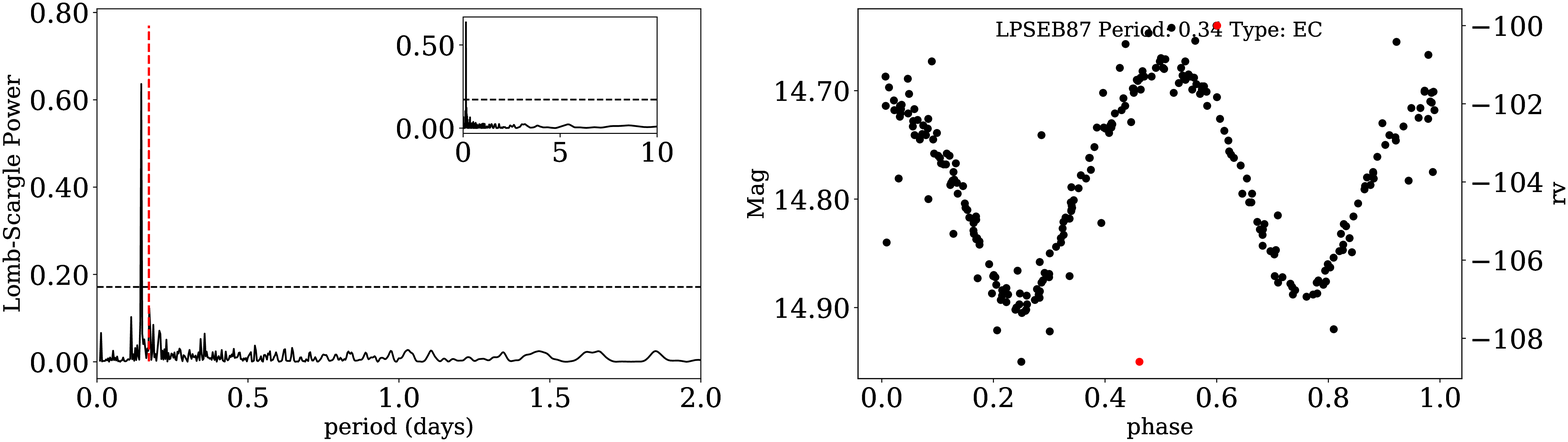}
  \includegraphics[width=4.5in]{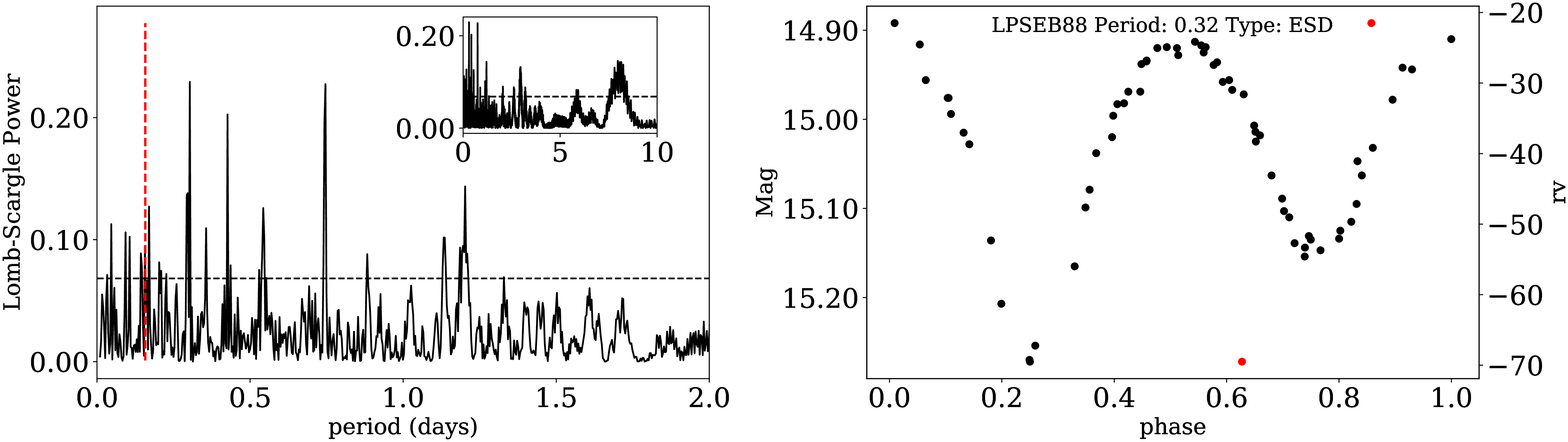}
\caption{(Continued) }
  \label{full}
\end{figure*}

\begin{acknowledgements}
We wish to thank the referee for their most useful comments which have greatly improved the paper. This work made use of astroML\footnote{\url{https:http://www.astroml.org/}} and IPAC database for PTF\footnote{\url{https://irsa.ipac.caltech.edu/cgi-bin/Gator/nph-dd}}. We acknowledge the use of LAMOST data and catalog. We would like to thank Lin He and Chang-Qing Luo for the useful discussion. We also thank Weiming Gu for the feedback on the work. Fan Yang and Ji-Feng Liu acknowledge funding from National Natural Science Foundation of China (NSFC.11988101), National Science Fund for Distinguished Young Scholars (No.11425313) and National Key Research and Development Program of China (No.2016YFA0400800).

\end{acknowledgements}

\clearpage
\newpage
\bibliographystyle{unsrt}
\bibliographystyle{aasjournal}

\begin{thebibliography}{dummy}

\bibitem[Abt \& Levy(1975)]{1975BAAS....7..268A} Abt, H.~A., \& Levy, S.~G.\ 1975, \baas, 7, 268
\bibitem[Abt \& Willmarth(2006)]{2006ApJS..162..207A} Abt, H.~A., \& Willmarth, D.\ 2006, \apjs, 162, 207
\bibitem[Andersen(1991)]{1991A&ARv...3...91A} Andersen, J.\ 1991, \aapr, 3, 91
\bibitem[Alcock et al.(1997)]{1997ApJ...486..697A} Alcock, C., Allsman, R.~A., Alves, D., et al.\ 1997, \apj, 486, 697
\bibitem[Ashcraft et al.(2012)]{2012MNRAS.424..620A} Ashcraft, T.~A., Hynes, R.~I., \& Robinson, E.~L.\ 2012, \mnras, 424, 620
\bibitem[Baade(1946)]{1946PASP...58..249B} Baade, W.\ 1946, \pasp, 58, 249 
\bibitem[Bonanos et al.(2006)]{2006ApJ...652..313B} Bonanos, A.~Z., Stanek, K.~Z., Kudritzki, R.~P., et al.\ 2006, \apj, 652, 313
\bibitem[Casares et al.(2014)]{2014Natur.505..378C} Casares, J., Negueruela, I., Rib{\'o}, M., et al.\ 2014, \nat, 505, 378 
\bibitem[Chabrier et al.(2007)]{2007A&A...472L..17C} Chabrier, G., Gallardo, J., \& Baraffe, I.\ 2007, \aap, 472, L17
\bibitem[Campbell \& Curtis(1905)]{1905LicOB...3..136C} Campbell, W.~W., \& Curtis, H.~D.\ 1905, Lick Observatory Bulletin, 3, 136
\bibitem[Claret et al.(1993)]{1993AdSpR..13..735C} Claret, A., Ballet, J., Goldwurm, A., et al.\ 1993, Advances in Space Research, 13, 735
\bibitem[Cui et al.(2012)]{2012RAA....12.1197C} Cui, X.-Q., Zhao, Y.-H., Chu, Y.-Q., et al.\ 2012, Research in Astronomy and Astrophysics, 12, 1197 
\bibitem[Daniel et al.(1967)]{1967Natur.213...21D} Daniel, R.~R., Joseph, G., Lavakare, P.~J., \& Sunderrajan, R.\ 1967, \nat, 213, 21
\bibitem[Deng et al.(2012)]{Deng2012} Deng, L.-C., Newberg, H.~J., Liu, C., et al.\ 2012, Research in Astronomy and Astrophysics, 12, 735  
\bibitem[Duquennoy \& Mayor(1991)]{1991A&A...248..485D} Duquennoy, A., \& Mayor, M.\ 1991, \aap, 248, 485
\bibitem[Ferwerda(1943)]{1943BAN.....9..337F} Ferwerda, J.~G.\ 1943, \bain, 9, 337 
\bibitem[Fu et al.(2008)]{2008AJ....135.1958F} Fu, J.~N., Khokhuntod, P., Rodr{\'{\i}}guez, E., et al.\ 2008, \aj, 135, 1958
\bibitem[Gao et al.(2014)]{2014ApJ...788L..37G} Gao, S., Liu, C., Zhang, X., et al.\ 2014, \apjl, 788, L37
\bibitem[Gao et al.(2016)]{2016ApJS..224...37G} Gao, Q., Xin, Y., Liu, J.-F., Zhang, X.-B., \& Gao, S.\ 2016, \apjs, 224, 37  
\bibitem[Guinan et al.(1998)]{1998ApJ...509L..21G} Guinan, E.~F., Fitzpatrick, E.~L., DeWarf, L.~E., et al.\ 1998, \apjl, 509, L21
\bibitem[Guinan \& Engle(2006)]{2006Ap&SS.304....5G} Guinan, E.~F., \& Engle, S.~G.\ 2006, \apss, 304, 5  
\bibitem[Grison et al.(1995)]{1995A&AS..109..447G} Grison, P., Beaulieu, J.-P., Pritchard, J.~D., et al.\ 1995, \aaps, 109, 447 
\bibitem[Herschel(1802)]{1802RSPT...92..477H} Herschel, W.\ 1802, Philosophical Transactions of the Royal Society of London Series I, 92, 477
  \bibitem[Hurley et al.(2002)]{2002MNRAS.329..897H} Hurley, J.~R., Tout, C.~A., \& Pols, O.~R.\ 2002, \mnras, 329, 897

\bibitem[Ivezi{\'c} et al.(2007)]{2007AJ....134..973I} Ivezi{\'c}, {\v Z}., Smith, J.~A., Miknaitis, G., et al.\ 2007, \aj, 134, 973
\bibitem[Jiang et al.(2012)]{2012MNRAS.421.2769J} Jiang, D., Han, Z., Ge, H., Yang, L., \& Li, L.\ 2012, \mnras, 421, 2769
\bibitem[Jackson et al.(2012)]{2012ApJ...751..112J} Jackson, B.~K., Lewis, N.~K., Barnes, J.~W., et al.\ 2012, \apj, 751, 112
\bibitem[Kallrath, \& Milone(2009)]{book2009} Kallrath, J., \& Milone, E.~F.\ 2009, Eclipsing Binary Stars: Modeling and Analysis: Astronomy and Astrophysics Library. ISBN 978-1-4419-0698-4. Springer-Verlag New York
  
\bibitem[Kao et al.(2016)]{2016MNRAS.461.2747K} Kao, W., Kaplan, D.~L., Prince, T.~A., et al.\ 2016, \mnras, 461, 2747 
\bibitem[Law et al.(2009)]{2009PASP..121.1395L} Law, N.~M., Kulkarni, S.~R., Dekany, R.~G., et al.\ 2009, \pasp, 121, 1395
\bibitem[Liu et al.(2013)]{2013Natur.503..500L} Liu, J.-F., Bregman, J.~N., Bai, Y., Justham, S., \& Crowther, P.\ 2013, \nat, 503, 500
\bibitem[Liu et al.(2015)]{2015Natur.528..108L} Liu, J.-F., Bai, Y., Wang, S., et al.\ 2015, \nat, 528, 108 
\bibitem[Lomb(1976)]{1976Ap&SS..39..447L} Lomb, N.~R.\ 1976, \apss, 39, 447
  \bibitem[Matijevi{\v{c}} et al.(2011)]{2011AJ....141..200M} Matijevi{\v{c}}, G., Zwitter, T., Bienaym{\'e}, O., et al.\ 2011, \aj, 141, 200
\bibitem[Matijevi{\v{c}} et al.(2010)]{2010AJ....140..184M} Matijevi{\v{c}}, G., Zwitter, T., Munari, U., et al.\ 2010, \aj, 140, 184
\bibitem[Mazeh \& Faigler(2010)]{2010A&A...521L..59M} Mazeh, T., \& Faigler, S.\ 2010, \aap, 521, L59
\bibitem[Milone et al.(2008)]{2008ASSL..352.....M} Milone, E.~F., Leahy, D.~A., \& Hobill, D.~W.\ 2008, Astrophysics and Space Science Library
 \bibitem[Matijevi{\v{c}} et al.(2012)]{2012AJ....143..123M} Matijevi{\v{c}}, G., Pr{\v{s}}a, A., Orosz, J.~A., et al.\ 2012, \aj, 143, 123
\bibitem[Otero et al.(2004)]{2004IBVS.5570....1O} Otero, S.~A., Wils, P., \& Dubovsky, P.~A.\ 2004, Information Bulletin on Variable Stars, 5570, 1
\bibitem[Paczynski \& Sasselov(1997)]{1997vsar.conf..309P} Paczynski, B., \& Sasselov, D.\ 1997, Variables Stars and the Astrophysical Returns of the Microlensing Surveys, 309 
\bibitem[Pourbaix et al.(2004)]{2004A&A...424..727P} Pourbaix, D., Tokovinin, A.~A., Batten, A.~H., et al.\ 2004, \aap, 424, 727 
\bibitem[Pr{\v s}a \& Zwitter(2005)]{2005ApJ...628..426P} Pr{\v s}a, A., \& Zwitter, T.\ 2005, \apj, 628, 426
\bibitem[Pr{\v s}a et al.(2011)]{2011AJ....141...83P} Pr{\v s}a, A., Batalha, N., Slawson, R.~W., et al.\ 2011, \aj, 141, 83 
\bibitem[Pr{\v s}a et al.(2016)]{2016ApJS..227...29P} Pr{\v s}a, A., Conroy, K.~E., Horvat, M., et al.\ 2016, \apjs, 227, 29 
\bibitem[Paczynski \& Pojmanski(2000)]{2000AAS...196.1001P} Paczynski, B., \& Pojmanski, G.\ 2000, Bulletin of the American Astronomical Society, 32, 10.01
\bibitem[Paczy{\'n}ski et al.(2006)]{2006MNRAS.368.1311P} Paczy{\'n}ski, B., Szczygie{\l}, D.~M., Pilecki, B., et al.\ 2006, \mnras, 368, 1311

\bibitem[Qian(2003)]{2003MNRAS.342.1260Q} Qian, S.\ 2003, \mnras, 342, 1260 
\bibitem[Raghavan et al.(2010)]{2010ApJS..190....1R} Raghavan, D., McAlister, H.~A., Henry, T.~J., et al.\ 2010, \apjs, 190, 1 
\bibitem[Rau et al.(2009)]{2009PASP..121.1334R} Rau, A., Kulkarni, S.~R., Law, N.~M., et al.\ 2009, \pasp, 121, 1334 
\bibitem[Remillard \& McClintock(2006)]{2006ARA&A..44...49R} Remillard, R.~A., \& McClintock, J.~E.\ 2006, \araa, 44, 49
  \bibitem[Ren et al.(2014)]{2014A&A...570A.107R} Ren, J.~J., Rebassa-Mansergas, A., Luo, A.~L., et al.\ 2014, \aap, 570, A107
  \bibitem[Rucinski(1992)]{1992AJ....103..960R} Rucinski, S.~M.\ 1992, \aj, 103, 960
  \bibitem[Rucinski(1993)]{1993PASP..105.1433R} Rucinski, S.~M.\ 1993, \pasp, 105, 1433
  \bibitem[Rucinski(1997)]{1997AJ....113..407R} Rucinski, S.~M.\ 1997, \aj, 113, 407
\bibitem[Samus' et al.(2017)]{2017ARep...61...80S} Samus', N.~N., Kazarovets, E.~V., Durlevich, O.~V., et al.\ 2017, Astronomy Reports, 61, 80
\bibitem[Slawson et al.(2011)]{2011AJ....142..160S} Slawson, R.~W., Pr{\v s}a, A., Welsh, W.~F., et al.\ 2011, \aj, 142, 160 
\bibitem[Soszy{\'n}ski et al.(2015)]{2015AcA....65...39S} Soszy{\'n}ski, I., St{\c e}pie{\'n}, K., Pilecki, B., et al.\ 2015, \actaa, 65, 39
\bibitem[Soszy{\'n}ski et al.(2016)]{2016AcA....66..405S} Soszy{\'n}ski, I., Pawlak, M., Pietrukowicz, P., et al.\ 2016, \actaa, 66, 405
\bibitem[Stepien(2006)]{2006AcA....56..347S} Stepien, K.\ 2006, \actaa, 56, 347 
\bibitem[Swope \& Shapley(1938)]{1938BHarO.909...14S} Swope, H.~H., \& Shapley, H.\ 1938, Harvard College Observatory Bulletin, 909, 14 
\bibitem[Torres \& Ribas(2002)]{2002ApJ...567.1140T} Torres, G., \& Ribas, I.\ 2002, \apj, 567, 1140 
\bibitem[Udalski et al.(1998)]{1998AcA....48..563U} Udalski, A., Soszynski,., Szymanski, M., et al.\ 1998, \actaa, 48, 563
\bibitem[VanderPlas et al.(2012)]{astroML} VanderPlas, J., Connolly, A.~J., Ivezic, Z., et al.\ 2012, Proceedings of Conference on Intelligent Data Understanding (CIDU, 47
\bibitem[VanderPlas \& Ivezi{\'c}(2015)]{2015ApJ...812...18V} VanderPlas, J.~T., \& Ivezi{\'c}, {\v Z}.\ 2015, \apj, 812, 18
\bibitem[Vogel(1890)]{1890AN....123..289V} Vogel, H.~C.\ 1890, Astronomische Nachrichten, 123, 289
\bibitem[Wilson \& Devinney(1971)]{1971ApJ...166..605W} Wilson, R.~E., \& Devinney, E.~J.\ 1971, \apj, 166, 605 
\bibitem[Wilson(2012)]{2012AJ....144...73W} Wilson, R.~E.\ 2012, \aj, 144, 73 
\bibitem[Wyithe \& Wilson(2001)]{2001ApJ...559..260W} Wyithe, J.~S.~B., \& Wilson, R.~E.\ 2001, \apj, 559, 260
\bibitem[Wyrzykowski et al.(2004)]{2004AcA....54....1W} Wyrzykowski, L., Udalski, A., Kubiak, M., et al.\ 2004, \actaa, 54, 1
\bibitem[Xiang et al.(2015)]{2015MNRAS.448..822X} Xiang, M.~S., Liu, X.~W., Yuan, H.~B., et al.\ 2015, \mnras, 448, 822  
\bibitem[Xiang et al.(2017)]{2017MNRAS.467.1890X} Xiang, M.-S., Liu, X.-W., Yuan, H.-B., et al.\ 2017, \mnras, 467, 1890
  \bibitem[Yang et al.(2005)]{2005PASJ...57..983Y} Yang, Y.-G., Qian, S.-B., Zhu, L.-Y., et al.\ 2005, \pasj, 57, 983
\bibitem[Zhao et al.(2012)]{2012RAA....12..723Z} Zhao, G., Zhao, Y.-H., Chu, Y.-Q., Jing, Y.-P., \& Deng, L.-C.\ 2012, Research in Astronomy and Astrophysics, 12, 723
\bibitem[Zhang et al.(2015)]{2015AJ....150...37Z} Zhang, X.~B., Luo, Y.~P., Wang, K., \& Luo, C.~Q.\ 2015, \aj, 150, 37
\bibitem[Zhang et al.(2019)]{zhang2019} Zhang, B., Liu, C., Li, C.-Q., et al.\ 2019, arXiv e-prints, arXiv:1910.13154
\bibitem[Zhu et al.(2014)]{2014AJ....147...42Z} Zhu, L.~Y., Qian, S.~B., Soonthornthum, B., et al.\ 2014, \aj, 147, 42
 
\end{thebibliography}

\label{lastpage}
\end{document}